\documentclass[aps,prd,showpacs,nofootinbib,floatfix,eqsecnum]{revtex4}
\usepackage{amsmath,amssymb,latexsym,graphicx,multirow,psfrag}

\newcommand{\Cs}{\mathcal{C}}
\newcommand{\Is}{\mathcal{I}}
\newcommand{\Js}{\mathcal{J}} 
\newcommand{\Ms}{\mathcal{M}}
\newcommand{\Ns}{\mathcal{N}}
\newcommand{\Rs}{\mathcal{R}}
\newcommand{\Ss}{\mathcal{S}}

\allowdisplaybreaks

\begin{document}
\title{Compact binary systems in scalar-tensor gravity. II. Tensor gravitational waves to second post-Newtonian order}
\author{Ryan N.\ Lang}
\email{lang@phys.ufl.edu}
\affiliation{Department of Physics, University of Florida, Gainesville, Florida 32611, USA}

\begin{abstract}
We derive the tensor gravitational waveform generated by a binary of nonspinning compact objects (black holes or neutron stars) in a general class of scalar-tensor theories of gravity.  The waveform is accurate to second post-Newtonian order beyond the leading order quadrupole approximation.  We use the direct integration of the relaxed Einstein equations formalism, appropriately adapted to scalar-tensor theories, along with previous results for the equations of motion in these theories.  The self-gravity of the compact objects is treated with an approach developed by Eardley.  The scalar field causes deviations from the general relativistic waveform that depend only on a small number of parameters.  Among the effects of the scalar field are new hereditary terms which depend on the past history of the source.  One of these, a dipole-dipole coupling, produces a zero-frequency ``gravitational-wave memory'' equivalent to the Christodoulou memory of general relativity.  In the special case of two black holes, the waveform reduces to the general relativistic waveform.  For a mixed (black hole-neutron star) system, the waveform is identical to that of Einstein's theory to first post-Newtonian order, with deviations at higher order depending only on a single parameter.  The behavior in these cases matches that found for the equations of motion.  
\end{abstract}
\maketitle

\section{Introduction}
\label{sec:intro}
Compact binaries are one of the most abundant and interesting sources of gravitational waves (GWs).  Systems comprising stellar-mass black holes ($m_\text{BH} \sim$ 1--100 $M_\odot$) and/or neutron stars, which lie in the ``high-frequency'' GW band (1--$10^3$ Hz), are likely to be the first detected.  Indeed, current rate estimates predict that ground-based detectors like Advanced LIGO \cite{h10a} and Advanced Virgo \cite{virgo} will see several to hundreds of these sources per year once they become operational (although rates are very uncertain) \cite{a10}.  Higher mass binaries, containing massive or supermassive black holes, can be detected by other means.  Systems with masses $10^4$--$10^7 M_\odot$ lie in the ``low-frequency'' band ($10^{-4}$--1 Hz) and will be detected by a space-based detector like the proposed eLISA mission \cite{a13}.  Heavier systems ($m_\text{BH} \sim 10^8$--$10^9 M_\odot$) could be detected very soon by the timing of radio pulsars \cite{h10b}.

Detection of the GWs is very challenging: The waves interact extremely weakly with matter, and there are many sources of noise.  The separation of signal from noise is achieved by the matched filtering process, which requires the generation of extremely accurate theoretical template waveforms for the expected GW signal.  For the inspiral of compact binary systems, templates are expressed in the post-Newtonian approximation to general relativity, an expansion in powers of $v/c \sim (Gm/rc^2)^{1/2}$.  Each power is considered to be one-half a post-Newtonian (PN) order.  Templates for nonspinning binary systems have been constructed to $O((v/c)^6)$, or 3PN order \cite{bfis08}, with partial results at 3.5PN order \cite{fmbi12}.  (For a comprehensive review, see \cite{b13}.)  Numerical relativity codes provide the ability to add waveforms for the final merger of the compact objects to the end of the inspiral signal \cite{p05,clmz06,bcckv06}.

The use of extremely accurate template waveforms also allows for the extraction of source parameters from the measured GW signal.  Parameter estimation studies show that properties like compact object mass and spins, source position, and luminosity distance can be determined with high precision \cite{lh06,lh08,lhc11}.  Alone, this information can probe astrophysical regimes heretofore unexplored.  In combination with a coincident electromagnetic detection, the utility of GW measurements increases.  For instance, an electromagnetically determined redshift and a gravitationally determined luminosity distance allow for a unique probe of cosmology \cite{hh05}.  

For these reasons, detectors like Advanced LIGO and eLISA are often considered to be exciting new {\em astrophysical} observatories.  However, they will also be extremely important {\em physics} experiments.  The comparison of measured GW signals to highly accurate template waveforms also allows for fundamental tests of the theory itself.  Einstein's general relativity (GR) has been tested extremely well in the regimes of the solar system and the binary pulsar \cite{w06}.  However, the environment of an inspiraling, merging compact binary represents a strong-field, dynamical regime in which GR has currently not been tested.  GW measurements will place constraints on the validity of GR in this regime.

There are many ways by which one may test GR with the GW signal of inspiraling compact binaries.  One approach is simply to check the self-consistency of terms in the post-Newtonian sequence \cite{aiqs06}.  However, such a test can only find deviations from GR, not characterize them.  Other tests may involve putting constraints on parameters in specific alternative theories of gravity \cite{w94,w98,sw02,wy04,bbw05a,bbw05b,sw09,aw09,sy09,ypc12,bgha12}.  A third approach parametrizes the waveform in terms of generic, theory-dependent parameters \cite{yp09,mais10,myw12,a12}, much like the parametrized post-Newtonian formalism did for solar-system tests \cite{w93,w06}.  Most of these analyses have only relied on the dominant, lowest order effects in the waveform model.

A particularly important alternative to GR is the collection of scalar-tensor theories of gravity \cite{fm03}.  They have a long history, dating back over 50 years, and represent one of the simplest possible modifications to Einstein's theory.  While solar system and binary pulsar tests put strong constraints on these theories \cite{w06}, they, like all theories, have not been tested in the strong-field, dynamical regime of inspiraling compact binaries.  Furthermore, they remain well motivated.  For instance, many so-called $f(R)$ theories, which modify the action of general relativity to allow arbitrary functions of the Ricci scalar, can be expressed in the form of a scalar-tensor theory \cite{dt10}.  These $f(R)$ theories may explain the acceleration of the universe without resorting to dark energy.  Scalar-tensor theories are also potential low-energy limits of string theory \cite{fm03}.

This paper is part of a series which seeks to develop the gravitational waveform for inspiraling compact binaries in scalar-tensor theories to high order in the post-Newtonian approximation.  Specifically, we are interested in theories described by the action

\begin{equation}
S = \frac{1}{16\pi}\int \left[\phi R-\frac{\omega(\phi)}{\phi}g^{\mu \nu}\partial_\mu \phi \partial_\nu \phi\right]\sqrt{-g}\ d^4x+S_m(\mathfrak{m},g_{\mu \nu}) \, ,
\label{eq:action}
\end{equation}
where $g_{\mu \nu}$ is the spacetime metric, $g$ is its determinant, $R$ is the Ricci scalar derived from this metric, $\phi$ is the scalar field, and $\omega$ is the scalar-tensor coupling.  Note that $\omega = \omega(\phi)$ is not a constant; that is, we are not restricting our attention to Brans-Dicke theory.  We do, however, restrict ourselves only to massless scalar fields (i.e., those without a potential).  We have also written $S_m$ to represent the matter action.  Note that it depends only on the matter fields $\mathfrak{m}$ and the metric; the scalar field $\phi$ does not couple directly to the matter.  This means that \eqref{eq:action} is expressed in the ``Jordan'' frame, in which standard rods and clocks measure distances and times.  All of our work will be done in this frame.  An alternative representation is the ``Einstein'' frame, related to the Jordan frame by a conformal transformation \cite{de92}.

 The first step in the construction of gravitational waveforms is to calculate the equations of motion for the compact objects.  This was the subject of the first paper in this series \cite{mw13}.  Mirshekari and Will (hereafter MW) computed the equations of motion to order $(v/c)^5$ (2.5PN) beyond the leading term.  They made use of a method known as direct integration of the relaxed Einstein equations (DIRE), based on the original framework of Epstein and Wagoner \cite{ew75} and then extended by Will, Wiseman, and Pati \cite{w92, ww96, pw00, pw02}.  This approach has been shown to give identical results to other methods, including the ``post-Minkowskian'' method \cite{b13}, the Hamiltonian approach \cite{js98}, the ``strong-field point-particle limit'' strategy \cite{fi07}, and the ``effective field theory'' method \cite{gr06}.  It is also easily adapted to scalar-tensor theories.  

In the adapted DIRE method, the scalar-tensor field equations are first rewritten in a ``relaxed'' form: flat-spacetime wave equations for a ``gravitational field'' $\tilde{h}^{\mu \nu}$ and a modified scalar field $\varphi$.  The wave equations are simplified by the choice of a particular coordinate system, represented by a gauge condition on $\tilde{h}^{\mu \nu}$.  Together, the wave equations and gauge condition contain all the content of the full field equations.  

The wave equations are then solved formally using a retarded Green's function, valid everywhere in spacetime.  To convert them to a more useful form, however, these formal solutions are evaluated differently in different regions of spacetime: In the ``near zone'' close to the source (defined in Sec.\ \ref{sec:solutions} below), the integrals are expanded using a slow-motion approximation.  Far away from the source, in the ``radiation zone,'' a special coordinate transformation is used to evaluate the solutions.  The total solution for each field ($\tilde{h}^{\mu \nu}$ and $\varphi$) is then the sum of the two separate solutions; any terms dependent on the arbitrary boundary between zones must cancel out. 

In this paper, we calculate the tensor gravitational waveform to order $(v/c)^4$ (2PN) beyond the leading-order ``Newtonian'' quadrupole.  In doing so, we also make use of the adapted DIRE approach.  The difference with MW is fundamentally simple: In that paper, the gravitational field $\tilde{h}^{\mu \nu}$ was evaluated in the near zone, in order to calculate how it affects the motion of the compact bodies.  We, by contrast, evaluate $\tilde{h}^{\mu \nu}$ (or more specifically, $\tilde{h}^{ij}$) in the radiation zone, where it will be measured by a GW detector.  Our procedure very closely follows that of Wiseman and Will \cite{ww96} (hereafter WW), although the notation is updated to match that of MW.  Along the way, we must make use of MW's results for the equations of motion.  

As discussed in MW, one key concern in this study is the treatment of the compact bodies' internal gravity.  Since scalar-tensor theories do not obey the strong equivalence principle, the motion and gravitational-wave emission of a binary depend on the internal composition of its constituent bodies.  To handle this effect, we have adopted the approach of Eardley \cite{e75}.  We treat the matter stress-energy tensor as a sum of delta functions located at the position of each compact object.  However, instead of assigning each body a constant mass, we let the mass be a function of the scalar field, $M_A = M_A(\phi)$.  This gives the matter action an indirect dependence on $\phi$, even though we still work in the Jordan frame.  In the final waveform, this dependence will appear as the ``sensitivity'' of the mass to variations in the scalar field,

\begin{equation}
s_A \equiv \left(\frac{d \ln M_A(\phi)}{d \ln \phi}\right)_0 \, ,
\end{equation}
as well as derivatives of this quantity.  (The subscript 0 means that the derivative should be evaluated using the asymptotic value of the scalar field, $\phi_0$.)  In the weak-field limit, the sensitivity is proportional to the Newtonian self-gravitational energy per unit mass of the body.  For neutron stars, the sensitivity depends on the mass and equation of state of the star, with typical values 0.1--0.3 \cite{wz89,z92}.  For black holes, $s = 0.5$, and all derivatives vanish.

Recently, numerical simulations of compact binaries in scalar-tensor gravity have shown that the sensitivities of neutron stars can change dramatically during the late inspiral \cite{bppl13,pbpl13,stob13}.  This ``dynamical scalarization'' effect is a generalization of the ``spontaneous scalarization'' of individual stars discovered by Damour and Esposito-Far{\`e}se \cite{de93,de96}.  Since we assume that the sensitivities are constant in time, our work does not capture this effect.  In any case, dynamical scalarization only becomes relevant during the late portion of the inspiral, when our use of the post-Newtonian approximation also becomes invalid.

We find that, at 0PN and 0.5PN order, the gravitational waves are identical to those in general relativity, except for two changes.  First, each explicit factor of total mass $m = m_1+m_2$ contained in the expression is modified to $\alpha m$, where
\begin{equation}
\alpha \equiv \frac{3+2\omega_0}{4+2\omega_0}+\frac{(1-2s_1)(1-2s_2)}{4+2\omega_0} \, .
\end{equation}
Here $\omega_0 \equiv \omega(\phi_0)$.  Second, the overall waveform is scaled by $(3+2\omega_0)/(4+2\omega_0)$.  At higher PN order, the deviations become more complicated.  However, they are still described by a relatively small number of parameters, the same as those used in MW to characterize differences between the GR and scalar-tensor equations of motion.  Almost every term in the waveform has a counterpart in the GR waveform with the same basic structure.  Terms which are entirely new result from the existence of a scalar dipole moment $\Is_s^i$ (defined in Sec.\ \ref{sec:surfacemoment} below).  Unlike the regular dipole moment $\Is^i$, the scalar dipole moment cannot be made to vanish by choosing center-of-mass coordinates.  This is a direct consequence of scalar-tensor theories violating the strong equivalence principle.  While the main impact of $\Is_s^i$ is in the generation of scalar dipole radiation, its presence also has a strong effect on the tensor waveform.

In general relativity, integration over the radiation zone produces terms which enter the final waveform beginning at 1.5PN order.  Some of these terms are ``instantaneous''; that is, they depend only on the binary's state at a particular time.  (Because the waves travel at the speed of light $c$, they depend on the binary's state not at the current time $t$, but at a retarded one $\tau = t-R/c$, where $R$ is the distance to the system.)  Other terms are ``hereditary'': They require an integration of the binary's behavior from an infinite time in the past until the moment the waves are emitted.  At 1.5PN order, the hereditary contribution involves a logarithmic factor and is known as the gravitational-wave ``tail.''  It arises from a coupling between monopole and mass quadrupole moments of the source and can be physically described as backscattering of the outgoing radiation on the curved spacetime of the binary.  Monopole-mass octupole and monopole-current quadrupole couplings create further tail terms at 2PN order.  Higher PN orders include more tails, as well as ``tails of tails'' arising from three-moment couplings.

The addition of the scalar field to the radiation-zone integrals produces no additional tail terms.  However, it does produce another kind of hereditary term, which involves an integral of moments of the system over its entire past history {\em without} a logarithmic multiplier.  At 1.5PN order, the hereditary term involves a mass dipole-mass dipole coupling, while at 2PN order, the coupling is between the mass dipole and the mass quadrupole.  Both of these produce oscillatory terms, like all the other pieces of the waveform.  However, the dipole-dipole integral also produces a zero-frequency (DC) term, which grows secularly throughout the inspiral.  Such terms are often referred to as ``nonlinear gravitational-wave memory,'' or ``Christodoulou memory'' \cite{p83,c91,bd92,f09}.  Unlike oscillatory terms, they cause a permanent change in the state of a GW detector.  In general relativity, the nonlinear memory does not appear until 2.5PN order, where it is driven by a mass quadrupole-mass quadrupole coupling.  Even with the new dipole-dipole effect, the nonlinear memory formally enters the waveform at relatively high post-Newtonian order.  However, because it is integrated over the past history of the binary, the memory's actual effect is comparable in magnitude to lower order post-Newtonian terms.  It is, therefore, potentially detectable by gravitational-wave detectors \cite{f09,f09b}.

One interesting limit of our results is the case where both compact objects are black holes.  In scalar-tensor theory, isolated black holes behave identically to those in general relativity.  MW verified that a similar statement is true about binary black holes.  Specifically, they showed that the 2.5PN equations of motion in scalar-tensor theory are identical to those in general relativity, except for an unmeasurable rescaling of masses.  They then conjectured that the same should hold true for the gravitational waves emitted by the binary.  Our work shows that this conjecture is correct for tensor gravitational waves, at least to 2PN order.  MW also found special results for mixed systems, those containing one black hole and one neutron star.  In that case, the equations of motion are identical to GR, with mass rescaling, up to 1PN order.  Beyond that order, they deviate from Einstein's theory, but the deviation depends only on a single parameter.  Unfortunately, this parameter is the same in Brans-Dicke and more generalized scalar-tensor theories.  We have found that all of these properties carry over to the tensor gravitational waves emitted by a mixed system.  

To find the gravitational-wave phasing, we require an expression for the energy loss.  That, in turn, requires the scalar gravitational waveform.  It will be considered in a follow-up paper.  While the procedure is the same as for the tensor waveform, the actual calculation is somewhat more complex and lengthy.  With both tensor and scalar pieces in hand, we can write down the full response of a detector to the inspiral of a circular binary.  Finally, we will be able to use the final waveform in a parameter estimation study.  We wish to investigate how well detectors like Advanced LIGO can measure differences between GR and scalar-tensor theories, as well as how the additional terms in the waveform influence the measurement of astrophysical parameters, like those described above.

The outline of the paper is as follows: Section \ref{sec:DIRE} presents the field equations for the tensor and scalar fields, as derived from the action \eqref{eq:action}.  We cast them into a relaxed form and show how they can be solved for field and source points in different regions of spacetime.  We also discuss how the tensor and scalar fields affect a gravitational-wave detector.  Section \ref{sec:NZsource} reviews the results from MW in the near zone surrounding the source.  Post-Newtonian expansions of the near-zone source are needed to calculate the gravitational waveform.  

Section \ref{sec:EWmoments} describes the calculation of the Epstein-Wagoner moments, the fundamental pieces needed to derive the near-zone contribution to the gravitational waveform.  The calculation of the two-index moment $I_{\text{EW}}^{ij}$ is described in some detail in order to clarify the major issues involved.  Section \ref{sec:twobody} shows how the moments can be converted from a generic $N$-body scenario to the specific two-body case we consider.  We also discuss how the equations of motion, taken from MW, are used to expand time derivatives of the moments.  

Section \ref{sec:RZintegrals} leaves these pieces aside and presents the other half of the puzzle: the radiation-zone contribution to the gravitational waves.  First, we must calculate the tensor and scalar fields in the radiation zone, far from the compact objects.  They have both near-zone and radiation-zone sources.  Then these fields are used to calculate the GWs.  Terms produced here enter the final waveform at 1.5PN and 2PN orders and include the hereditary effects described earlier (tails and memory).  Finally, in Sec.\ \ref{sec:results}, we present the full 2PN tensor gravitational waveform for a nonspinning compact binary in massless scalar-tensor theory.  We also discuss some features of the results in more detail.

In this paper, we use units in which $c = 1$.  We do not set $G = 1$; as we shall see, the effective Newtonian gravitational constant depends on the asymptotic value of the scalar field.  Greek indices run over four spacetime values (0, 1, 2, 3), while Latin indices run over three spatial values (1, 2, 3).  We use the Einstein summation convention, in which repeated indices are summed over.  We use a multi-index notation for products of vector components: $x^{ijk} \equiv x^ix^jx^k$.  A capital letter superscript denotes a product of that dimensionality: $x^L \equiv x^{k_1}x^{k_2}\cdots x^{k_l}$.  Angular brackets around indices denote symmetric, trace-free (STF) products (see Appendix \ref{app:unitvectors} for details).  Finally, we use standard notation for symmetrized and antisymmetrized indices, e.g. $x^{(i}y^{j)} \equiv (x^iy^j+x^jy^i)/2$ and $x^{[i}y^{j]} \equiv (x^iy^j-x^jy^i)/2$.  At times we use a bar to separate indices which should be symmetrized from those which should not, e.g. $x^{(i|}y^jz^{|k)} \equiv (x^iy^jz^k+x^ky^jz^i)/2$.

\section{Construction and solution of the relaxed scalar-tensor equations}
\label{sec:DIRE}
\subsection{Field equations and relaxed form}
\label{sec:fieldeqns}
The field equations for theories described by \eqref{eq:action} are given by 

\begin{subequations}
\begin{align}
G_{\mu \nu} &= \frac{8\pi}{\phi}T_{\mu \nu}+\frac{\omega(\phi)}{\phi^2}\left(\phi_{,\mu}\phi_{,\nu}-\frac{1}{2}g_{\mu\nu}\phi_{,\lambda}\phi^{,\lambda}\right)+\frac{1}{\phi}(\phi_{;\mu\nu}-g_{\mu\nu}\Box_g \phi) \, ,
\label{eq:tensoreqn}\\
\Box_g \phi &= \frac{1}{3+2\omega(\phi)}\left(8\pi T - 16\pi\phi\frac{\partial T}{\partial \phi}-\frac{d\omega}{d\phi}\phi_{,\lambda}\phi^{,\lambda}\right) \, .
\label{eq:scalareqn}
\end{align}
\end{subequations}
As stated above, we work in the Jordan representation of the theory, so that $g_{\mu \nu}$ is the physical metric.  The quantity $G_{\mu \nu}$ is the Einstein tensor constructed from this metric, while $\phi$ is the scalar field, $\omega(\phi)$ is the coupling function, $T_{\mu \nu}$ is the stress energy of matter and nongravitational fields, and $T \equiv g^{\alpha \beta}T_{\alpha \beta}$ is its trace.  We use commas to denote ordinary derivatives.  Semicolons denote covariant derivatives (taken using $g_{\mu \nu}$ in the usual way), and $\Box_g \equiv g^{\alpha \beta}\partial_\alpha \partial_\beta$ is the d'Alembertian with indices raised by the metric.  In the Jordan representation, the derivative $\partial T/\partial \phi$ is not present for normal matter, but will be present for gravitationally bound bodies, as we will discuss in Sec.\ \ref{sec:eardley}.

We assume that far away from the sources, the metric reduces to the Minkowski metric, $\eta^{\mu \nu}$, and the scalar field tends to a constant $\phi_0$.  We introduce a rescaled scalar field,

\begin{equation}
\varphi \equiv \frac{\phi}{\phi_0} \, ,
\end{equation}
a conformally transformed metric,

\begin{equation}
\tilde{g}_{\mu \nu} \equiv \varphi g_{\mu \nu} \, ,
\label{eq:gtilde}
\end{equation}
and a ``gothic'' version of that metric with its indices raised,

\begin{equation}
\tilde{\mathfrak{g}}^{\mu \nu} \equiv \sqrt{-\tilde{g}}\tilde{g}^{\mu \nu} \, .
\end{equation}
Here $\tilde{g}$ is the determinant of $\tilde{g}_{\mu \nu}$.  We can then define the gravitational field as

\begin{equation}
\tilde{h}^{\mu \nu} \equiv \eta^{\mu \nu} - \tilde{\mathfrak{g}}^{\mu \nu} \, .
\label{eq:htilde}
\end{equation}
We use a tilde to differentiate this field from the gravitational field defined in general relativity, $h^{\mu \nu}$, which has the same definition but with $\mathfrak{g}^{\mu \nu} \equiv \sqrt{-g}g^{\mu \nu}$ in place of $\tilde{\mathfrak{g}}^{\mu \nu}$.  [See, for instance, WW (2.2).]  We impose the Lorenz gauge condition

\begin{equation}
{\tilde{h}^{\mu \nu}}_{\hphantom{\mu\nu}, \nu} = 0 \, .
\end{equation}
Then the field equation \eqref{eq:tensoreqn} reduces to

\begin{equation}
\Box_\eta \tilde{h}^{\mu \nu} = -16\pi \tau^{\mu \nu} \, ,
\label{eq:tensorwave}
\end{equation}
where $\Box_\eta \equiv \eta^{\alpha \beta}\partial_\alpha \partial_\beta$ is the flat-spacetime wave operator and the source is

\begin{equation}
\tau^{\mu\nu} \equiv (-g)\frac{\varphi}{\phi_0}T^{\mu \nu} + \frac{1}{16\pi}(\Lambda^{\mu\nu} + \Lambda_s^{\mu \nu}) \, .
\end{equation}
Here $T^{\mu \nu}$ is the stress energy of matter and nongravitational fields.  In our case, we have no other fields, so this represents the {\em compact} piece of the source.  The quantity $\Lambda^{\mu\nu}$ represents the gravitational-field contribution to the stress energy:

\begin{equation}
\Lambda^{\mu \nu} \equiv 16\pi(-\tilde{g}) \tilde{t}_\text{LL}^{\mu \nu}+\tilde{h}^{\mu \alpha}_{\hphantom{\mu\alpha} , \beta}\tilde{h}^{\nu \beta}_{\hphantom{\nu\beta},\alpha}-\tilde{h}^{\alpha \beta}\tilde{h}^{\mu \nu}_{\hphantom{\mu\nu},\alpha\beta} \, ,
\label{eq:Lambdamunu}
\end{equation}
where 
\begin{equation}
\begin{split}
(-\tilde{g})\tilde{t}_\text{LL}^{\mu \nu} &\equiv \frac{1}{16\pi}\left[\tilde{g}_{\lambda \alpha}\tilde{g}^{\beta \rho}\tilde{h}^{\mu\lambda}_{\hphantom{\mu\lambda},\beta}\tilde{h}^{\nu\alpha}_{\hphantom{\mu\alpha},\rho}+\frac{1}{2}\tilde{g}_{\lambda\alpha}\tilde{g}^{\mu \nu}\tilde{h}^{\lambda\beta}_{\hphantom{\lambda\beta},\rho}\tilde{h}^{\rho\alpha}_{\hphantom{\rho\alpha},\beta}-2\tilde{g}_{\alpha\beta}\tilde{g}^{\lambda(\mu}\tilde{h}^{\nu)\beta}_{\hphantom{\nu)\beta},\rho}\tilde{h}^{\rho\alpha}_{\hphantom{\rho\alpha},\lambda}\right.\\
&\left.\quad+\frac{1}{8}(2\tilde{g}^{\mu\lambda}\tilde{g}^{\nu\alpha}-\tilde{g}^{\mu\nu}\tilde{g}^{\lambda\alpha})(2\tilde{g}_{\beta\rho}\tilde{g}_{\sigma\tau}-\tilde{g}_{\rho\sigma}\tilde{g}_{\beta\tau})\tilde{h}^{\beta\tau}_{\hphantom{\beta\tau},\lambda}\tilde{h}^{\rho\sigma}_{\hphantom{\rho\sigma},\alpha}\right]
\end{split}
\end{equation}
is the Landau-Lifshitz tensor evaluated with conformal (tilde) variables \eqref{eq:gtilde} and \eqref{eq:htilde} instead of standard ones.  With this definition, $\Lambda^{\mu \nu}$ will have the same fundamental form as it does in general relativity, except that the gravitational field is $\tilde{h}^{\mu \nu}$ instead of $h^{\mu \nu}$.  The final piece is the scalar contribution to the stress energy,

\begin{equation}
\Lambda_s^{\mu \nu} \equiv \frac{(3+2\omega)}{\varphi^2}\varphi_{,\alpha}\varphi_{,\beta}\left(\tilde{\mathfrak{g}}^{\mu\alpha}\tilde{\mathfrak{g}}^{\nu\beta}-\frac{1}{2}\tilde{\mathfrak{g}}^{\mu\nu}\tilde{\mathfrak{g}}^{\alpha\beta}\right) \, .
\end{equation}
The scalar field equation \eqref{eq:scalareqn} can also be written as a flat-spacetime wave equation,

\begin{equation}
\Box_\eta \varphi = -8\pi \tau_s \, ,
\label{eq:scalarwave}
\end{equation}
with source

\begin{equation}
\tau_s \equiv -\frac{1}{3+2\omega}\sqrt{-g}\frac{\varphi}{\phi_0}\left(T-2\varphi\frac{\partial T}{\partial \varphi}\right)-\frac{1}{8\pi}\tilde{h}^{\alpha\beta}\varphi_{,\alpha\beta}+\frac{1}{16\pi}\frac{d}{d\varphi}\left[\ln\left(\frac{3+2\omega}{\varphi^2}\right)\right]\varphi_{,\alpha}\varphi_{,\beta}\tilde{\mathfrak{g}}^{\alpha\beta} \, .
\end{equation}

\subsection{Solution of the wave equations}
\label{sec:solutions}
The wave equations \eqref{eq:tensorwave} and \eqref{eq:scalarwave} can be solved formally in all spacetime by using a retarded Green's function,

\begin{subequations}
\begin{align}
\tilde{h}^{\mu \nu}(t,\mathbf{x}) &= 4\int \frac{\tau^{\mu \nu}(t',\mathbf{x}')\delta(t'-t+|\mathbf{x}-\mathbf{x}'|)}{|\mathbf{x}-\mathbf{x}'|}d^4x' \, ,\label{eq:hintegral}\\
\varphi(t,\mathbf{x}) &= 2\int \frac{\tau_s(t',\mathbf{x}')\delta(t'-t+|\mathbf{x}-\mathbf{x}'|)}{|\mathbf{x}-\mathbf{x}'|}d^4x' \, .
\label{eq:phiintegral}
\end{align}
\end{subequations}
The delta function in both these equations restricts the integration to being over the past flat-spacetime null cone $\Cs$ emanating from the field point $(t,\mathbf{x})$.  To obtain explicit solutions, we divide the spacetime into two regions.  Define the characteristic size of the source as $\Ss$.  We assume the bodies move at velocities $v \ll 1$.  Then the {\em near zone} is defined as the area with $|\mathbf{x}| = R < \Rs$, where $\Rs \sim \Ss/v$ is the characteristic wavelength of gravitational radiation from the system.  (We use capital $R$ to denote the distance from the binary's center of mass to a field point in order to avoid confusion later with $r$, the orbital radius of the binary.)  Everything outside the near zone ($R > \Rs$) is the {\em radiation zone}.

We carry out the integrals \eqref{eq:hintegral} and \eqref{eq:phiintegral} in two pieces: one integral over the near zone and one over the radiation zone.  Each is done by a fundamentally different method.  In the end, the final solution is the sum of the two pieces.  WW and \cite{pw00} showed explicitly that any terms dependent on the boundary radius $\Rs$ in one-half of the integral would be exactly cancelled by pieces in the other half of the integral, leaving the final answer, as expected, independent of this arbitrary parameter.  In our work, we simply assume this property and ignore any terms which depend on $\Rs$.

The integrals are also evaluated differently depending on what field point $\mathbf{x}$ we are interested in.  For instance, MW calculated $\tilde{h}^{\mu \nu}$ and $\varphi$ at field points $\mathbf{x}$ in the near zone, where the bodies are located.  The near-zone fields were then used to calculate the equations of motion for the bodies.  In this paper, we are interested in the gravitational waves, so we will ultimately want the fields evaluated in the radiation zone.  However, we will still need the near-zone fields as source terms (see $\Lambda^{ij}$ and $\Lambda_s^{ij}$ above).

With two integration regions and two possibilities for field points, there are a total of four distinct ways to evaluate \eqref{eq:hintegral} and \eqref{eq:phiintegral}.  For a complete description of these methods, refer to WW and \cite{pw00}.  Here we give only a brief summary.  For field points in the near zone and integration over the near zone, we can treat the retardation as a small perturbation.  Expanding $\tau^{\mu \nu}(t-|\mathbf{x}-\mathbf{x}'|)$ in powers of $|\mathbf{x}-\mathbf{x}'|$, we find

\begin{subequations}
\begin{align}
\tilde{h}_\Ns ^{\mu \nu}(t,\mathbf{x}) &= 4\sum_{m=0}^\infty \frac{(-1)^m}{m!}\frac{\partial^m}{\partial t^m}\int_\Ms \tau^{\mu \nu}(t,\mathbf{x}')|\mathbf{x}-\mathbf{x}'|^{m-1}d^3x' \label{eq:hNN} \, , \\
\varphi_\Ns (t,\mathbf{x}) &= 2\sum_{m=0}^\infty \frac{(-1)^m}{m!}\frac{\partial^m}{\partial t^m}\int_\Ms \tau_s(t,\mathbf{x}')|\mathbf{x}-\mathbf{x}'|^{m-1}d^3x' \, .
\label{eq:phiNN}
\end{align}
\end{subequations}
Here $\Ns $ is the three-dimensional hypersurface representing the intersection of the past null cone $\Cs$ and the near-zone world tube.  After the expansion, the actual integration takes place over $\Ms$, the intersection of the hypersurface $t = \text{const}$ with the near-zone world tube.  Note that in the near zone, the slow motion approximation $v \ll 1$ means that each time derivative corresponds to an increase of one-half post-Newtonian order.

For field points in the near zone and integration over the radiation zone, we recognize that the source contains only field terms which are themselves retarded.  Therefore, it is prudent to change variables and integrate over $\tau' = t'-R'$, where $R' \equiv |\mathbf{x}'|$.  We find

\begin{subequations}
\begin{align}
\begin{split}
\tilde{h}_{\Cs-\Ns }^{\mu\nu}(t,\mathbf{x}) &= 4\int_{\tau-2\Rs}^{\tau-2\Rs+2R}d\tau'\int_0^{2\pi}d\phi' \int_{1-\xi}^1 \frac{\tau^{\mu\nu}(\tau'+R',\mathbf{x}')}{t-\tau'-\mathbf{\hat{N}}'\cdot \mathbf{x}}[R'(\tau',\Omega')]^2d\cos\theta'\\
&\quad +4\int_{-\infty}^{\tau-2\Rs}d\tau'\oint\frac{\tau^{\mu\nu}(\tau'+R',\mathbf{x}')}{t-\tau'-\mathbf{\hat{N}}'\cdot\mathbf{x}}[R'(\tau',\Omega')]^2d^2\Omega'
\label{eq:hNR} \, ,
\end{split}\\
\begin{split}
\varphi_{\Cs-\Ns }(t,\mathbf{x}) &= 2\int_{\tau-2\Rs}^{\tau-2\Rs+2R}d\tau'\int_0^{2\pi}d\phi' \int_{1-\xi}^1 \frac{\tau_s(\tau'+R',\mathbf{x}')}{t-\tau'-\mathbf{\hat{N}}'\cdot \mathbf{x}}[R'(\tau',\Omega')]^2d\cos\theta'\\
&\quad +2\int_{-\infty}^{\tau-2\Rs}d\tau'\oint\frac{\tau_s(\tau'+R',\mathbf{x}')}{t-\tau'-\mathbf{\hat{N}}'\cdot\mathbf{x}}[R'(\tau',\Omega')]^2d^2\Omega' \, ,
\label{eq:phiNR}
\end{split}
\end{align}
\end{subequations}
where $\mathbf{\hat{N}}' \equiv \mathbf{x}'/R'$ and $\xi \equiv (\tau-\tau')(2R-2\Rs+\tau-\tau')/(2R\Rs)$.  Here, the notation $\Cs-\Ns $ denotes that this is the integration over all pieces of the past null cone which do not intersect the near zone; that is, they are in the radiation zone.  For each $\tau'$, the inner pieces integrate over the intersection of $\Cs$ with the future null cone emanating from the center of mass of the system at time $\tau'$.  The $\tau'$ integration is then a summation over all such future-directed null cones, starting from the infinite past and continuing until the cones no longer overlap.  It turns out that the contributions from these integrals only come into play at higher post-Newtonian order than was considered in MW.

For field points in the radiation zone and integration over the near zone, we expand the entire integrand in powers of $|\mathbf{x}'|/R$ and find

\begin{subequations}
\begin{align}
\tilde{h}_\Ns ^{\mu\nu}(t,\mathbf{x}) &= 4\sum_{q=0}^\infty\frac{(-1)^q}{q!}\left(\frac{1}{R}M^{\mu\nu k_1\cdots k_q}\right)_{,k_1\cdots k_q} \label{eq:hRN}\, ,\\
\varphi_\Ns ^{\mu\nu}(t,\mathbf{x}) &= 2\sum_{q=0}^\infty\frac{(-1)^q}{q!}\left(\frac{1}{R}M_s^{k_1\cdots k_q}\right)_{,k_1\cdots k_q} \label{eq:phiRN}\, ,
\end{align}
where
\end{subequations}

\begin{subequations}
\begin{align}
M^{\mu\nu k_1\cdots k_q}(\tau) &\equiv \int_\Ms\tau^{\mu\nu}(\tau,\mathbf{x}')x^{\prime k_1}\cdots x^{\prime k_q}d^3x' \, ,
\label{eq:hM}\\
M_s^{k_1\cdots k_q}(\tau) &\equiv \int_\Ms\tau_s(\tau,\mathbf{x}')x^{\prime k_1}\cdots x^{\prime k_q}d^3x' \, .
\label{eq:phiM}
\end{align}
\end{subequations}
Here $\Ms$ is again the intersection of the near-zone world tube with a constant-time hypersurface; however, in this case, that time is the retarded time $\tau = t-R$.  

For gravitational waves, we can simplify \eqref{eq:hRN} and \eqref{eq:phiRN} in two ways: (1) We are only interested in the spatial piece of the tensor, $\tilde{h}^{ij}$.  (2) Gravitational-wave detectors operate in the regime $R \gg \Rs$, which we call the {\em far-away zone}.  (It is, of course, a subset of the radiation zone.)  Therefore, we can expand \eqref{eq:hRN} and \eqref{eq:phiRN} in powers of $1/R$, keeping only the lowest order term.  The results are

\begin{subequations}
\begin{align}
\tilde{h}_\Ns ^{ij}(t,\mathbf{x}) &= \frac{4}{R}\sum_{m=0}^\infty \frac{1}{m!}\frac{\partial^m}{\partial t^m}\int_\Ms\tau^{ij}(\tau,\mathbf{x}')(\mathbf{\hat{N}}\cdot\mathbf{x'})^m d^3x' \, ,
\label{eq:tensorEW}\\
\varphi_\Ns (t,\mathbf{x}) &= \frac{2}{R}\sum_{m=0}^\infty \frac{1}{m!}\frac{\partial^m}{\partial t^m}\int_\Ms\tau_s(\tau,\mathbf{x}')(\mathbf{\hat{N}}\cdot\mathbf{x'})^m d^3x' \, ,
\label{eq:scalarEW}
\end{align}
\end{subequations}
where $\mathbf{\hat{N}} \equiv \mathbf{x}/R$ is the direction from the source to the detector.  The Lorenz gauge condition implies a conservation law for the source,

\begin{equation}
{\tau^{\mu \nu}}_{,\nu} = 0 \, .
\end{equation}
Using this, we can rewrite the tensor equation as

\begin{equation}
\tilde{h}_\Ns ^{ij}(t,\mathbf{x}) = \frac{2}{R}\frac{d^2}{dt^2}\sum_{m=0}^\infty\hat{N}^{k_1}\cdots\hat{N}^{k_m}I_{\text{EW}}^{ijk_1\cdots k_m}(\tau) \, ,
\label{eq:EWwaveform}
\end{equation}
where the $I_{\text{EW}}^{M+2}$ are known as ``Epstein-Wagoner'' (EW) moments \cite{ew75}.  They are given by

\begin{subequations}
\begin{align}
I_{\text{EW}}^{ij} &\equiv \int_\Ms \tau^{00}x^{ij}d^3x + I_{\text{EW (surf)}}^{ij} \, ,\\
I_{\text{EW}}^{ijk} &\equiv \int_\Ms(2\tau^{0(i}x^{j)k}-\tau^{0k}x^{ij})\ d^3x + I_{\text{EW (surf)}}^{ijk} \, ,\\
I_{\text{EW}}^{ijk_1\cdots k_m} &\equiv \frac{2}{m!}\frac{d^{m-2}}{dt^{m-2}}\int_\Ms\tau^{ij}x^{k_1\cdots k_m}d^3x \quad (m \geq 2) \, .
\label{eq:EWgt4}
\end{align}
\end{subequations}
The largest piece of the work in this paper is the calculation of these EW moments to the necessary post-Newtonian order.  The use of the conservation law to rearrange the two- and three-index EW moments results in the ``surface'' moments $I_{\text{EW (surf)}}^{ij}$ and $I_{\text{EW (surf)}}^{ijk}$,

\begin{subequations}
\begin{align}
\frac{d^2}{dt^2}I_{\text{EW (surf)}}^{ij} &= \oint_{\partial \Ms}[4\tau^{l(i}x^{j)}-(\tau^{kl}x^{ij})_{,k}]\Rs^2 \hat{n}^l d^2\Omega \, , \label{eq:IijS}\\
\frac{d}{dt}I_{\text{EW (surf)}}^{ijk} &= \oint_{\partial \Ms}(2\tau^{l(i}x^{j)k}-\tau^{kl}x^{ij})\Rs^2 \hat{n}^l d^2\Omega \, .
\label{eq:IijkS}
\end{align}
\end{subequations}
They are evaluated on $\partial \Ms$, a sphere of radius $\Rs$ bounding the hypersurface $\Ms$; $\hat{n}^l$ is a radial unit vector pointing outward from this boundary.  

Finally, if the field point is in the radiation zone and the integration is over the radiation zone, we can again use a change of variable.  The results are identical to \eqref{eq:hNR} and \eqref{eq:phiNR}, except with a different limit of integration,

\begin{subequations}
\begin{align}
\begin{split}
\tilde{h}_{\Cs-\Ns }^{\mu\nu}(t,\mathbf{x}) &= 4\int_{\tau-2\Rs}^\tau d\tau'\int_0^{2\pi}d\phi' \int_{1-\xi}^1 \frac{\tau^{\mu\nu}(\tau'+R',\mathbf{x}')}{t-\tau'-\mathbf{\hat{N}}'\cdot \mathbf{x}}[R'(\tau',\Omega')]^2d\cos\theta'\\
&\quad +4\int_{-\infty}^{\tau-2\Rs}d\tau'\oint\frac{\tau^{\mu\nu}(\tau'+R',\mathbf{x}')}{t-\tau'-\mathbf{\hat{N}}'\cdot\mathbf{x}}[R'(\tau',\Omega')]^2d^2\Omega' \, ,
\label{eq:hRR}
\end{split}\\
\begin{split}
\varphi_{\Cs-\Ns }(t,\mathbf{x}) &= 2\int_{\tau-2\Rs}^\tau d\tau'\int_0^{2\pi}d\phi' \int_{1-\xi}^1 \frac{\tau_s(\tau'+R',\mathbf{x}')}{t-\tau'-\mathbf{\hat{N}}'\cdot \mathbf{x}}[R'(\tau',\Omega')]^2d\cos\theta'\\
&\quad +2\int_{-\infty}^{\tau-2\Rs}d\tau'\oint\frac{\tau_s(\tau'+R',\mathbf{x}')}{t-\tau'-\mathbf{\hat{N}}'\cdot\mathbf{x}}[R'(\tau',\Omega')]^2d^2\Omega' \, .
\label{eq:phiRR}
\end{split}
\end{align}
\end{subequations}
To find the gravitational waves, we merely ignore all nonspatial components (for the tensor waves) and consider only terms with $1/R$ dependence.  These radiation-zone integrals will produce hereditary terms in the final GW signal, including tail and memory effects.

\subsection{Effect on GW detectors}
\label{sec:detectors}
To find the effect of the (tensor) gravitational field and the scalar field on a GW detector, we first need to convert back to the physical metric.  The inverse metric is

\begin{equation}
g^{\mu \nu} = \varphi \tilde{g}^{\mu \nu} = \frac{\varphi}{\sqrt{-\tilde{g}}}\mathfrak{\tilde{g}}^{\mu \nu} \, .
\end{equation}
The determinant $\tilde{g}$ is given by

\begin{equation}
\tilde{g} = \det \tilde{\mathfrak{g}}^{\mu \nu} = -1+\tilde{h}+O(\tilde{h}^2) \, ,
\end{equation}
where $\tilde{h} \equiv \eta_{\mu \nu} \tilde{h}^{\mu \nu}$ is the trace of $\tilde{h}^{\mu \nu}$.  We are not concerned with terms of order $\tilde{h}^2$ and higher because we are working in the far-away zone, where $\tilde{h} \sim 1/R$.  Putting everything together, we get

\begin{equation}
g^{\mu \nu} = \eta^{\mu \nu} - \tilde{h}^{\mu \nu} + \frac{1}{2}\tilde{h} \eta^{\mu \nu} + \Psi \eta^{\mu \nu} + O\left(\frac{1}{R^2}\right)\, .
\end{equation}
Here $\Psi \equiv \varphi - 1$, and we have also made use of the fact that $\Psi \sim 1/R$ in the far-away zone.  The physical metric is therefore
\begin{equation}
g_{\mu \nu} = \eta_{\mu \nu} + \tilde{h}_{\mu \nu} - \frac{1}{2}\tilde{h} \eta_{\mu \nu} - \Psi \eta_{\mu \nu} + O\left(\frac{1}{R^2}\right)\, ,
\end{equation}
where $\tilde{h}_{\mu \nu} = \eta_{\mu \alpha}\eta_{\nu \beta} \tilde{h}^{\alpha \beta}$ is lowered using the Minkowski metric.  

A gravitational-wave detector works by measuring the separation $\xi$ between test masses.  If the distance between the test masses is small compared to the wavelength of the GWs, and the masses move slowly, the separation obeys the equation

\begin{equation}
\ddot{\xi}^i = -R_{0i0j}\xi^j \, ,
\end{equation}
where dots denote time derivatives and $R_{0i0j}$ are components of the Riemann curvature tensor.  We can use the metric to calculate them,

\begin{equation}
R_{0i0j} = -\frac{1}{2} \ddot{\tilde{h}}^{ij}_\text{TT} -\frac{1}{2}\ddot{\Psi}(\hat{N}^i\hat{N}^j-\delta^{ij}) \, .
\end{equation}
Here ``TT'' designates the transverse-traceless projection of the gravitational field tensor, which satisfies the conditions

\begin{equation}
\hat{N}^i\tilde{h}^{ij}_\text{TT} = \hat{N}^j\tilde{h}^{ij}_\text{TT} = 0 = \delta^{ij}\tilde{h}^{ij}_\text{TT} \, .
\end{equation}
We can find the TT part of a tensor by using the projection operator,

\begin{equation}
\tilde{h}^{ij}_\text{TT} = \left(P^{ip}P^{jq}-\frac{1}{2}P^{ij}P^{pq}\right)\tilde{h}^{ij} \, ,
\label{eq:TTproj}
\end{equation}
where $P^{pq} = \delta^{pq}-\hat{N}^p\hat{N}^q$ is the transverse projection operator.  The fact that only the TT piece of $\tilde{h}^{ij}$ contributes to the measured GWs will allow us to simplify our calculation by prematurely dropping terms which cannot possibly produce a TT contribution at the end of the day.  Note that the scalar field $\Psi$ will contribute a transverse ``breathing'' mode to the GW signal; it will be treated in a subsequent paper in the series.

\section{Calculation of the near-zone source}
\label{sec:NZsource}
In this section, we review results from MW for the source $\tau^{\mu \nu}$ in the near zone.  These results are needed to calculate the Epstein-Wagoner moments and thus the near-zone contribution to the gravitational waveform.

\subsection{Potentials, fields, and field source}
\label{sec:NZfields}
Following MW, the compact source can be described in terms of densities \cite{bd89},

\begin{subequations}
\begin{align}
\sigma &\equiv T^{00} + T^{ii} \, ,\\
\sigma^i &\equiv T^{0i} \, , \\
\sigma^{ij} &\equiv T^{ij} \, , \\
\sigma_s &\equiv -T + 2\varphi \frac{\partial T}{\partial \varphi} \, .
\end{align}
\end{subequations}
We can then define a number of Poisson-like potentials.  For example, given a generic Poisson integral for a function $f(t,\mathbf{x})$,

\begin{equation}
P(f) \equiv \frac{1}{4\pi}\int_\Ms \frac{f(t,\mathbf{x}')}{|\mathbf{x}-\mathbf{x}'|} d^3x' \, ,
\label{eq:Ppot}
\end{equation}
then the basic ``Newtonian'' potential is

\begin{equation}
U_\sigma \equiv P(4\pi\sigma) = \int_\Ms \frac{\sigma(t,\mathbf{x}')}{|\mathbf{x}-\mathbf{x}'|}d^3x' \, .
\end{equation}
We also have a scalar equivalent,

\begin{equation}
U_{s\sigma} \equiv P(4\pi\sigma_s) = \int_\Ms \frac{\sigma_s(t,\mathbf{x}')}{|\mathbf{x}-\mathbf{x}'|}d^3x' \, .
\end{equation}
The $\sigma$ subscript clarifies that these potentials use the densities defined in this section.  Later, we will convert to a new density which reflects the specific source (compact binaries) we are studying.  We delay defining the rest of the potentials until then.  Expressions for all other $\sigma$-density potentials can be found in MW (3.12) and (3.13).  Note that the generic Poisson integral has the property

\begin{equation}
\nabla^2 P(f) = -f \, .
\label{eq:delsquaredP}
\end{equation}
This will be very useful throughout the calculation.

For convenience, we can rewrite the fields as

\begin{subequations}
\begin{align}
\tilde{h}^{00} &\equiv N \, ,\\
\tilde{h}^{0i} &\equiv K^i \, ,\\
\tilde{h}^{ij} &\equiv B^{ij} \, ,\\
\tilde{h}^{ii} &\equiv B \, , \\
\varphi &\equiv 1+\Psi \, .
\end{align}
\end{subequations}
We use this notation in all spacetime.  In the near zone, $N\sim O(\epsilon)$, $K^i \sim O(\epsilon^{3/2})$, $B^{ij} \sim B \sim O(\epsilon^2)$, and $\Psi \sim O(\epsilon)$, where the post-Newtonian counting parameter $\epsilon \sim v^2 \sim m/r$.  Here $m$ is the mass of the system, $r$ is a typical distance scale, and $v$ is its characteristic speed.  (Later, $r$ will be the orbital radius of a binary, and $v$ will be the magnitude of its relative velocity.)

To obtain expressions for the near-zone fields, we solve \eqref{eq:hNN} and \eqref{eq:phiNN} iteratively.  At lowest order, we only solve for $N$ and $\Psi$, the other quantities being intrinsically higher order.  The sources only include compact terms: $\tau^{00} = \sigma/\phi_0+O(\rho\epsilon)$ and $\tau_s = \sigma_s/[\phi_0(3+2\omega_0)]+O(\rho\epsilon)$, where $\omega_0 \equiv \omega(\phi_0)$.  Using the definitions of the potentials, we find

\begin{subequations}
\begin{align}
N &= \frac{4U_\sigma}{\phi_0} + O(\epsilon^2) = 4G(1-\zeta)U_\sigma + O(\epsilon^2) \, , \label{eq:N0}\\
\Psi &= \frac{2U_{s\sigma}}{\phi_0(3+2\omega_0)} + O(\epsilon^2) = 2G\zeta U_{s\sigma} + O(\epsilon^2) \, .
\label{eq:Psi0}
\end{align}
\end{subequations}
Here 

\begin{equation}
G \equiv \frac{1}{\phi_0}\frac{4+2\omega_0}{3+2\omega_0}
\end{equation}
is the effective gravitational constant.  The definition is chosen so that, for a perfect fluid with no internal gravitational binding energy, the metric component $g_{00}$ matches the result from general relativity, $g_{00} = -1+2GU_\sigma$.  We do not set $G$ equal to 1, since it depends on the asymptotic value of the scalar field $\phi$, which could potentially vary in time.  The other parameter is

\begin{equation}
\zeta \equiv \frac{1}{4+2\omega_0} \, .
\end{equation}
For the next order, we need to begin evaluating the field terms of the source, $\Lambda^{\mu \nu}$ and $\Lambda_s^{\mu \nu}$.  These are given explicitly in terms of $N$, $K^i$, $B^{ij}$, $B$, and $\Psi$ in MW (3.4) and (3.5).  We can plug in \eqref{eq:N0} for $N$ and \eqref{eq:Psi0} for $\Psi$.  This will be enough to get the fields at next order, as shown in MW (4.10).  This procedure is iterated until the fields and sources are obtained to the necessary order.  In the next section, we will need the expressions for $\tau^{\mu \nu}$ in order to calculate the Epstein-Wagoner moments.

\subsection{Matter source}
\label{sec:eardley}
The previous section describes a generic matter source characterized by densities $\sigma$, $\sigma^i$, $\sigma^{ij}$, and $\sigma_s$ (the ``$\sigma$ densities'').  All of the expressions for near-zone fields and sources in MW are written in terms of these generic densities.  We now convert to a more realistic source for the situation we are considering: one made up of an arbitrary number of compact objects.  (We later specialize to the two-body case, but many steps of the calculation are valid for a general system.)  Since a compact object is gravitationally bound, its total mass depends on its internal gravitational energy.  This, in turn, depends on the effective local value of the gravitational coupling.  In scalar-tensor theory, the coupling is controlled by the value of the scalar field $\phi$ in the vicinity of the body.  

To deal with this complication, we use the approach of Eardley \cite{e75}.  In his method, we consider the compact objects to be point masses, with a mass $M(\phi)$ that is a function of the scalar field.  The stress-energy tensor is then given by

\begin{equation}
\begin{split}
T^{\mu \nu}(x^\alpha) &= (-g)^{-1/2}\sum_A \int d\tau \, M_A(\phi)u_A^\mu u_A^\nu \delta^4(x_A^\alpha(\tau)-x^\alpha) \\
&= (-g)^{-1/2}\sum_A M_A(\phi) u_A^\mu u_A^\nu (u_A^0)^{-1}\delta^3(\mathbf{x}-\mathbf{x}_A) \, .
\end{split}
\end{equation}
Here $u_A^\mu$ is the four-velocity of body $A$, and $\tau$ is the proper time measured along its world line.  (This is the only instance in which we use the symbol $\tau$ for this purpose.)  The dependence of mass on $\phi$ is what leads to the $\partial T/\partial \phi$ term in \eqref{eq:scalareqn}, which would not normally be present in the Jordan frame.  (Remember that in the Jordan representation, the scalar field does not directly couple to the matter.  This indirect coupling is merely a way of treating the complexity of the source in scalar-tensor theory.)

We expand $M_A(\phi)$ about the asymptotic value $\phi_0$,

\begin{equation}
\begin{split}
M_A(\phi) &= M_{A0} + \left(\frac{dM_A}{d\phi}\right)_0 \delta \phi + \frac{1}{2}\left(\frac{d^2M_A}{d\phi^2}\right)_0 \delta \phi^2 + \frac{1}{6}\left(\frac{d^3M_A}{d\phi^3}\right)_0 \delta \phi^3 +\cdots\\
&= m_A\left[1+s_A\Psi+\frac{1}{2}(s_A^2+s_A'-s_A)\Psi^2+\frac{1}{6}(s_A''+3s_A' s_A-3s_A'+s_A^3-3s_A^2+2s_A)\Psi^3+O(\Psi^4)\right]\\
&\equiv m_A[1+\Ss(s_A;\Psi)] \, ,
\end{split}
\end{equation}
where $m_A \equiv M_{A0}$.  We define the sensitivity and its derivatives as

\begin{subequations}
\begin{align}
s_A &\equiv \left(\frac{d\ln M_A(\phi)}{d\ln \phi}\right)_0 \, ,\\
s_A' &\equiv \left(\frac{d^2\ln M_A(\phi)}{d(\ln \phi)^2}\right)_0 \, ,\\
s_A'' &\equiv \left(\frac{d^3\ln M_A(\phi)}{d(\ln \phi)^3}\right)_0 \, ,\\
\end{align}
\end{subequations}
and so on.  Note that $s'$ has the opposite sign of the equivalent quantity in \cite{w93} and \cite{abwz12}.  If we define a new density

\begin{equation}
\rho^* \equiv \sum_A m_A\delta^3(\mathbf{x}-\mathbf{x}_A) \, ,
\end{equation}
the stress energy becomes

\begin{equation}
T^{\mu \nu} = \rho^*(-g)^{-1/2}u^0v^\mu v^\nu[1+\Ss(s;\Psi)] \, ,
\end{equation}
where $v^\mu = (1,\mathbf{v})$ is the ordinary velocity.  The various velocities and the sensitivity $s$ technically should have body labels, but they will each pick one up when multiplied by the delta function in $\rho^*$.  We use this convention frequently in the steps to follow.  Returning to the definitions of the $\sigma$ densities, we find

\begin{subequations}
\begin{align}
\sigma &= \rho^*(-g)^{-1/2}u^0(1+v^2)[1+\Ss(s;\Psi)] \, , \label{eq:sigma}\\
\sigma^i &= \rho^*(-g)^{-1/2}u^0v^i[1+\Ss(s;\Psi)] \, , \label{eq:sigmai}\\
\sigma^{ij} &= \rho^*(-g)^{-1/2}u^0v^{ij}[1+\Ss(s;\Psi)] \, , \label{eq:sigmaij}\\
\sigma_s &= \rho^*(-g)^{-1/2}(u^0)^{-1}[(1-2s)+\Ss_s(s;\Psi)] \, ,
\label{eq:sigmas}
\end{align}
\end{subequations}
where
\begin{equation}
\Ss_s(s;\Psi) \equiv -2a_s\Psi-b_s\Psi^2+O(\Psi^3) 
\end{equation}
and
\begin{subequations}
\begin{align}
a_s &\equiv s^2+s'-\frac{1}{2}s \, ,\\
a_s' &\equiv s''+2ss'-\frac{1}{2}s' \, , \\
b_s &\equiv a_s'-a_s+sa_s \, .
\end{align}
\end{subequations}
With these quantities, we can also rewrite

\begin{equation}
\Ss(s;\Psi) = s\Psi+\frac{1}{4}(2a_s-s)\Psi^2+O(\Psi^3) \, .
\end{equation}
By substituting the metric into \eqref{eq:sigma}-\eqref{eq:sigmas}, the $\sigma$ densities can be written in terms of the $\rho^*$ density as a post-Newtonian expansion.  These expressions are given in MW (5.12).  We will need them to translate $\tau^{\mu \nu}$ from the expressions in MW to the versions we need.

We also define new potentials based on the $\rho^*$ density.  For instance,

\begin{subequations}
\begin{align}
U &\equiv \int_\Ms\frac{\rho^*(t,\mathbf{x}')}{|\mathbf{x}-\mathbf{x}'|}d^3x' \, ,\\
U_s &\equiv \int_\Ms\frac{(1-2s(\mathbf{x}'))\rho^*(t,\mathbf{x}')}{|\mathbf{x}-\mathbf{x}'|}d^3x' \, .
\end{align}
\end{subequations}
More generally,

\begin{subequations}
\begin{align}
\Sigma(f) &\equiv \int_\Ms\frac{\rho^*(t,\mathbf{x}')f(t,\mathbf{x}')}{|\mathbf{x}-\mathbf{x}'|}d^3x' = P(4\pi\rho^*f) \, , \label{eq:sigmapot}\\
\Sigma^i(f) &\equiv \int_\Ms\frac{\rho^*(t,\mathbf{x}')v^{\prime i}f(t,\mathbf{x}')}{|\mathbf{x}-\mathbf{x}'|}d^3x' = P(4\pi\rho^*v^if) \, , \label{eq:sigmaipot}\\
\Sigma^{ij}(f) &\equiv \int_\Ms\frac{\rho^*(t,\mathbf{x}')v^{\prime ij}f(t,\mathbf{x}')}{|\mathbf{x}-\mathbf{x}'|}d^3x' = P(4\pi\rho^*v^{ij}f) \, , \label{eq:sigmaijpot}\\
\Sigma_s(f) &\equiv \int_\Ms\frac{(1-2s(\mathbf{x}'))\rho^*(t,\mathbf{x}')f(t,\mathbf{x}')}{|\mathbf{x}-\mathbf{x}'|}d^3x' = P(4\pi(1-2s)\rho^*f) \, , \label{eq:sigmaspot}\\
X(f) &\equiv \int_\Ms\rho^*(t,\mathbf{x}')f(t,\mathbf{x}')|\mathbf{x}-\mathbf{x}'| d^3x' \, , \label{eq:Xpot} \\
Y(f) &\equiv \int_\Ms \rho^*(t,\mathbf{x}')f(t,\mathbf{x}')|\mathbf{x}-\mathbf{x}'|^3 d^3x' \, ,
\label{eq:Ypot}
\end{align}
\end{subequations}
plus natural generalizations of $X$ and $Y$ like $X^i$, $X_s$, etc.  All the potentials listed in this paper can be expressed in terms of $P(f)$ \eqref{eq:Ppot} or these basic forms.  They are listed in Appendix \ref{app:potentials}.  MW (5.13)--(5.22) show how to convert between many $\sigma$-density and $\rho$-density potentials (e.g., $U_\sigma$ and $U$).  These expressions will also be useful in converting $\tau^{\mu \nu}$.

\section{Epstein-Wagoner moments}
\label{sec:EWmoments}
In this section, we calculate the Epstein-Wagoner moments which are needed to generate the near-zone contribution to the gravitational waveform.  

\subsection{Two-index moment $I_{\text{EW}}^{ij}$}
\label{sec:2index}
We begin with the two-index moment.  Its lowest order piece will produce the lowest order gravitational waves, which we define as 0PN.  It is given by

\begin{equation}
\begin{split}
I^{ij}_{\text{EW}} &= \int_\Ms \tau^{00} x^{ij} d^3x + I^{ij}_{\text{EW (surf)}}\\
&= I^{ij}_C + I^{ij}_F + I^{ij}_S \, .
\end{split}
\end{equation}
For clarity, the moment has been split into three pieces: (1) $I_C^{ij}$, in which the integral is taken over the compact part of $\tau^{00}$; (2) $I_F^{ij}$, in which the integral is taken over the field part of $\tau^{00}$; and (3) $I_S^{ij} \equiv I^{ij}_{\text{EW (surf)}}$, the surface moment.  To calculate the first two pieces, we need the source $\tau^{00}$ evaluated to 2PN order, or $O(\rho \epsilon^2)$.  This is found in MW (4.14a), but it is expressed using the $\sigma$ density.  Converting to the $\rho^*$ density, we find

\begin{align}
\tau^{00} &= \rho^* G(1-\zeta)\left\{1+\frac{1}{2}v^2+3G(1-\zeta)U-G\zeta (1-2s)U_s+\frac{1}{2}G(1-\zeta)\Phi_1-3G^2(1-\zeta)^2\Phi_2+\frac{1}{2}G\zeta(1-2s)\Phi^s_1 \right.\nonumber \\
&\qquad+ G^2\zeta(1-\zeta)(1-2s)\Phi^s_2+4G^2\zeta^2(1-2s)\Sigma(a_sU_s)+\frac{3}{8}v^4+\frac{9}{2}G^2(1-\zeta)^2U^2+\frac{7}{2}G(1-\zeta)v^2 U\nonumber \\
&\qquad- 4G(1-\zeta)v^kV^k-\frac{1}{2}G\zeta(1-2s)v^2 U_s-3G^2\zeta(1-\zeta)(1-2s)UU_s+\frac{3}{2}G(1-\zeta)\ddot{X}-\frac{1}{2}G\zeta(1-2s)\ddot{X}_s\nonumber \\
&\left.\qquad +G^2[-3\zeta(1-\zeta)+\zeta(\zeta+2\lambda_1)(1-2s)]\Phi^s_{2s}+G^2\left[\frac{1}{2}\zeta(\zeta+2\lambda_1)(1-2s)+2\zeta^2 a_s\right]U_s^2\right\}\nonumber \\
&\quad+G^2(1-\zeta)^2\left[-\frac{7}{8\pi}(\nabla U)^2-\frac{5}{8\pi}\nabla U \cdot \nabla \Phi_1+\frac{3}{4\pi}G(1-\zeta)\nabla U \cdot \nabla \Phi_2+\frac{3}{4\pi}G\zeta \nabla U \cdot \nabla \Phi^s_{2s} \right. \nonumber \\
&\qquad+\frac{5}{8\pi}\dot{U}^2-\frac{1}{\pi}U\ddot{U}-\frac{2}{\pi}\dot{U}^{,k}V^k+\frac{3}{2\pi}V^{k,l}V^{l,k}+\frac{1}{2\pi}V^{k,l}V^{k,l}+\frac{1}{\pi}\dot{V}^kU^{,k}-\frac{1}{\pi}U^{,kl}\Phi_1^{kl}-\frac{7}{8\pi}\nabla U \cdot \nabla \ddot{X}\nonumber \\
&\left.\qquad-\frac{5}{2\pi}G(1-\zeta)U(\nabla U)^2-\frac{1}{\pi}G(1-\zeta)U^{,kl}P_2^{kl}\right]\nonumber \\
&\quad+G^2\zeta(1-\zeta)\left[\frac{1}{8\pi}(\nabla U_s)^2-\frac{1}{8\pi}\nabla U_s \cdot \nabla \Phi^s_1-\frac{1}{4\pi}G(1-\zeta)\nabla U_s \cdot \nabla \Phi^s_2 - \frac{1}{4\pi}G(2\lambda_1+\zeta)\nabla U_s \cdot \nabla \Phi^s_{2s} \right.\nonumber \\
&\qquad-\frac{1}{\pi}G\zeta \nabla U_s \cdot \nabla \Sigma(a_sU_s)+\frac{1}{2\pi}G(1-\zeta)U(\nabla U_s)^2+\frac{1}{\pi}G(1-\zeta)U_s\nabla U \cdot \nabla U_s-\frac{1}{\pi}G(1-\zeta)U^{,kl}P_{2s}^{kl}+\frac{1}{8\pi}\dot{U}_s^2\nonumber \\
&\left.\qquad+\frac{1}{8\pi}\nabla U_s \cdot \nabla \ddot{X}_s\right] +O(\rho\epsilon^{5/2})\, .
\label{eq:tau00}
\end{align}
Recall that all the potentials are defined in Appendix \ref{app:potentials}.  We have introduced the quantity

\begin{equation}
\lambda_1 \equiv \frac{(d\omega/d\varphi)_0\zeta}{3+2\omega_0} \, .
\end{equation}
Later, we will also need

\begin{equation}
\lambda_2 \equiv \frac{(d^2\omega/d\varphi^2)_0\zeta^2}{3+2\omega_0} \, .
\end{equation}
The compact moment can be written down by inspection,

\begin{align}
I_C^{ij} &= G(1-\zeta)\sum_A m_A x_A^{ij} \left\{1+\frac{1}{2}v_A^2+\sum_{B\neq A}\frac{Gm_B}{r_{AB}}[3(1-\zeta)-\zeta(1-2s_A)(1-2s_B)]\right\} \nonumber \\
&\quad+\frac{3}{8}G(1-\zeta)\sum_A m_A x_A^{ij}v_A^4 \nonumber \\
&\quad+G^2(1-\zeta)\sum_A\sum_{B\neq A}\frac{m_Am_B}{r_{AB}} x_A^{ij}\left\{\frac{1}{2}[7(1-\zeta)-\zeta(1-2s_A)(1-2s_B)]v_A^2+2(1-\zeta)v_B^2\right. \nonumber \\
&\qquad-4(1-\zeta)\mathbf{v}_A \cdot \mathbf{v}_B-\frac{1}{2}[3(1-\zeta)-\zeta(1-2s_A)(1-2s_B)][\mathbf{a}_B\cdot \mathbf{x}_{AB}+(\mathbf{v}_B\cdot \mathbf{\hat{n}}_{AB})^2] \nonumber \\
&\qquad+\sum_{C\neq A}\frac{Gm_C}{r_{AC}}\left[\frac{9}{2}(1-\zeta)^2-3\zeta(1-\zeta)(1-2s_A)(1-2s_C)+2\zeta^2a_{sA}(1-2s_B)(1-2s_C)\right. \nonumber \\
&\left.\qquad\quad+\frac{1}{2}\zeta(\zeta+2\lambda_1)(1-2s_A)(1-2s_B)(1-2s_C)\right]  \nonumber \\
&\qquad+\sum_{C\neq B}\frac{Gm_C}{r_{BC}}\left[-3(1-\zeta)^2+\zeta(1-\zeta)(1-2s_A)(1-2s_B)+4\zeta^2(1-2s_A)a_{sB}(1-2s_C)\right. \nonumber \\
&\left.\left.\qquad\quad-3\zeta(1-\zeta)(1-2s_B)(1-2s_C)+\zeta(\zeta+2\lambda_1)(1-2s_A)(1-2s_B)(1-2s_C)\vphantom{(1-\zeta)^2}\right]\vphantom{\frac{1}{2}}\right\} \, .
\end{align}
Here body $A$ (for example) has position $\mathbf{x}_A$, velocity $\mathbf{v}_A$, and acceleration $\mathbf{a}_A$.  The distance between bodies $A$ and $B$ is $r_{AB}$, and the unit vector $\mathbf{\hat{n}}_{AB} = \mathbf{x}_{AB}/r_{AB}$ points from body $B$ to body $A$.

To calculate the field moment $I_F^{ij}$, we must evaluate a series of integrals involving the 1PN and 2PN potentials.  Examining \eqref{eq:tau00}, we see that there are 24 different integrals.  However, integrals such as $\int_\Ms (\nabla U)^2 x^{ij}d^3x$ and $\int_\Ms (\nabla U_s)^2 x^{ij} d^3x$ are essentially identical: The functional form of the integrand is the same, with the only difference being the addition of sensitivity factors $1-2s$.  If we count such integrals together, there turns out to be 15 unique pieces to compute.  (Many of these also share similar fundamental components, but we count them separately due to different factors of velocity or acceleration.)

\subsubsection{1PN field integral}
\label{sec:1PNintegral}
The simplest integral is the 1PN term $\int_\Ms (\nabla U)^2 x^{ij} d^3x$.  (When discussing these integrals, we will ignore the constants in front.)  We will calculate it explicitly in order to point out some techniques which will be used throughout the calculation.  First, we integrate by parts,

\begin{equation}
\int_\Ms (\nabla U)^2 x^{ij} d^3x = \oint_{\partial \Ms}UU^{,k}x^{ij} d^2 S^k-\int_\Ms UU^{,kk} x^{ij} d^3x - 2\int_\Ms UU^{,(i}x^{j)} d^3x \, .
\end{equation}
The surface integral is evaluated on the boundary of the near zone at a constant (retarded) time; therefore, $\partial \Ms$ is a sphere with radius $\Rs$.  We can write $x^i = \Rs\hat{n}^i$ and $d^2S^k = \Rs^2 \hat{n}^k$, where $\hat{n}^i$ is a unit vector normal to the surface.  We then expand $U$ and $U^{,k}$ in inverse powers of $\Rs$ and look for any $\Rs$-independent terms in the surface integral.  We find that

\begin{equation}
U = \sum_A m_A\left(\frac{1}{\Rs}+\frac{\hat{n}^ax_A^a}{\Rs^2}+\frac{1}{2}\frac{(3\hat{n}^{ab}-\delta^{ab})x_A^{ab}}{\Rs^3}+\cdots\right) \sim \frac{1}{\Rs}+\epsilon\frac{\hat{n}^a}{\Rs^2}+\frac{1+\hat{n}^{ab}}{\Rs^3}+\cdots \, .
\label{eq:Uexpansion}
\end{equation}
The center of mass (CM) of the system is defined as $\mathbf{x}_{\text{CM}} = \sum_A m_A x_A^i + O(\rho\epsilon)$.  If we choose to work in coordinates where $\mathbf{x}_{\text{CM}} = 0$, then the second term in the above expansion is one post-Newtonian order higher than the others.  We use the order parameter $\epsilon$ to mark this.  With that exception, we only care about the $\Rs$ scaling and the number of unit vectors $\hat{n}^i$.  The derivative of the potential scales like

\begin{equation}
U^{,k} \sim \frac{\hat{n}^k}{\Rs^2}+\epsilon\frac{1+\hat{n}^{ck}}{\Rs^3}+\frac{\hat{n}^c+\hat{n}^{cdk}}{\Rs^4}+\cdots \, .
\end{equation}
So the only $\Rs$-independent terms in the surface integrand scale like $\epsilon \hat{n}^{aij}$.  When integrated over the surface, an odd number of $\hat{n}^i$ vanishes.  (See Appendix \ref{app:unitvectors}.)  Therefore, the surface integral makes no contribution.

The first volume integral can be simplified using $\nabla^2 U = -4\pi \rho^*$ [see \eqref{eq:delsquaredP}], making it essentially a compact integral.  The solution is easily seen to be

\begin{equation}
-\int_\Ms UU^{,kk}x^{ij} d^3x = 4\pi \sum_A \sum_{B\neq A} \frac{m_Am_B}{r_{AB}}x_A^{ij} \, .
\label{eq:1PNintegral}
\end{equation}
The second volume integral can be integrated by parts again,

\begin{equation}
-2\int_\Ms UU^{,(i}x^{j)}d^3x = -\oint_{\partial \Ms}U^2x^{(i}d^2S^{j)} + \int_\Ms U^2\delta^{ij}d^3x \, .
\end{equation}
The surface integral again vanishes by virtue of integrating an odd number of unit vectors.  Meanwhile, the volume integral can be ignored because it will not survive the transverse-traceless projection.  So \eqref{eq:1PNintegral} alone is the value of the 1PN integral.

Throughout this work, we routinely drop terms which will not, in the end, survive the transverse-traceless projection.  For convenience, we refer to these terms as ``non-TT.''  Non-TT terms are easy to identify by sight: Any term containing $\delta^{ij}$, $\hat{N}^i$, or $\hat{N}^j$  is non-TT.  [This can be checked by applying the TT projection \eqref{eq:TTproj} to such terms.]  For the two-index moment, $\delta^{ij}$ pieces are the only ones we drop.  For higher order moments, terms like $\delta^{ik}$, $\delta^{jk}$, $\delta^{il}$, $\delta^{jl}$, ... are also dropped.  This is because the final expression for the waveform \eqref{eq:EWwaveform} contracts the EW moments with direction vectors $\hat{N}^k$, $\hat{N}^l$, ... for all dummy indices (i.e., all those besides $i$ and $j$).  Therefore, a term like $\delta^{ik}$ will introduce $\hat{N}^i$ in the final waveform, and that is non-TT.  Note that terms like $\delta^{kl}$ (or any other involving two dummy indices) must be kept.

As discussed earlier, the integral $\int_\Ms(\nabla U_s)^2 x^{ij} d^3x$ is essentially the same as the one we have just calculated.  One slight difference does occur in the surface integrals: Because $\sum_A m_A(1-2s_A)x_A^i$ does not vanish, even to lowest post-Newtonian order, the $1/\Rs^2$ ($1/\Rs^3$) term in the expansion of $U_s$ ($U_s^{,k}$) will be at the same order as all the other terms.  However, this makes no difference: The surface integrals still vanish.  The final answer is exactly the same as \eqref{eq:1PNintegral}, only with sensitivity factors added,

\begin{equation}
\int_\Ms (\nabla U_s)^2 x^{ij} d^3x = 4\pi \sum_A \sum_{B\neq A} \frac{m_A(1-2s_A)m_B(1-2s_B)}{r_{AB}}x_A^{ij} \, .
\end{equation}

\subsubsection{2PN two-potential field integrals}
\label{sec:mainintegrals}
The other integrals enter at 2PN order.  Eleven of them involve only two potentials; of these, all but the one involving $P_{2}^{kl}$ (or $P_{2s}^{kl}$) can be solved using a straightforward procedure.  Consider the following example:

\begin{equation}
\int_\Ms\nabla U \cdot \nabla \Phi_1 x^{ij} d^3x = \oint_{\partial \Ms} U \Phi_1^{,k} x^{ij} d^2S^k - \int_\Ms U \Phi_1^{,kk} x^{ij} d^3x - 2\int_\Ms U\Phi_1^{,(i}x^{j)}d^3x \, .
\end{equation}
Here the surface integral vanishes for the same reason as above (i.e., an odd number of $\hat{n}^i$).  The first volume integral is evaluated easily using $\nabla^2\Phi_1 = -4\pi\rho^*v^2$.  The tricky part is the second volume integral.  We can write it as

\begin{equation}
- 2\int_\Ms U\Phi_1^{,(i}x^{j)}d^3 x = 2 \sum_{A,B}m_Av_A^2m_B\int_\Ms\frac{1}{|\mathbf{x}-\mathbf{x}_B|}\frac{(x-x_A)^{(i}}{|\mathbf{x}-\mathbf{x}_A|^3}x^{j)}d^3x \, .
\end{equation}
It can be evaluated using techniques developed in WW.  We first change integration variables from $\mathbf{x}$ to $\mathbf{y} = \mathbf{x}-\mathbf{x}_A$.  Looking just at the main piece of the integral, this gives, in our particular example,

\begin{equation}
\begin{split}
\int_\Ms\frac{1}{|\mathbf{x}-\mathbf{x}_B|}\frac{(x-x_A)^{(i}}{|\mathbf{x}-\mathbf{x}_A|^3}x^{j)}d^3x &= \int_{\Ms_y}\frac{1}{|\mathbf{y}+\mathbf{x}_{AB}|}\frac{\hat{y}^{(i}}{y^2}(y\hat{y}^{j)}+x_A^{j)})d^3y \\
&\quad- \oint_{\partial\Ms_y}\frac{1}{|\mathbf{y}+\mathbf{x}_{AB}|}\frac{\hat{y}^{(i}}{y^2}(y\hat{y}^{j)}+x_A^{j)}) (\mathbf{\hat{y}}\cdot \mathbf{x}_A) \Rs^2 d^2\Omega_y \\
&\quad+ \frac{1}{2}\oint_{\partial\Ms_y}\mathbf{x}_A\cdot \nabla\left[\frac{1}{|\mathbf{y}+\mathbf{x}_{AB}|}\frac{\hat{y}^{(i}}{y^2}(y\hat{y}^{j)}+x_A^{j)})\right](\mathbf{\hat{y}}\cdot\mathbf{x}_A) \Rs^2 d^2\Omega_y + \cdots \, .
\end{split}
\end{equation}
There are two cases to consider: $A = B$ and $A \neq B$.  In both cases, the infinite series of surface integrals vanishes: The terms either depend on $\Rs$ or average to zero because of an odd number of unit vectors.  When $A = B$, $\mathbf{x}_{AB} = 0$, making the volume integral easy to evaluate. It also vanishes, for the same reasons as the surface integrals.  When $A \neq B$, we make use of the following expansion:

\begin{equation}
\frac{1}{|\mathbf{y}+\mathbf{x}_{AB}|}= \sum_{l,m}\frac{4\pi}{2l+1}\frac{(-r_<)^l}{r_>^{l+1}}Y_{l m}^*(\mathbf{\hat{n}}_{AB})Y_{l m}(\mathbf{\hat{y}}) \, ,
\label{eq:harmonicexpansion}
\end{equation}
where $Y_{l m}$ are the spherical harmonics, and $r_{<(>)}$ denotes the lesser (greater) of $r_{AB}$ and $y$.  We substitute this expansion into the volume integral and then express all products of $\hat{y}^i$ in terms of symmetric, trace-free (STF) products $\hat{y}^{\langle L'\rangle}$.  (See Appendix B.)  Here $\langle L'\rangle$ denotes an $l'$-dimensional STF combination.  We can then perform the angular integration using

\begin{equation}
\sum_m \int Y^*_{l m}(\mathbf{\hat{n}}_{AB})Y_{l m}(\mathbf{\hat{y}})\hat{y}^{\langle L'\rangle}d^2\Omega_y = \hat{n}_{AB}^{\langle L\rangle}\delta^{l l'} \, .
\label{eq:angularintegral}
\end{equation}
In our example, we find

\begin{equation}
\int_\Ms\frac{1}{|\mathbf{x}-\mathbf{x}_B|}\frac{(x-x_A)^{(i}}{|\mathbf{x}-\mathbf{x}_A|^3}x^{j)}d^3x = 4\pi\int_0^\Rs \left[\frac{1}{5}\frac{r_<^2}{r_>^3}y\hat{n}_{AB}^{\langle ij\rangle}+\frac{1}{3}\frac{1}{r_>}y\delta^{ij}-\frac{1}{3}\frac{r_<}{r_>^2}\hat{n}_{AB}^{(i}x_A^{j)}\right]dy \, .
\end{equation}
Finally, the radial integral is evaluated using

\begin{equation}
\int_0^\Rs \frac{r_<^l}{r_>^{l+1}}y^q dy = \frac{2l+1}{(l+q+1)(l-q)}r_{AB}^q \, ,
\label{eq:radialintegral}
\end{equation}
where we have dropped terms dependent on $\Rs$.  For our example, we find

\begin{equation}
\begin{split}
\int_\Ms\frac{1}{|\mathbf{x}-\mathbf{x}_B|}\frac{(x-x_A)^{(i}}{|\mathbf{x}-\mathbf{x}_A|^3}x^{j)}d^3x &= 4\pi\left[\frac{1}{4}r_{AB}\hat{n}_{AB}^{\langle ij\rangle}-\frac{1}{6}r_{AB}\delta^{ij}-\frac{1}{2}\hat{n}_{AB}^{(i}x_A^{j)}\right]\\
&= 4\pi\left[\frac{1}{4}r_{AB}\hat{n}_{AB}^{ij}-\frac{1}{2}\hat{n}_{AB}^{(i}x_A^{j)}\right] \, ,
\end{split}
\end{equation}
where in the last step we have converted back to non-STF notation and discarded the non-TT $\delta^{ij}$ terms.  This expression can then be used to evaluate the original integral, with the mass and velocity factors added back in.

As stated above, ten of the two-potential integrals can be evaluated with this step-by-step procedure: First, integrate by parts so that one potential is undifferentiated.  (In some cases, this step is unnecessary because one potential is already undifferentiated.)  Next, if any piece can be converted to a compact integral through relations like $\nabla^2 \Phi_1 = -4\pi \rho^*v^2$, evaluate that piece.  Then write out the potentials and their derivatives explicitly in terms of masses, positions, velocities, and accelerations.  When two derivatives are taken, care must be taken to add appropriate delta functions,

\begin{subequations}
\begin{align}
U^{,ij} &= U^{,ij}_{\text{norm}} - \frac{4\pi}{3}\sum_A m_A \delta^{ij} \delta^3(\mathbf{x}-\mathbf{x}_A) \, , \\
\dot{U}^{,i} &= \dot{U}^{,i}_{\text{norm}} + \frac{4\pi}{3}\sum_A m_A v_A^i \delta^3(\mathbf{x}-\mathbf{x}_A) \, , \\
\ddot{U} &= \ddot{U}_{\text{norm}} - \frac{4\pi}{3}\sum_A m_A v_A^2 \delta^3(\mathbf{x}-\mathbf{x}_A) \, , \\
\ddot{X}^{,ij} &= \ddot{X}^{,ij}_{\text{norm}} - \frac{8\pi}{15}\sum_A m_A (v_A^2\delta^{ij}+2v_A^{ij}) \delta^3(\mathbf{x}-\mathbf{x}_A) \, ,
\end{align}
\end{subequations}
where ``norm'' denotes the derivative computed from the definitions of the potentials.  The extra terms are needed to ensure the right answer when the doubly differentiated potentials are integrated in a sphere around the point mass position $\mathbf{x}_A$.  [Compare $\nabla^2 (1/|\mathbf{x}-\mathbf{x}_A|) = -4\pi\delta^3(\mathbf{x}-\mathbf{x}_A)$.]

Next, carefully examine any surface integrals.  It turns out that no surface integral contributes to the final answer, for one of three reasons: (1) it has no $\Rs$-independent terms, (2) the $\Rs$-independent terms vanish upon integrating over the surface, or (3) the $\Rs$-independent terms average to something proportional to $\delta^{ij}$.  The last type of term does not vanish, but it is non-TT, and we can ignore it.  

For the remaining volume integrals, change variables from $\mathbf{x}$ to $\mathbf{y} = \mathbf{x}-\mathbf{x}_A$, where $A$ is the label on the differentiated potential.  Check that the surface integrals so generated and $A = B$ volume integrals contribute nothing, for one of the three reasons above.  Then integrate the $A \neq B$ volume integral over angle and radius using \eqref{eq:harmonicexpansion}, \eqref{eq:angularintegral}, and \eqref{eq:radialintegral}, keeping only TT terms at the end of the calculation.  Many of the $A \neq B$ integrals which arise in this process appear in several of the ten ``main'' integrals, and the results can be reused with the appropriate coefficients multiplied in.

\subsubsection{$P_2^{kl}$ field integral}
\label{sec:P2integral}
The final two-potential integral is $\int_\Ms U^{,kl}P_2^{kl} x^{ij} d^3x$ (and its counterpart with $P_{2s}^{kl}$),

\begin{equation}
\begin{split}
\int_\Ms U^{,kl}P_2^{kl}x^{ij} d^3x &= \sum_A m_A\int_\Ms \left[3\frac{(x-x_A)^{kl}}{|\mathbf{x}-\mathbf{x}_A|^5}-\frac{\delta^{kl}}{|\mathbf{x}-\mathbf{x}_A|^3}-\frac{4\pi}{3}\delta^{kl}\delta^3(\mathbf{x}-\mathbf{x}_A)\right]\\
&\quad \times \left[\frac{1}{4\pi}\int_\Ms\frac{d^3x'}{|\mathbf{x}-\mathbf{x}'|}U^{\prime ,k}U^{\prime ,l}d^3x'\right]x^{ij} d^3x \, ,
\end{split}
\end{equation}
where $U'$ is the usual potential written as a function of $\mathbf{x}'$.  We can integrate over $x$ (unprimed) first, using the techniques of the previous section, except with $\mathbf{x}_B\rightarrow \mathbf{x}'$.  Dropping the primes on the remaining integration variable, the result is

\begin{equation}
\begin{split}
\int_\Ms U^{,kl}P_2^{kl}x^{ij} d^3x  &= \sum_A m_A\int_\Ms d^3x\ U^{,k}U^{,l}\left[\frac{1}{6}\frac{(x_A-x)^{kl ij}}{|\mathbf{x}_A-\mathbf{x}|^3}+\frac{1}{6}\frac{(x_A-x)^{k(i}}{|\mathbf{x}_A-\mathbf{x}|}\delta^{j)l}+\frac{1}{6}\frac{(x_A-x)^{l(i}}{|\mathbf{x}_A-\mathbf{x}|}\delta^{j)k}\right.\\
&\quad -\frac{1}{6}\frac{(x_A-x)^{ij}}{|\mathbf{x}_A-\mathbf{x}|}\delta^{kl}-\frac{1}{3}|\mathbf{x}_A-\mathbf{x}|\delta^{k(i}\delta^{j)l}-\frac{1}{2}\frac{(x_A-x)^{kl (i}}{|\mathbf{x}_A-\mathbf{x}|^3}x_A^{j)}-\frac{(x_A-x)^{(k}}{|\mathbf{x}_A-\mathbf{x}|}\delta^{l)(i}x_A^{j)}\\
&\left.\quad+\frac{1}{2}\frac{(x_A-x)^{(i}}{|\mathbf{x}_A-\mathbf{x}|}x_A^{j)}\delta^{kl}+\frac{1}{2}\frac{(x_A-x)^{kl}}{|\mathbf{x}_A-\mathbf{x}|^3}x_A^{ij}-\frac{1}{2}\frac{1}{|\mathbf{x}_A-\mathbf{x}|}\delta^{kl}x_A^{ij}\right] \, .
\end{split}
\end{equation}
Following WW (D5), we can rewrite this as

\begin{equation}
\begin{split}
\int_\Ms U^{,kl}P_2^{kl}x^{ij} d^3x &= \sum_A m_A\int_\Ms d^3x\ U^{,k}U^{,l}\left[-\frac{1}{6}\Phi^{A,ijkl}+\Psi^{A,k(i}\delta^{j)l}-\frac{1}{2}\Psi^{A,kl(i}x_A^{j)}+2X^{A,k}\delta^{l(i}x_A^{j)}\right.\\
&\left.\quad-\frac{1}{2}X^{A,kl}x_A^{ij}-X^A\delta^{k(i}\delta^{j)l}\right] \, ,
\end{split}
\end{equation}
where $\Phi^A \equiv |\mathbf{x}-\mathbf{x}_A|^5/15$, $\Psi^A \equiv |\mathbf{x}-\mathbf{x}_A|^3/3$, and $X^A \equiv |\mathbf{x}-\mathbf{x}_A|$.  These six terms can be evaluated individually.  For the first four, we first integrate by parts and find a vanishing surface integral, a simple-to-compute volume integral (i.e., one involving a Laplacian), and a more difficult volume integral.  This last piece is integrated by parts again, leading to another vanishing surface integral and a final volume integral.  The final volume integrals in each case seem difficult to evaluate, but fortunately they cancel in pairs when the four terms are combined.

The fifth term behaves similarly, except it is possible to evaluate all the integrals eventually.  It involves four integrations by parts and three volume integrals which convert to compact integrals by means of a Laplacian.  The sixth piece is the most difficult,

\begin{equation}
-\sum_A m_A \int_\Ms U^{,i}U^{,j}X^A d^3x = -\sum_{A,B,C}m_Am_Bm_C\int_\Ms U^{B,i}U^{C,j}X^A \, ,
\label{eq:threebodyint}
\end{equation}
where $U^B \equiv 1/|\mathbf{x}-\mathbf{x}_B|$ and $U^C \equiv 1/|\mathbf{x}-\mathbf{x}_C|$.  (This definition, excluding the mass, is equivalent to the earlier definitions of $\Phi^A$, $\Psi^A$, and $X^A$.)  There are four cases to consider.  For $A = B = C$, we change variables to $\mathbf{y} = \mathbf{x}-\mathbf{x}_A$ and find no contributions from either the main term or the surface terms.  For $A = C \neq B$ (and $A = B \neq C$), we use the same substitution (twice) after integrating by parts.  The evaluation proceeds much as in Sec. \ref{sec:mainintegrals} above.

For $B = C \neq A$, we make a slightly different substitution (since $B$, not $A$, labels the differentiated potentials): $\mathbf{y} = \mathbf{x}-\mathbf{x}_B$.  This gives

\begin{equation}
-\sum_{A,B,C}m_Am_Bm_C\int_\Ms U^{B,i}U^{C,j}X^A = -\sum_A \sum_{B\neq A}m_Am_B^2\int_{\Ms_y}\frac{\hat{y}^{ij}}{y^2}|\mathbf{y}+\mathbf{x}_{BA}|\ dy\ d^2\Omega_y \, ,
\end{equation}
plus vanishing surface terms.  To evaluate this, we use a new expansion in spherical harmonics,

\begin{equation}
|\mathbf{y}+\mathbf{x}_{BA}|=\sum_{l,m}\frac{4\pi}{2l+1}\left[\frac{1}{2l+3}\frac{(-r_<)^{l+2}}{r_>^{l+1}}-\frac{1}{2l-1}\frac{(-r_<)^l}{r_>^{l-1}}\right]Y_{l m}^*(\mathbf{\hat{n}}_{BA})Y_{l m}(\mathbf{\hat{y}}) \, .
\end{equation}
The final case, $A \neq B \neq C$, is very complicated.  It is worked out in Appendix D of WW.  Switching body labels to match their notation, we get

\begin{equation}
-\sum_{A,B,C}m_A m_B m_C \int_\Ms U^{A,i}U^{B,j}X^C = -\pi\sum_{A \neq B \neq C}m_A m_B m_C \partial_A^i \partial_B^j F(\mathbf{x}_{AC},\mathbf{x}_{BC}) \, ,
\end{equation}
where $\partial_A^i \equiv \partial/\partial x_A^i$ and

\begin{equation}
F(\mathbf{x}_{AC},\mathbf{x}_{BC}) \equiv -\frac{2}{3}[(r_{AC}+r_{BC})r_{AB}-r_{AC}r_{BC}+2\mathbf{x}_{AC}\cdot\mathbf{x}_{BC}\ln(r_{AC}+r_{BC}+r_{AB})] \, .
\end{equation}
This term is irrelevant for compact binaries, but we keep it for completeness.  The final answer for this field integral is

\begin{equation}
\begin{split}
\int_\Ms U^{,kl}P_2^{kl}x^{ij} d^3x &= \sum_A \sum_{B\neq A}m_Am_B\left[r_{AB}^2\left(\frac{2\pi}{3}\hat{n}_{AB}^{ijk}-\frac{8\pi}{3}\hat{n}_{AB}^{(i}\delta^{j)k}\right)\right.\\
&\left.\qquad +r_{AB}(-2\pi \hat{n}_{AB}^{k(i}+6\pi\delta^{k(i})x_A^{j)}+2\pi x_A^{ij}\hat{n}_{AB}^k \vphantom{\frac{2\pi}{3}}\right]\left(-\sum_{C\neq B}\frac{m_C\hat{n}_{BC}^k}{r_{BC}^2}\right)\\
&\quad-2\pi \sum_A \sum_{B\neq A}\sum_{C\neq B}\frac{m_Am_Bm_C}{r_{AB}r_{BC}}x_A^{ij}+\pi\sum_A \sum_{B\neq A}\sum_{C\neq A}\frac{m_Am_Bm_C}{r_{AB}r_{AC}}x_A^{ij} \\
&\quad+3\pi \sum_A\sum_{B\neq A}m_A^2m_B\hat{n}_{AB}^{ij}-\pi \sum_{A \neq B \neq C} m_A m_B m_C \partial_A^i \partial_B^j F(\mathbf{x}_{AC},\mathbf{x}_{BC}) \, .
\end{split}
\label{eq:integral13}
\end{equation}

\subsubsection{Three-potential field integrals}
\label{sec:threepotintegrals}
There are three field integrals involving the product of three potentials.  The simplest is

\begin{equation}
\int_\Ms U(\nabla U)^2 x^{ij} d^3x = \frac{1}{2}\oint_{\partial \Ms} U^2 U^{,k} x^{ij} d^2S^k -\frac{1}{2}\int_\Ms U^2U^{,kk} x^{ij} d^3x -\int_\Ms U^2 U^{,(i}x^{j)}d^3x \, .
\end{equation}
The surface integral is non-TT, and the first volume integral is trivial.  Integrating the second volume integral by parts gives another surface integral and another volume integral, both of which are non-TT.

The other two field integrals are similar but involve combinations of $U$ and $U_s$.  To solve each individually, we would have to compute a three-potential subintegral in the manner of \eqref{eq:threebodyint} above.  However, when the two field integrals are appropriately combined in the EW moment, this subintegral cancels and can therefore be ignored.  Ignoring non-TT contributions, we find

\begin{equation}
\int_\Ms [U(\nabla U_s)^2+2U_s\nabla U \cdot \nabla U_s]x^{ij}d^3x = -\frac{1}{2}\int_\Ms U^{,kk} U_s^2 x^{ij} d^3x - \int_\Ms U U_s U_s^{,kk} x^{ij} d^3x \, .
\end{equation}
Each of the subintegrals is trivial to evaluate.  With the evaluation of the three-body integrals, we have now completed all 15 field integrals necessary to compute the two-index EW moment.

\subsubsection{Surface moment}
\label{sec:surfacemoment}
We can rewrite the surface moment \eqref{eq:IijS} as

\begin{equation}
\frac{d^2}{dt^2}I_{\text{EW (surf)}}^{ij} = \frac{1}{16\pi}\oint_{\partial \Ms}[2\Lambda_T^{k(i}\hat{n}^{j)k}\Rs^3-{\Lambda^{kl}_T}_{,l}\hat{n}^{ijk}\Rs^4]d^2\Omega \, ,
\label{eq:twoindexsurf}
\end{equation}
where $\Lambda_T^{ij} = \Lambda^{ij}+\Lambda_s^{ij}$.  (Recall that the boundary of the near zone is far beyond the compact source: $\Rs \gg \Ss$.)  We are again only interested in terms which do not depend on $\Rs$.  Since the first term of the integrand multiplies $\Lambda_T^{ij}$ by $\Rs^3$, we look for pieces of $\Lambda_T^{ij}$ with $\Rs^{-3}$ dependence.  The second term of the integrand adds an extra factor of $\Rs$, but the derivative on $\Lambda_T^{ij}$ reduces the overall scaling by $\Rs$ to compensate.  To survive the angular integration, a piece of $\Lambda_T^{ij}$ must contain an even number of unit vectors.  The two unit vectors in the first term retain the $\mathbf{\hat{n}}$ parity of $\Lambda_T^{ij}$; the three unit vectors in the second term flip it, but the derivative flips it back.

We begin with the first piece of the source, $\Lambda^{ij}$.  Because of the two time derivatives, we need to know it to 3PN order [i.e., $O(\rho\epsilon^3)$] in order to find the final moment to 2PN order.  Unfortunately, MW does not contain an expression for $\Lambda^{ij}$ to this order.  However, because of the way we have defined our quantities, $\Lambda^{ij}$ will have exactly the same form as the general relativistic version found in Eq.\ (4.4c) of \cite{pw00}.  The only difference is that the fields $N$, $K^i$, $B^{ij}$, and $B$ we plug into (4.4c) are pieces of our new gravitational field $\tilde{h}^{\mu \nu}$ instead of the GR field $h^{\mu \nu}$.

We must expand $\Lambda^{ij}$ in the vicinity of $r = \Rs$ and look for terms with $\Rs^{-3}$ dependence.  The first step is to expand the individual fields $\tilde{h}^{\mu \nu}$.  For $\tilde{h}^{00} = N$, the lowest order piece is given above by \eqref{eq:N0}, and its expansion is just the expansion of $U$ given in \eqref{eq:Uexpansion}.  Higher order pieces of $N$ are given in MW (4.10a), (4.10e), and (4.15a).  The various potentials in these expressions, when converted to $\rho^*$ density, can be expanded similarly to $U$.  As with earlier surface integrals, we only care about a term's post-Newtonian order, its dependence on $\Rs$, and its number of unit vectors $\hat{n}^i$.  With this restriction, we see that all potentials of the same family [i.e., $P$ (or $\Sigma$), $X$, and $Y$] share the same basic expansion.  The only relevant differences occur for terms containing $\sum_A m_A x_A^i$ or its derivatives.  As seen in the expansion of $U$ itself, these terms are one post-Newtonian order higher than the others in the same expansion.  In the end, we find

\begin{equation}
\begin{split}
N &\sim \epsilon\left[\frac{1}{\Rs}+\frac{\hat{n}^{(e)}}{\Rs^3}+\cdots\right]+\epsilon^2\left[\frac{\hat{n}^{(e)}}{\Rs}+\frac{1+\hat{n}^{(o)}}{\Rs^2}+\frac{\hat{n}^{(e+o)}}{\Rs^3}+\cdots\right]+\epsilon^{5/2}\\
&\quad+ \epsilon^3\left[\hat{n}^{(e)}\Rs+\hat{n}^{(o)}+\frac{\hat{n}^{(e+o)}}{\Rs}+\frac{\hat{n}^{(e+o)}}{\Rs^2}+\frac{\hat{n}^{(e+o)}}{\Rs^3}+\cdots\right]+O(\epsilon^{7/2}) \, .
\end{split}
\end{equation}
For simplicity, we have omitted the exact factors of $\mathbf{\hat{n}}$.  Instead, we just use the superscript $(e)$ to represent a sum of one or more terms with even parity $\mathbf{\hat{n}}$ (e.g., $1+\hat{n}^{ab}$), $(o)$ for one or more terms with odd parity $\mathbf{\hat{n}}$, and $(e+o)$ when terms are present of both parities.  Note that the $O(\epsilon^{5/2})$ term is different from the others; as seen in MW (4.10e), it contains no potentials and is, in fact, independent of position $\mathbf{x}$.

The other fields can be expanded as

\begin{equation}
K^i \sim \epsilon^{3/2}\left[\frac{\hat{n}^{(o)}}{\Rs^2}+\frac{\hat{n}^{(e)}}{\Rs^3}+\cdots\right]+\epsilon^{5/2}\left[\hat{n}^{(o)}+\frac{\hat{n}^{(e)}}{\Rs}+\frac{\hat{n}^{(o)}}{\Rs^2}+\frac{\hat{n}^{(e+o)}}{\Rs^3}+\cdots\right]+O(\epsilon^3) \, ,
\end{equation}

\begin{equation}
B \sim \epsilon^2\left[\frac{1}{\Rs}+\frac{1+\hat{n}^{(o)}}{\Rs^2}+\frac{\hat{n}^{(e+o)}}{\Rs^3}+\cdots\right] + \epsilon^{5/2}+ \epsilon^3\left[\Rs+\hat{n}^{(o)}+\frac{\hat{n}^{(e+o)}}{\Rs}+\frac{\hat{n}^{(e+o)}}{\Rs^2}+\frac{\hat{n}^{(e+o)}}{\Rs^3}+\cdots\right]+O(\epsilon^{7/2}) \, .
\end{equation}
The expansion for $B^{ij}$, which we need only to $O(\epsilon^2)$, is the same as for $B$.  The $O(\epsilon^{5/2})$ piece of $B$, like that of $N$, is independent of position.  [See MW (4.10f).]  To calculate $\Lambda^{ij}$, we need to take spatial and time derivatives of these fields.  For spatial derivatives, we merely divide each term (excepting those independent of position) by $\Rs$ and change the parity of $\mathbf{\hat{n}}$.  Time derivatives add an extra factor of $\epsilon^{1/2}$ each, while also affecting the coefficients we have chosen to ignore.

Only some of the pieces of $\Lambda^{ij}$ produce terms which scale like $\Rs^{-3}$.  Ignoring coefficients, these are $N^{,i}N^{,j}$, $\delta^{ij}(\nabla N)^2$, $N^{,(i}\dot{K}^{j)}$, $\delta^{ij}N^{,k}\dot{K}^k$, $N^{,(i}B^{,j)}$, $\delta^{ij}\nabla N\cdot \nabla B$, $\delta^{ij}\dot{N}^2$, $N\ddot{B}^{ij}$, $\delta^{ij}\dot{N}\dot{B}$, and $\delta^{ij}N\dot{N}^2$.  The first six of these scale like $\hat{n}^{(o)}/\Rs^3$, the next three like $(\delta^{ij}+\hat{n}^{(o)})/\Rs^3$, and the last like $\delta^{ij}/\Rs^3$.   (For $N\ddot{B}^{ij}$, the $\delta^{ij}$ factor is hidden inside $B^{ij}$.)  Terms with an odd number of unit vectors will vanish trivially during the angular integration of \eqref{eq:twoindexsurf}.  The even parity terms will not vanish.  However, plugging into \eqref{eq:twoindexsurf}, we see that the angular integral reduces to the average of $\hat{n}^{ij}$. This produces $\delta^{ij}$, and so these terms are non-TT.  We conclude that $\Lambda^{ij}$ makes no relevant contributions to the surface moment.

We turn next to potential contributions from $\Lambda_s^{ij}$.  We will need to expand it to $O(\rho \epsilon^3)$ first.  The result is

\begin{equation}
\begin{split}
\Lambda_s^{ij} &= (3+2\omega_0)\left\{\Psi^{,i}\Psi^{,j}-\frac{1}{2}\delta^{ij}(\nabla \Psi)^2\right\}\\
&\quad -(3+2\omega_0)\left\{2\left(1-\frac{\omega_0'}{3+2\omega_0}\right)\Psi\left[\Psi^{,i}\Psi^{,j}-\frac{1}{2}\delta^{ij}(\nabla \Psi)^2\right]-\frac{1}{2}\delta^{ij}\dot{\Psi}^2\right\}\\
&\quad +(3+2\omega_0)\left\{\left(3-\frac{4\omega_0'}{3+2\omega_0}+\frac{\omega_0''}{3+2\omega_0}\right)\Psi^2\left[\Psi^{,i}\Psi^{,j}-\frac{1}{2}\delta^{ij}(\nabla \Psi)^2\right]-\left(1-\frac{\omega_0'}{3+2\omega_0}\right)\delta^{ij}\Psi\dot{\Psi}^2\right.\\
&\left.\quad -2\Psi^{,k}\Psi^{,(i}B^{j)k}+\frac{1}{2}(\nabla \Psi)^2B^{ij}+\frac{1}{2}\delta^{ij}\Psi^{,k}\Psi^{,l}B^{kl}-2\dot{\Psi}\Psi^{,(i}K^{j)}+\delta^{ij}\dot{\Psi}\Psi^{,k}K^k+\frac{1}{2}\delta^{ij}\dot{\Psi}^2N\right\}+O(\rho\epsilon^4) \, .
\end{split}
\label{eq:Lambdas}
\end{equation}
The expansion of $\Psi$ is

\begin{equation}
\begin{split}
\Psi &\sim \epsilon\left[\frac{1}{\Rs}+\frac{\hat{n}^{(o)}}{\Rs^2}+\frac{\hat{n}^{(e)}}{\Rs^3}+\cdots\right]+\epsilon^2\left[\hat{n}^{(o)}+\frac{\hat{n}^{(e)}}{\Rs}+\frac{1+\hat{n}^{(o)}}{\Rs^2}+\frac{\hat{n}^{(e+o)}}{\Rs^3}+\cdots\right]+\epsilon^{5/2}[\hat{n}^{(o)}\Rs+1]\\
&\quad +\epsilon^3\left[\hat{n}^{(o)}\Rs^2+\hat{n}^{(e)}\Rs+\hat{n}^{(o)}+\frac{\hat{n}^{(e+o)}}{\Rs}+\frac{\hat{n}^{(e+o)}}{\Rs^2}+\frac{\hat{n}^{(e+o)}}{\Rs^3}+\cdots\right] +O(\epsilon^{7/2})\, .
\end{split}
\end{equation}
As with $N$ and $B$, the $O(\epsilon^{5/2})$ term does not contain a potential; in this case, it has a term linearly dependent on the position (so that $\hat{n}^{(o)} = \hat{n}^a$) in addition to one independent of the position.

We again find a number of terms which scale like $\Rs^{-3}$.  They derive from the $\Psi^{,i}\Psi^{,j}$, $\delta^{ij}(\nabla \Psi)^2$, $\delta^{ij}\dot{\Psi}^2$, $\delta^{ij}\dot{\Psi}^2N$, and $\delta^{ij}\Psi \dot{\Psi}^2$ pieces of $\Lambda_s^{ij}$.  When inserted into \eqref{eq:twoindexsurf}, all of these but one either vanish because of odd $\mathbf{\hat{n}}$ parity or are ignored because they produce something which is non-TT.  The surviving term is $(3+2\omega_0)[\Psi^{,i}\Psi^{,j}-\delta^{ij}(\nabla \Psi)^2/2]$, applied to the $O(\epsilon)$ piece of $\Psi$ and the linear-in-position $O(\epsilon^{5/2})$ piece of $\Psi$.  Notably, the latter piece depends on three time derivatives of the scalar dipole moment,

\begin{equation}
\Is_s^i \equiv \int_\Ms \tau_s x^i d^3x \, .
\label{eq:scalardipole}
\end{equation}
Plugging into \eqref{eq:twoindexsurf}, we find that the only contribution to the surface moment is

\begin{equation}
\frac{d^2}{dt^2}I_{\text{EW (surf)}}^{ij} = -\frac{2}{9}G^2\zeta(1-\zeta)\sum_{A,B} m_A(1-2s_A)m_B(1-2s_B)x_A^{(i}\dot{a}_B^{j)} \, .
\end{equation}

\subsubsection{Final two-index EW moment}
\label{sec:final2index}
We now add up the results for the 15 field integrals (and their variations), multiplied by the appropriate coefficients, to find the total field moment $I_F^{ij}$.  We then add this to the compact moment $I_C^{ij}$ and the surface moment $I_S^{ij}$ to form the total two-index EW moment.  It contains pieces at 0PN, 1PN, 1.5PN, and 2PN order,

\begin{align}
I_{\text{EW}}^{ij} &= G(1-\zeta)\sum_A m_Ax_A^{ij}\left\{1+\frac{1}{2}v_A^2-\frac{1}{2}\sum_{B\neq A}\frac{Gm_B}{r_{AB}}[1-\zeta+\zeta(1-2s_A)(1-2s_B)]\right\} \nonumber \\
&\quad-\frac{2}{9}G^2\zeta(1-\zeta)\left(\frac{d}{dt}\right)^{-2}\left[\sum_{A,B}m_A(1-2s_A)m_B(1-2s_B)x_A^{(i}\dot{a}_B^{j)}\right] \nonumber \\
&\quad+\frac{3}{8}G(1-\zeta)\sum_A m_Ax_A^{ij}v_A^4 \nonumber \\
&\quad+G^2(1-\zeta)\sum_A\sum_{B\neq A}\frac{m_Am_B}{r_{AB}}x_A^{ij}\left\{\frac{1}{3}[7(1-\zeta)-2\zeta(1-2s_A)(1-2s_B)]v_A^2\right. \nonumber \\
&\qquad-\frac{1}{12}[11(1-\zeta)-\zeta(1-2s_A)(1-2s_B)]v_B^2-\frac{1}{6}[11(1-\zeta)-\zeta(1-2s_A)(1-2s_B)](\mathbf{v}_A\cdot\mathbf{v}_B) \nonumber \\
&\qquad-\frac{1}{12}[1-\zeta+\zeta(1-2s_A)(1-2s_B)](\mathbf{v}_A\cdot \mathbf{\hat{n}}_{AB})^2+\frac{1}{6}[1-\zeta+\zeta(1-2s_A)(1-2s_B)](\mathbf{v}_B\cdot \mathbf{\hat{n}}_{AB})^2 \nonumber \\
&\qquad-\frac{1}{6}[1-\zeta+\zeta(1-2s_A)(1-2s_B)](\mathbf{v}_A\cdot \mathbf{\hat{n}}_{AB})(\mathbf{v}_B\cdot \mathbf{\hat{n}}_{AB})+\frac{1}{6}[1-\zeta+\zeta(1-2s_A)(1-2s_B)](\mathbf{a}_A+\mathbf{a}_B)\cdot\mathbf{x}_{AB}\nonumber \\
&\qquad+\sum_{C\neq B}\frac{Gm_C}{r_{BC}}\left[\frac{1}{2}(1-\zeta)^2+\frac{1}{2}\zeta(1-\zeta)(1-2s_A)(1-2s_B)+2\zeta^2(1-2s_A)a_{sB}(1-2s_C)\right. \nonumber \\
&\left.\left.\qquad\quad+\frac{1}{2}\zeta(1-\zeta)(1-2s_B)(1-2s_C)+\frac{1}{2}\zeta(\zeta+2\lambda_1)(1-2s_A)(1-2s_B)(1-2s_C)\right]\right\} \nonumber \\
&\quad+G^2(1-\zeta)\sum_A\sum_{B\neq A}\frac{m_Am_B}{r_{AB}}\left\{-\frac{1}{12}
[7(1-\zeta)-\zeta(1-2s_A)(1-2s_B)]v_A^2x_A^{(i}x_B^{j)}\right. \nonumber \\
&\qquad+\frac{2}{3}(1-\zeta)v_B^2x_A^{(i}x_B^{j)}+\frac{1}{12}[1-\zeta+\zeta(1-2s_A)(1-2s_B)](\mathbf{v}_A\cdot\mathbf{v}_B)x_A^{(i}x_B^{j)} \nonumber \\
&\qquad+\frac{1}{12}[7(1-\zeta)-\zeta(1-2s_A)(1-2s_B)](\mathbf{v}_A\cdot \mathbf{\hat{n}}_{AB})^2x_A^{(i}x_B^{j)} \nonumber \\
&\qquad-\frac{2}{3}(1-\zeta)(\mathbf{v}_B\cdot \mathbf{\hat{n}}_{AB})^2x_A^{(i}x_B^{j)}-\frac{1}{12}[1-\zeta+\zeta(1-2s_A)(1-2s_B)](\mathbf{v}_A\cdot \mathbf{\hat{n}}_{AB})(\mathbf{v}_B\cdot \mathbf{\hat{n}}_{AB})x_A^{(i}x_B^{j)} \nonumber \\
&\qquad-\frac{1}{3}[5(1-\zeta)-\zeta(1-2s_A)(1-2s_B)][(\mathbf{v}_A+\mathbf{v}_B)\cdot\mathbf{x}_{AB}]v_A^{(i}x_A^{j)} \nonumber \\
&\qquad-\frac{1}{6}[11(1-\zeta)-\zeta(1-2s_A)(1-2s_B)][(\mathbf{v}_A+\mathbf{v}_B)\cdot\mathbf{x}_{AB}]v_A^{(i}x_B^{j)} \nonumber \\
&\left.\qquad+\frac{1}{6}[13(1-\zeta)+\zeta(1-2s_A)(1-2s_B)]r_{AB}^2v_A^{ij}-\frac{1}{12}[49(1-\zeta)+\zeta(1-2s_A)(1-2s_B)]r_{AB}^2v_A^{(i}v_B^{j)}\right\} \nonumber \\
&\quad+G^2(1-\zeta)\sum_A\sum_{B\neq A}m_Am_Br_{AB}\left\{-\frac{1}{12}[1-\zeta+\zeta(1-2s_A)(1-2s_B)](\mathbf{a}_A\cdot \mathbf{\hat{n}}_{AB})x_A^{(i}\hat{n}_{AB}^{j)}\right. \nonumber \\
&\left.\qquad-\frac{1}{6}[11(1-\zeta)-\zeta(1-2s_A)(1-2s_B)]a_A^{(i}x_A^{j)}-\frac{1}{12}[23(1-\zeta)-\zeta(1-2s_A)(1-2s_B)]a_B^{(i}x_A^{j)}\right\} \nonumber \\
&\quad+G^3(1-\zeta)^2\sum_A\sum_{B\neq A}m_A^2m_B\left[-3(1-\zeta)-\frac{1}{3}\zeta(1-2s_A)^2-\frac{8}{3}\zeta(1-2s_A)(1-2s_B)\right]\hat{n}_{AB}^{ij} \nonumber \\
&\quad+G^3(1-\zeta)^2\sum_{A \neq B \neq C}m_Am_Bm_C[1-\zeta+\zeta(1-2s_A)(1-2s_B)]\partial_A^i\partial_B^jF(\mathbf{x}_{AC},\mathbf{x}_{BC})\, .
\label{eq:2indexmoment}
\end{align}
The notation $(d/dt)^{-2}$ means that two antiderivatives must be taken of the following expression.  In practice, we do not need to worry about doing this, since we will eventually take two time derivatives of $I_{\text{EW}}^{ij}$ when evaluating \eqref{eq:EWwaveform}.

In the GR limit $\zeta = 0$ (and for completeness, $G = 1$), this expression reduces to WW (4.17).  [Note that (4.17) of WW has a sign error on the term containing $\{(\mathbf{v}_A+\mathbf{v}_B)^2-[(\mathbf{v}_A+\mathbf{v}_B)\cdot\mathbf{\hat{n}}_{AB}]\}/2$.]  There are two important caveats about this expression.  First, the surface moment has been written using the lowest order piece of $\Is_s^i$.  Note also that the terms $-\sum_{C\neq B} m_C \hat{n}_{BC}^k/r_{BC}^2$ from \eqref{eq:integral13} do not appear in that form.  Similar terms occur in the related integral involving $P_{2s}^{kl}$.  The two sets of terms can be combined and then simplified using

\begin{equation}
a_B^i = -\sum_{C\neq B}[1-\zeta+\zeta(1-2s_B)(1-2s_C)]\frac{Gm_C \hat{n}_{BC}^i}{r_{BC}^2} \, .
\end{equation}
This is the Newtonian equation of motion as defined in MW (6.1).  With this substitution, $I_{\text{EW}}^{ij}$ is simpler and easier to compare to the WW results (which use the same trick).  Because of these two choices, \eqref{eq:2indexmoment} is good only for calculations at 2PN order.  When going to higher order, the expression should be reverted to its more generic form (not shown) before adding the explicit higher order contributions; otherwise, the final answer will not be accurate.

\subsection{Three-index moment $I_{\text{EW}}^{ijk}$}
\label{sec:3index}
The rest of the moments are calculated in much the same way as $I_{\text{EW}}^{ij}$.  For the three-index moment, it is useful to rewrite it as

\begin{equation}
I_{\text{EW}}^{ijk} = \tilde{I}_{\text{EW}}^{ijk}+\tilde{I}_{\text{EW}}^{jik}-\tilde{I}_{\text{EW}}^{kij} \, ,
\label{eq:Iijkchunks}
\end{equation}
where

\begin{equation}
\tilde{I}_{\text{EW}}^{ijk} = \int_\Ms\tau^{0i}x^{jk}d^3x + \tilde{I}_S^{ijk} \, .
\end{equation}
Here the volume integral can again be split into compact and field components, $\tilde{I}_C^{ijk}$ and $\tilde{I}_F^{ijk}$.  The surface moment is given by

\begin{equation}
\frac{d}{dt}\tilde{I}_S^{ijk} = \oint_{\partial \Ms}\Lambda_T^{l i}\hat{n}^{jkl}\Rs^4 d^2\Omega \, .
\end{equation}
To get $I_{\text{EW}}^{ijk}$ to 2PN order, we technically need to know the source $\tau^{0i}$ to $O(\rho \epsilon^2)$.  However, there are no terms at that order, so really we need it only to $O(\rho \epsilon^{3/2})$.  Converting MW (4.12b) to the $\rho^*$ density, we find

\begin{equation}
\begin{split}
\tau^{0i} &= \rho^*v^iG(1-\zeta)\left[1+\frac{1}{2}v^2+3G(1-\zeta)U-G\zeta(1-2s) U_s\right]\\
&\quad + G(1-\zeta)\left[G(1-\zeta)\left(\frac{2}{\pi}U^{,j}V^{[j,i]}+\frac{3}{4\pi}\dot{U}U^{,i}\right)-\frac{1}{4\pi}G\zeta\dot{U}_sU_s^{,i}\right] + O(\rho\epsilon^{5/2}) \, .
\end{split}
\label{eq:tau0i}
\end{equation}
The compact moment is given by

\begin{equation}
\tilde{I}_{C}^{ijk} = G(1-\zeta)\sum_A m_Av_A^ix_A^{jk}\left\{1+\frac{1}{2}v_A^2+\sum_{B\neq A}\frac{Gm_B}{r_{AB}}[3(1-\zeta)-\zeta(1-2s_A)(1-2s_B)]\right\} \, .
\end{equation}
There are only two types of field integral, and both can be calculated by the method of Sec.\ \ref{sec:mainintegrals}.  

Finally, we must investigate the surface moment.  The procedure is similar to that for the two-index surface moment, with three essential differences.  First, there is only one time derivative, compared to two in the two-index case.  This means that we only need to consider $\Lambda_T^{ij}$ to $O(\rho \epsilon^{5/2})$.  Second, we are interested in terms in $\Lambda_T^{ij}$ which have $\Rs^{-4}$ dependence.  Finally, if a piece of $\Lambda_T^{ij}$ is to survive the angular integration, it must have an odd number of unit vectors.  Again, only one term contributes to the surface moment: the one involving the scalar dipole \eqref{eq:scalardipole}.  The final expression is given by

\begin{equation}
\frac{d}{dt}\tilde{I}_S^{ijk} = \frac{1}{15}G^2\zeta(1-\zeta)\sum_{A,B}m_A(1-2s_A)m_B(1-2s_B)(2x_A^{i(j}\dot{a}_B^{k)}-3\dot{a}_B^ix_A^{jk}) \, .
\end{equation}
Like the two-index surface moment, this expression uses only the lowest order form of $\Is_s^i$.  This is sufficient for our purposes, but care must be taken in any future work to higher post-Newtonian order.

Adding everything up, we get a final expression for $\tilde{I}_{\text{EW}}^{ijk}$.  It contains pieces at 0.5PN, 1.5PN, and 2PN order,

\begin{equation}
\begin{split}
\tilde{I}_{\text{EW}}^{ijk} &= G(1-\zeta)\sum_Am_Av_A^ix_A^{jk}\left\{1+\frac{1}{2}v_A^2-\frac{1}{2}\sum_{B\neq A}\frac{Gm_B}{r_{AB}}\left[1-\zeta+\zeta(1-2s_A)(1-2s_B)\right]\right\}\\
&\quad-\frac{1}{2}G^2(1-\zeta)\sum_A\sum_{B\neq A}\frac{m_Am_B}{r_{AB}}[1-\zeta+\zeta(1-2s_A)(1-2s_B)](\mathbf{v}_A\cdot \mathbf{\hat{n}}_{AB})\hat{n}_{AB}^ix_A^{jk}\\
&\quad-\frac{1}{12}G^2(1-\zeta)\sum_A\sum_{B\neq A}m_A m_Br_{AB}\left\{2[1-\zeta+\zeta(1-2s_A)(1-2s_B)](\mathbf{v}_A\cdot\mathbf{\hat{n}}_{AB})\hat{n}_{AB}^{ijk}\right.\\
&\left.\qquad+[11(1-\zeta)-\zeta(1-2s_A)(1-2s_B)](2v_A^i\hat{n}_{AB}^{jk}-v_A^j\hat{n}_{AB}^{ik}-v_A^k\hat{n}_{AB}^{ij})\right\}\\
&\quad+\frac{1}{2}G^2(1-\zeta)\sum_A\sum_{B\neq A}m_A m_B\left\{[1-\zeta+\zeta(1-2s_A)(1-2s_B)](\mathbf{v}_A\cdot\mathbf{\hat{n}}_{AB})\hat{n}_{AB}^{i(j}x_A^{k)}\right.\\
&\left.\qquad+[7(1-\zeta)-\zeta(1-2s_A)(1-2s_B)](v_A^ix_A^{(j}\hat{n}_{AB}^{k)}-v_A^{(j}x_A^{k)}\hat{n}_{AB}^i)\right\} \\
&\quad+\frac{1}{15}G^2\zeta(1-\zeta)\left(\frac{d}{dt}\right)^{-1}\left[\sum_{A,B}m_A(1-2s_A)m_B(1-2s_B)(2x_A^{i(j}\dot{a}_B^{k)}-3\dot{a}_B^ix_A^{jk})\right] \, .
\end{split}
\end{equation}
This reduces to WW (4.22) in the GR limit.

\subsection{Four-index moment $I_{\text{EW}}^{ijkl}$}
\label{sec:4index}
To evaluate the four-index moment, we need $\tau^{ij}$ to $O(\rho \epsilon^2)$, which is not given in MW.  We use Eq.\ (4.4c) of \cite{pw00} and \eqref{eq:Lambdas} to find

\begin{equation}
\begin{split}
\tau^{ij} &= (-g)\frac{\varphi}{\phi_0}T^{ij}+\frac{1}{16\pi}\Lambda^{ij}+\frac{1}{16\pi}\Lambda_s^{ij}\\
&= G(1-\zeta)\sigma^{ij}(1+N-3\Psi) \\
&\quad + \frac{1}{16\pi}\left\{\frac{1}{4}\left[N^{,i}N^{,j}-\frac{1}{2}\delta^{ij}(\nabla N)^2\right]+2K^{k,(i}K^{j),k}-K^{k,i}K^{k,j}-K^{i,k}K^{j,k}+2N^{,(i}\dot{K}^{j)}+\frac{1}{2}N^{,(i}B^{,j)} \right.\\
&\left.\qquad -\frac{1}{2}N\left[N^{,i}N^{,j}-\frac{1}{2}\delta^{ij}(\nabla N)^2\right]-\delta^{ij}\left[K^{l,k}K^{[k,l]}+N^{,k}\dot{K}^{k}+\frac{3}{8}\dot{N}^2+\frac{1}{4}\nabla N\cdot \nabla B\right]\right\}\\
&\quad + \frac{1}{16\pi}\left\{(3+2\omega_0)\left[\Psi^{,i}\Psi^{,j}-\frac{1}{2}\delta^{ij}(\nabla \Psi)^2\right]-(3+2\omega_0)\left[2\left(1-\frac{\omega'_0}{3+2\omega_0}\right)\Psi\left(\Psi^{,i}\Psi^{,j}-\frac{1}{2}\delta^{ij}(\nabla \Psi)^2\right)-\frac{1}{2}\delta^{ij}\dot{\Psi}^2\right]\right\}\\
&\quad + O(\rho \epsilon^{5/2}) \, .
\end{split}
\end{equation}
We can ignore all terms with $\delta^{ij}$, because they will be non-TT.  Substituting for $N$, $K^i$, $B$, and $\Psi$, and changing to the $\rho^*$ density, we find an ``effective'' source,

\begin{align}
\tau^{ij}_{\text{eff}} &= \rho^*v^{ij} G(1-\zeta)\left[1+\frac{1}{2}v^2+3G(1-\zeta)U-G\zeta(1-2s)U_s\right] \nonumber \\
&\quad + G^2(1-\zeta)^2\left[\frac{1}{4\pi}U^{,i}U^{,j}+\frac{3}{4\pi}U^{,(i}\Phi_1^{,j)}-\frac{1}{2\pi}G(1-\zeta)U^{,(i}\Phi_2^{,j)}-\frac{1}{2\pi}G\zeta U^{,(i}\Phi^{s,j)}_{2s}+\frac{1}{4\pi}U^{,(i}\ddot{X}^{,j)}\right. \nonumber \\
&\left.\qquad + \frac{2}{\pi}V^{k,(i}V^{j),k}-\frac{1}{\pi}V^{k,i}V^{k,j}-\frac{1}{\pi}V^{i,k}V^{j,k}+\frac{2}{\pi}U^{,(i}\dot{V}^{j)}\right] \nonumber \\
&\quad + G^2\zeta(1-\zeta)\left[\frac{1}{4\pi}U_s^{,i}U_s^{,j}-\frac{1}{4\pi}U_s^{,(i}\Phi^{s,j)}_1-\frac{1}{2 \pi}G(1-\zeta)U_s^{,(i}\Phi^{s,j)}_2-\frac{1}{2\pi}G(2\lambda_1+\zeta)U_s^{,(i}\Phi^{s,j)}_{2s}\right. \nonumber \\
&\left.\qquad-\frac{2}{\pi}G\zeta U_s^{,(i}(\Sigma(a_sU_s))^{,j)}+ \frac{1}{4\pi}U_s^{,(i}\ddot{X}_s^{,j)}\right]+O(\rho\epsilon^{5/2}) \, .
\end{align}
The compact moment is

\begin{equation}
I_C^{ijkl} = G(1-\zeta)\sum_A m_Av_A^{ij}x_A^{kl}\left\{1+\frac{1}{2}v_A^2+\sum_{B\neq A}\frac{Gm_B}{r_{AB}}[3(1-\zeta)-\zeta(1-2s_A)(1-2s_B)]\right\} \, .
\end{equation}
There are eight different field integrals, although some are very similar.  They can also be evaluated using the methods of Sec. \ref{sec:mainintegrals}.  Adding everything up, we find the final result, which contains 1PN and 2PN terms,

\begin{align}
I_{\text{EW}}^{ijkl} &= G(1-\zeta)\sum_A m_Ax_A^{kl}\left\{v_A^{ij}-\frac{1}{2}\sum_{B\neq A}\frac{Gm_B}{r_{AB}}[1-\zeta+\zeta(1-2s_A)(1-2s_B)]\hat{n}_{AB}^{ij}\right\} \nonumber \\
&\quad + \frac{1}{12}G^2(1-\zeta)\sum_A\sum_{B\neq A}[1-\zeta+\zeta(1-2s_A)(1-2s_B)]m_Am_Br_{AB}\hat{n}_{AB}^{ij}(\hat{n}_{AB}^{kl}-\delta^{kl}) \nonumber \\
&\quad +\frac{1}{2}G(1-\zeta)\sum_A m_Av_A^2 v_A^{ij}x_A^{kl} \nonumber \\
&\quad + G^2(1-\zeta)\sum_A\sum_{B\neq A}\frac{m_Am_B}{r_{AB}}x_A^{kl}\left\{-\frac{1}{2}[1-\zeta+\zeta(1-2s_A)(1-2s_B)]v_A^{ij}\right. \nonumber \\
&\qquad -\frac{1}{4}[7(1-\zeta)-\zeta(1-2s_A)(1-2s_B)]v_A^2\hat{n}_{AB}^{ij} \nonumber \\
&\qquad +\frac{1}{2}\hat{n}_{AB}^{ij}\sum_{C\neq A}\frac{Gm_C}{r_{AC}}[(1-\zeta)^2+\zeta(1-\zeta)(1-2s_A)(1-2s_C)+\zeta(1-\zeta)(1-2s_A)(1-2s_B) \nonumber \\
&\qquad \quad+ \zeta(2\lambda_1+\zeta)(1-2s_A)(1-2s_B)(1-2s_C)+4\zeta^2a_{sA}(1-2s_B)(1-2s_C)] \nonumber \\
&\qquad +\frac{1}{2}\hat{n}_{AB}^{ij}\sum_{C\neq B}\frac{Gm_C}{r_{BC}}[(1-\zeta)^2+\zeta(1-\zeta)(1-2s_B)(1-2s_C)+\zeta(1-\zeta)(1-2s_A)(1-2s_B) \nonumber \\
&\qquad \quad+ \zeta(2\lambda_1+\zeta)(1-2s_A)(1-2s_B)(1-2s_C)+4\zeta^2(1-2s_A)a_{sB}(1-2s_C)] \nonumber \\
&\qquad + [3(1-\zeta)-\zeta(1-2s_A)(1-2s_B)](\mathbf{v}_A\cdot\mathbf{\hat{n}}_{AB})v_A^{(i}\hat{n}_{AB}^{j)}+\frac{3}{4}[1-\zeta+\zeta(1-2s_A)(1-2s_B)](\mathbf{v}_A\cdot\mathbf{\hat{n}}_{AB})^2\hat{n}_{AB}^{ij} \nonumber \\
&\qquad-\frac{1}{2}[7(1-\zeta)-\zeta(1-2s_A)(1-2s_B)]a_A^{(i}x_{AB}^{j)}-\frac{1}{4}[1-\zeta+\zeta(1-2s_A)(1-2s_B)](\mathbf{a}_A\cdot\mathbf{x}_{AB})\hat{n}_{AB}^{ij} \nonumber \\
&\left.\qquad-4(1-\zeta)(\mathbf{v}_B\cdot\mathbf{\hat{n}}_{AB})v_A^{(i}\hat{n}_{AB}^{j)}+2(1-\zeta)(\mathbf{v}_A\cdot\mathbf{v}_B)\hat{n}_{AB}^{ij}\vphantom{\frac{1}{2}}\right\} \nonumber \\
&\quad +G^2(1-\zeta)\sum_A\sum_{B\neq A}m_Am_Br_{AB}\delta^{kl}\left\{-\frac{1}{24}[7(1-\zeta)-\zeta(1-2s_A)(1-2s_B)]v_A^2\hat{n}_{AB}^{ij}\right. \nonumber \\
&\qquad +\frac{1}{6}\hat{n}_{AB}^{ij}\sum_{C\neq A}\frac{Gm_C}{r_{AC}}[(1-\zeta)^2+\zeta(1-\zeta)(1-2s_A)(1-2s_C)+\zeta(1-\zeta)(1-2s_A)(1-2s_B) \nonumber \\
&\qquad\quad+\zeta(2\lambda_1+\zeta)(1-2s_A)(1-2s_B)(1-2s_C)+4\zeta^2a_{sA}(1-2s_B)(1-2s_C)] \nonumber \\
&\qquad +\frac{1}{12}[7(1-\zeta)-\zeta(1-2s_A)(1-2s_B)]v_A^{ij}+\frac{1}{6}[3(1-\zeta)-\zeta(1-2s_A)(1-2s_B)](\mathbf{v}_A\cdot\mathbf{\hat{n}}_{AB})v_A^{(i}\hat{n}_{AB}^{j)} \nonumber \\
&\qquad +\frac{1}{24}[1-\zeta+\zeta(1-2s_A)(1-2s_B)](\mathbf{v}_A\cdot\mathbf{\hat{n}}_{AB})^2\hat{n}_{AB}^{ij}+\frac{1}{12}[7(1-\zeta)-\zeta(1-2s_A)(1-2s_B)]a_A^{(i}x_{AB}^{j)} \nonumber \\
&\qquad -\frac{1}{24}[1-\zeta+\zeta(1-2s_A)(1-2s_B)](\mathbf{a}_A\cdot\mathbf{x}_{AB})\hat{n}_{AB}^{ij}-\frac{2}{3}(1-\zeta)(\mathbf{v}_B\cdot\mathbf{\hat{n}}_{AB})v_A^{(i}\hat{n}_{AB}^{j)} \nonumber \\
&\left.\qquad +\frac{4}{3}(1-\zeta)v_A^{(i}v_B^{j)}+\frac{1}{3}(1-\zeta)(\mathbf{v}_A\cdot\mathbf{v}_B)\hat{n}_{AB}^{ij}\right\} \nonumber \\
&\quad +G^2(1-\zeta)\sum_A\sum_{B\neq A}m_Am_Br_{AB}\hat{n}_{AB}^{ijkl}\left\{-\frac{1}{24}[15(1-\zeta)-\zeta(1-2s_A)(1-2s_B)]v_A^2\right. \nonumber \\
&\qquad -\frac{1}{6}\sum_{C\neq A}\frac{Gm_C}{r_{AC}}[(1-\zeta)^2+\zeta(1-\zeta)(1-2s_A)(1-2s_C)+\zeta(1-\zeta)(1-2s_A)(1-2s_B) \nonumber \\
&\qquad \quad+\zeta(2\lambda_1+\zeta)(1-2s_A)(1-2s_B)(1-2s_C)+4\zeta^2a_{sA}(1-2s_B)(1-2s_C)]\nonumber \\
&\qquad +\frac{3}{8}[1-\zeta+\zeta(1-2s_A)(1-2s_B)](\mathbf{v}_A\cdot\mathbf{\hat{n}}_{AB})^2-\frac{1}{8}[1-\zeta+\zeta(1-2s_A)(1-2s_B)](\mathbf{a}_A\cdot\mathbf{x}_{AB}) \nonumber \\
&\left.\qquad -\frac{1}{3}(1-\zeta)(\mathbf{v}_A\cdot\mathbf{v}_B)\right\} \nonumber \\
&\quad +G^2(1-\zeta)\sum_A\sum_{B\neq A}m_Am_B\hat{n}_{AB}^{(k}x_A^{l)}\left\{\frac{1}{6}[11(1-\zeta)-\zeta(1-2s_A)(1-2s_B)]v_A^2\hat{n}_{AB}^{ij}\right. \nonumber \\
&\qquad +\frac{2}{3}[5(1-\zeta)-\zeta(1-2s_A)(1-2s_B)]v_A^{ij}-\frac{4}{3}[2(1-\zeta)-\zeta(1-2s_A)(1-2s_B)](\mathbf{v}_A\cdot\mathbf{n}_{AB})v_A^{(i}\hat{n}_{AB}^{j)} \nonumber \\
&\qquad -[1-\zeta+\zeta(1-2s_A)(1-2s_B)](\mathbf{v}_A\cdot\mathbf{\hat{n}}_{AB})^2\hat{n}_{AB}^{ij}+\frac{2}{3}[5(1-\zeta)-\zeta(1-2s_A)(1-2s_B)]a_A^{(i}x_{AB}^{j)} \nonumber \\
&\left.\qquad +\frac{1}{3}[1-\zeta+\zeta(1-2s_A)(1-2s_B)](\mathbf{a}_A\cdot\mathbf{x}_{AB})\hat{n}_{AB}^{ij}+4(1-\zeta)(\mathbf{v}_B\cdot\mathbf{\hat{n}}_{AB})v_A^{(i}\hat{n}_{AB}^{j)}\right\} \nonumber \\
&\quad +G^2(1-\zeta)\sum_A\sum_{B\neq A}m_Am_Br_{AB}\hat{n}_{AB}^{ij}\left\{\frac{1}{6}[1-\zeta+\zeta(1-2s_A)(1-2s_B)](\mathbf{v}_A\cdot\mathbf{\hat{n}}_{AB})v_A^{(k}\hat{n}_{AB}^{l)}\right. \nonumber \\
&\qquad +\frac{1}{12}[1-\zeta+\zeta(1-2s_A)(1-2s_B)]v_A^{kl}-\frac{1}{12}[1-\zeta+\zeta(1-2s_A)(1-2s_B)]a_A^{(k}x_{AB}^{l)} \nonumber \\
&\left.\qquad +\frac{1}{6}[1-\zeta+\zeta(1-2s_A)(1-2s_B)]a_A^{(k}x_A^{l)}\right\} \nonumber \\
&\quad +G^2(1-\zeta)\sum_A\sum_{B\neq A}m_Am_Br_{AB}\hat{n}_{AB}^{kl}\left\{-\frac{1}{12}[13(1-\zeta)-3\zeta(1-2s_A)(1-2s_B)]v_A^{ij}\right. \nonumber \\
&\qquad +\frac{1}{6}[5(1-\zeta)-3\zeta(1-2s_A)(1-2s_B)](\mathbf{v}_A\cdot\mathbf{\hat{n}}_{AB})v_A^{(i}\hat{n}_{AB}^{j)} \nonumber \\
&\left.\qquad -\frac{1}{12}[13(1-\zeta)-3\zeta(1-2s_A)(1-2s_B)]a_A^{(i}x_{AB}^{j)}-\frac{4}{3}(1-\zeta)(\mathbf{v}_B\cdot\mathbf{\hat{n}}_{AB})v_A^{(i}\hat{n}_{AB}^{j)}-\frac{2}{3}(1-\zeta)v_A^{(i}v_B^{j)}\right\} \nonumber \\
&\quad +G^2(1-\zeta)\sum_A\sum_{B\neq A}m_Am_Br_{AB}\left\{\frac{1}{3}[3(1-\zeta)-\zeta(1-2s_A)(1-2s_B)]v_A^{(i}\hat{n}_{AB}^{j)}v_A^{(k}\hat{n}_{AB}^{l)}\right. \nonumber \\
&\left.\qquad -\frac{4}{3}(1-\zeta)v_A^{(i}\hat{n}_{AB}^{j)}v_B^{(k}\hat{n}_{AB}^{l)}\right\} \nonumber \\
&\quad +G^2(1-\zeta)\sum_A\sum_{B\neq A}m_Am_B\left\{-\frac{2}{3}[5(1-\zeta)-\zeta(1-2s_A)(1-2s_B)]v_A^{(i}\hat{n}_{AB}^{j)}v_A^{(k}x_A^{l)}\right. \nonumber \\
&\left.\qquad -\frac{1}{3}[1-\zeta+\zeta(1-2s_A)(1-2s_B)](\mathbf{v}_A\cdot\mathbf{\hat{n}}_{AB})\hat{n}_{AB}^{ij}v_A^{(k}x_A^{l)}+4(1-\zeta)v_A^{(i}\hat{n}_{AB}^{j)}v_B^{(k}x_A^{l)}\right\} \, .
\end{align}
This reduces to WW (4.26) in the GR limit.

\subsection{Five-index moment $I_{\text{EW}}^{ijkl m}$}
\label{sec:5index}
Because of the time derivative in \eqref{eq:EWgt4}, we only need the source to $O(\rho \epsilon^{3/2})$.  However, there is no contribution at that order, so $O(\rho\epsilon)$ will suffice,

\begin{equation}
\tau^{ij} = \rho^*v^{ij}G(1-\zeta)+\frac{1}{4\pi}G^2(1-\zeta)^2\left[U^{,i}U^{,j}-\frac{1}{2}\delta^{ij}(\nabla U)^2\right]+\frac{1}{4\pi}G^2\zeta(1-\zeta)\left[U_s^{,i}U_s^{,j}-\frac{1}{2}\delta^{ij}(\nabla U_s)^2\right]+O(\rho\epsilon^2) \, .
\label{eq:tauijlowest}
\end{equation}
We can, of course, discard the delta function terms, which are non-TT.  There is only one type of field integral to calculate.  The final five-index moment is entirely 1.5PN order,

\begin{equation}
\begin{split}
I_{\text{EW}}^{ijkl m} &= \frac{1}{3}G(1-\zeta)\frac{d}{dt}\left\{\sum_Am_Ax_A^{kl m}\left[v_A^{ij}-\frac{1}{2}\sum_{B\neq A}\frac{Gm_B}{r_{AB}}[1-\zeta+\zeta(1-2s_A)(1-2s_B)]\hat{n}_{AB}^{ij}\right]\right.\\
&\left.\quad + \frac{1}{4}\sum_A\sum_{B\neq A}Gm_Am_Br_{AB}[1-\zeta+\zeta(1-2s_A)(1-2s_B)]\hat{n}_{AB}^{ij}x_A^{(k}(\hat{n}_{AB}^{l m)}-\delta^{l m)})\right\} \, .
\end{split}
\end{equation}
This reduces to WW (4.27a) in the GR limit.

\subsection{Six-index moment $I_{\text{EW}}^{ijkl mn}$}
\label{sec:6index}
With two time derivatives, we only need the source to $O(\rho \epsilon)$, \eqref{eq:tauijlowest}.  There is again only one field integral to calculate, although it is the most difficult by far.  The final moment is entirely 2PN order,

\begin{equation}
\begin{split}
I_{\text{EW}}^{ijkl mn} &= \frac{1}{12}G(1-\zeta)\frac{d^2}{dt^2}\left\{\sum_Am_Ax_A^{kl mn}\left[v_A^{ij}-\frac{1}{2}\sum_{B\neq A}\frac{Gm_B}{r_{AB}}[1-\zeta+\zeta(1-2s_A)(1-2s_B)]\hat{n}_{AB}^{ij}\right]\right.\\
&\quad + \frac{1}{2}\sum_A\sum_{B\neq A}Gm_Am_Br_{AB}[1-\zeta+\zeta(1-2s_A)(1-2s_B)]\left[\vphantom{\frac{1}{10}}\hat{n}_{AB}^{ij}x_A^{(kl}(\hat{n}_{AB}^{mn)}-\delta^{mn)})\right.\\
&\left.\left.\quad -\frac{1}{10}x_{AB}^{ij}(2\hat{n}_{AB}^{kl mn}-2\hat{n}_{AB}^{(kl}\delta^{mn)}-\delta^{(kl}\delta^{mn)})\right]
\vphantom{\sum_Am_Ax_A^{kl mn}\left[v_A^{ij}-\frac{1}{2}\sum_{B\neq A}\frac{Gm_B}{r_{AB}}[1-\zeta+\zeta(1-2s_A)(1-2s_B)]\hat{n}_{AB}^{ij}\right]}
\right\} \, .
\end{split}
\end{equation}
This reduces to WW (4.27b) in the GR limit.

\section{Two-body EW moments}
\label{sec:twobody}
So far we have written down expressions for the EW moments which are fully general and can be applied to a system with any number of compact objects.  Now we convert the moments to the two-body case relevant for compact binaries.  We define the masses $m_A$ of the bodies as $m_1$ and $m_2$.  They have positions $\mathbf{x}_1$ and $\mathbf{x}_2$ and velocities $\mathbf{v}_1$ and $\mathbf{v}_2$.  

It will be useful to have the moments expressed in terms of relative variables, $r = r_{12}$, $\mathbf{x} = \mathbf{x}_{1}-\mathbf{x}_2$, $\mathbf{\hat{n}} = \mathbf{\hat{n}}_{12} = \mathbf{x}/r$, and $\mathbf{v} = \mathbf{v}_1-\mathbf{v}_2$.  (Note that we no longer use $\mathbf{\hat{n}}$ to represent unit normals at the surface of $\Ms$.)  In general relativity, the relationship between the individual variables and the relative variables can be fixed by writing down the conserved linear momentum.  The loss of momentum to gravitational waves occurs at 2.5PN order, beyond what we need to worry about.  However, in scalar-tensor theory, dipole radiation reaction enters at 1.5PN order, so it must be taken into account.  From MW (6.9) and (6.10), we find that the individual and relative velocities are related by

\begin{subequations}
\begin{align}
v_1^i &= \frac{m_2}{m}v^i + \delta^i \, ,\\
v_2^i &= -\frac{m_1}{m}v^i + \delta^i \, ,
\end{align}
\end{subequations}
with

\begin{equation}
\delta^i = \frac{1}{2}\eta\frac{\delta m}{m}\left[\left(v^2-\frac{G\alpha m}{r}\right)v^i-\frac{G\alpha m}{r^2}\dot{r}x^i\right]-\frac{2}{3}\zeta\eta\Ss_-\left(\Ss_++\frac{\delta m}{m}\Ss_-\right)\left(\frac{G\alpha m}{r}\right)^2\hat{n}^i+O(\epsilon^2) \, .
\label{eq:vdelta}
\end{equation} 
Here

\begin{equation}
\alpha \equiv 1-\zeta + \zeta(1-2s_1)(1-2s_2)
\label{eq:alpha}
\end{equation}
is a scalar-tensor parameter that enters the equations of motion at Newtonian order.  We also have defined

\begin{subequations}
\begin{align}
\Ss_+ &\equiv \alpha^{-1/2}(1-s_1-s_2) \, ,\\
\Ss_- &\equiv \alpha^{-1/2}(s_2-s_1) \, .
\end{align}
\label{eq:S+S-}
\end{subequations}
Finally, we have the usual variables $m \equiv m_1+m_2$, $\mu \equiv m_1m_2/m$, $\eta \equiv \mu/m$, and $\delta m \equiv m_1-m_2$.  The first term in \eqref{eq:vdelta} is of relative 1PN order and the second term, representing the dipole radiation, is of relative 1.5PN order.  We can find an antiderivative and write down the corresponding position relations,

\begin{subequations}
\begin{align}
x_1^i &= \frac{m_2}{m}x^i + \delta_x^i \, ,\\
x_2^i &= -\frac{m_1}{m}x^i + \delta_x^i \, ,
\end{align}
\end{subequations}
with

\begin{equation}
\delta_x^i = \frac{1}{2}\eta\frac{\delta m}{m}\left(v^2-\frac{G\alpha m}{r}\right)x^i+\frac{2}{3}\zeta\eta\Ss_-\left(\Ss_++\frac{\delta m}{m}\Ss_-\right)G\alpha mv^i+O(\epsilon^2) \, .
\label{eq:deltax}
\end{equation}
It may seem that we would need the $O(\epsilon^2)$ terms in $\delta^i$ and $\delta_x^i$ in order to calculate 2PN gravitational waves.  As it turns out, the 2PN piece of $\delta^i$ is not needed anywhere, while the 2PN piece of $\delta_x^i$ is only needed in the first (0PN) term of $I_{\text{EW}}^{ij}$, where it cancels exactly.

To simplify our expressions, we introduce other scalar-tensor parameters from MW,

\begin{subequations}
\begin{align}
\bar{\gamma} &\equiv -2\alpha^{-1}\zeta(1-2s_1)(1-2s_2) \, , \\
\bar{\beta}_1 &\equiv \alpha^{-2}\zeta(1-2s_2)^2(\lambda_1(1-2s_1)+2\zeta s_1') \, , \\
\bar{\beta}_2 &\equiv \alpha^{-2}\zeta(1-2s_1)^2(\lambda_1(1-2s_2)+2\zeta s_2') \, , \\
\bar{\delta}_1 &\equiv \alpha^{-2}\zeta(1-\zeta)(1-2s_1)^2 \, ,\\
\bar{\delta}_2 &\equiv \alpha^{-2}\zeta(1-\zeta)(1-2s_2)^2 \, ,\\
\bar{\chi}_1 &\equiv \alpha^{-3}\zeta(1-2s_2)^3[(\lambda_2-4\lambda_1^2+\zeta\lambda_1)(1-2s_1)-6\zeta\lambda_1s_1'+2\zeta^2s_1''] \, , \\
\bar{\chi}_2 &\equiv \alpha^{-3}\zeta(1-2s_1)^3[(\lambda_2-4\lambda_1^2+\zeta\lambda_1)(1-2s_2)-6\zeta\lambda_1s_2'+2\zeta^2s_2''] \, .
\end{align}
\label{eq:STparameters}
\end{subequations}
We also use the notation 

\begin{subequations}
\begin{align}
\xi_+ &\equiv \frac{1}{2}(\xi_1+\xi_2) \, ,\\
\xi_- &\equiv \frac{1}{2}(\xi_1-\xi_2) \, ,
\end{align}
\end{subequations}
where $\xi$ is one of $\bar{\beta}$, $\bar{\delta}$, or $\bar{\chi}$.  (This notation should not be confused with $\Ss_+$ and $\Ss_-$.)  Note that the $\bar{\chi}$ parameters do not occur in the EW moments; however, they do appear in the equations of motion, which we will need shortly.

Since acceleration terms only appear at 1.5PN (in $I_{\text{EW}}^{ij}$) or 2PN order (in $I_{\text{EW}}^{ij}$, $I_{\text EW}^{ijk}$, and $I_{\text{EW}}^{ijkl}$), we can replace them with the Newtonian expressions,

\begin{subequations}
\begin{align}
a_1^i &= -\frac{G\alpha m_2}{r^2}\hat{n}^i \, ,\\
a_2^i &= \frac{G\alpha m_1}{r^2}\hat{n}^i \, .
\end{align}
\end{subequations}
This will bring the expressions in line with WW (6.6) and allow an easy comparison.  Finally, we evaluate \eqref{eq:Iijkchunks} to get the total three-index moment.  The result is a set of simplified, two-body EW moments expressed in relative coordinates,

\begin{subequations}
\begin{equation}
\begin{split}
I_{\text{EW}}^{ij} &= G(1-\zeta)\mu x^{ij}\left\{1+\frac{1}{2}(1-3\eta)v^2-\frac{1}{2}(1-2\eta)\frac{G\alpha m}{r}\right\}\\
&\quad-\frac{8}{3}G(1-\zeta)\mu\eta\zeta \Ss_-^2\left(\frac{d}{dt}\right)^{-2}\left[\left(\frac{G\alpha m}{r}\right)^2\left(\dot{r}\hat{n}^{ij}-\frac{1}{3}\hat{n}^{(i}v^{j)}\right)\right]\\
&\quad+G(1-\zeta)\mu x^{ij}\left\{\frac{3}{8}(1-7\eta+13\eta^2)v^4+\left[\frac{1}{12}(28-79\eta-54\eta^2)+\frac{1}{2}\bar{\gamma}(3-10\eta)\right]v^2\frac{G\alpha m}{r}\right.\\
&\qquad+\left[-\frac{1}{4}(5+27\eta-4\eta^2)-\frac{2}{3}\bar{\gamma}(1+6\eta)-\frac{1}{12}\bar{\gamma}^2-\frac{1}{3}\left(\bar{\delta}_++\frac{\delta m}{m}\bar{\delta}_-\right)+\eta\left(\bar{\beta}_++\frac{\delta m}{m}\bar{\beta}_-\right)\right]\left(\frac{G\alpha m}{r}\right)^2\\
&\left.\qquad-\frac{1}{12}(1-13\eta+30\eta^2)\dot{r}^2\frac{G\alpha m}{r}\right\}\\
&\quad+G^2(1-\zeta)\alpha\mu mr\left\{\left[\frac{1}{6}(13+23\eta)+\bar{\gamma}(1+2\eta)\right]v^{ij}-\left(\frac{5}{3}+\bar{\gamma}\right)(1-4\eta)\dot{r}v^{(i}\hat{n}^{j)}\right\} \, ,
\end{split}
\label{eq:IijTB}
\end{equation}

\begin{equation}
\begin{split}
I_{\text{EW}}^{ijk} &= G(1-\zeta)\mu\frac{\delta m}{m}\left\{x^{ij}v^k-2v^{(i}x^{j)k}-v^{(i}x^{j)k}\left[(1-5\eta)v^2+\frac{1}{3}(7+12\eta+6\bar{\gamma})\frac{G\alpha m}{r}\right]\right.\\
&\left.\qquad+\frac{1}{2}x^{ij}v^k\left[(1-5\eta)v^2+\frac{1}{3}(17+12\eta+12\bar{\gamma})\frac{G\alpha m}{r}\right]+\frac{1}{6}(1-6\eta)\frac{G\alpha m}{r^2}\dot{r}x^{ijk}\right\}\\
&\quad+\frac{2}{3}G^2(1-\zeta)\alpha \mu^2\zeta\Ss_-\left(\Ss_++\frac{\delta m}{m}\Ss_-\right)\left(2v^{ij}x^k-\frac{G\alpha m}{r^3}x^{ijk}\right) \\
&\quad+\frac{2}{15}G(1-\zeta)\mu\eta\zeta\Ss_-\left(\Ss_+-\frac{\delta m}{m}\Ss_-\right)\left(\frac{d}{dt}\right)^{-1}\left[\frac{(G\alpha m)^2}{r}(-3\dot{r}\hat{n}^{ijk}+6v^{(i}\hat{n}^{j)k}-5\hat{n}^{ij}v^k)\right] \, ,
\end{split}
\label{eq:IijkTB}
\end{equation}

\begin{align}
I_{\text{EW}}^{ijkl} &= G(1-\zeta)\mu x^{kl}(1-3\eta)\left(v^{ij}-\frac{1}{3}\hat{n}^{ij}\frac{G\alpha m}{r}\right)-\frac{1}{6}G^2(1-\zeta)\alpha \mu mr\hat{n}^{ij}\delta^{kl} \nonumber \\
&\quad+G(1-\zeta)\mu x^{kl}\left\{\vphantom{\left(\frac{G\alpha m}{r}\right)^2}\frac{1}{2}(1-9\eta+21\eta^2)v^2v^{ij}+\left[-\frac{1}{24}(13-46\eta+36\eta^2)-\frac{1}{3}\bar{\gamma}(1-3\eta)\right]v^2\hat{n}^{ij}\frac{G\alpha m}{r}\right. \nonumber \\
&\qquad+\left[\frac{1}{4}(7-10\eta-36\eta^2)+\frac{4}{3}\bar{\gamma}(1-3\eta)\right]v^{ij}\frac{G\alpha m}{r}+\left[\frac{1}{6}(7-12\eta-36\eta^2)+\frac{2}{3}\bar{\gamma}(1-3\eta)\right]\dot{r}v^{(i}\hat{n}^{j)}\frac{G\alpha m}{r} \nonumber \\
&\qquad+\frac{1}{8}(1-6\eta+12\eta^2)\dot{r}^2\hat{n}^{ij}\frac{G\alpha m}{r} \nonumber \\
&\left.\qquad+\left[\frac{1}{24}(37-122\eta+48\eta^2)+\frac{2}{3}\left(\bar{\gamma}+\bar{\beta}_+-\frac{\delta m}{m}\bar{\beta}_-\right)(1-3\eta)\right]\hat{n}^{ij}\left(\frac{G\alpha m}{r}\right)^2\right\} \nonumber \\
&\quad+G^2 (1-\zeta)\alpha \mu mr \delta^{kl} \left\{\left[\frac{1}{12}(7-46\eta)+\frac{1}{3}\bar{\gamma}(1-6\eta)\right]v^{ij}+\left[-\frac{1}{24}(7+2\eta)-\frac{1}{6}\bar{\gamma}\right]v^2\hat{n}^{ij}\right. \nonumber \\
&\left.\qquad+\left[\frac{1}{6}(3+2\eta)+\frac{1}{3}\bar{\gamma}\right]\dot{r}v^{(i}\hat{n}^{j)}+\frac{1}{24}(1-2\eta)\dot{r}^2\hat{n}^{ij}+\left[-\frac{3}{8}+\frac{1}{3}\left(-\bar{\gamma}+\bar{\beta}_+-\frac{\delta m}{m}\bar{\beta}_-\right)\right]\hat{n}^{ij}\frac{G\alpha m}{r}\right\} \nonumber \\
&\quad+G^2(1-\zeta)\alpha \mu mr\left\{\frac{1}{12}(1-2\eta)\hat{n}^{ij}v^{kl}-\frac{1}{6}(1-4\eta)\dot{r}\hat{n}^{ij}v^{(k}\hat{n}^{l)}\right. \nonumber \\
&\left.\qquad+\left[-\frac{1}{3}(7-20\eta)-\frac{4}{3}\bar{\gamma}(1-3\eta)\right]v^{(i}\hat{n}^{j)}v^{(k}\hat{n}^{l)}\right\} \, ,
\label{eq:IijklTB}
\end{align}

\begin{equation}
I_{\text{EW}}^{ijkl m} = -\frac{1}{3}G(1-\zeta)\mu\frac{\delta m}{m}\frac{d}{dt}\left\{(1-2\eta)\left(v^{ij}-\frac{1}{4}\hat{n}^{ij}\frac{G\alpha m}{r}\right)x^{kl m}-\frac{1}{4}G\alpha mr\hat{n}^{ij}x^{(k}\delta^{l m)}\right\} \, ,
\label{eq:IijklmTB}
\end{equation}

\begin{equation}
\begin{split}
I_{\text{EW}}^{ijkl mn} &= \frac{1}{12}G(1-\zeta)\mu\frac{d^2}{dt^2}\left\{(1-5\eta+5\eta^2)\left(v^{ij}-\frac{1}{5}\hat{n}^{ij}\frac{G\alpha m}{r}\right)x^{kl mn}-\frac{1}{10}(3-10\eta)G\alpha mr\hat{n}^{ij}x^{(kl}\delta^{mn)}\right.\\
&\left.\quad+\frac{1}{10}G\alpha mrx^{ij}\delta^{(kl}\delta^{mn)}\right\} \, .
\end{split}
\label{eq:IijklmnTB}
\end{equation}
\end{subequations}
For the most part, these moments have the same basic forms as those in WW, with the coefficients altered to include the scalar-tensor parameters \eqref{eq:alpha}, \eqref{eq:S+S-}, and \eqref{eq:STparameters}.  As we saw in Sec. \ref{sec:EWmoments}, both $I_{\text{EW}}^{ij}$ and $I_{\text{EW}}^{ijk}$ have new terms at 1.5PN and 2PN order, respectively, resulting from the nonvanishing surface moments.  The conversion to relative coordinates introduces another new term in $I_{\text{EW}}^{ijk}$ at 2PN order.  It arises because of the interaction between the lowest order (0.5PN) piece of $I_{\text{EW}}^{ijk}$ and the 1.5PN (dipole radiation) terms in \eqref{eq:vdelta} and \eqref{eq:deltax}.  No such new term is created in the two-index moment: The 1.5PN piece of \eqref{eq:deltax} cancels out just like the (unshown) 2PN piece.

The final step is to take time derivatives.  The equation for $\tilde{h}^{ij}$, \eqref{eq:EWwaveform}, shows that we must take two derivatives of each moment.  The five- and six-index moments also contain their own time derivatives.  Along the way, we need to substitute the relative equation of motion for each acceleration $a^i$.  We take this result from MW, keeping terms up to 2PN order,

\begin{equation}
a^i = -\frac{G\alpha m}{r^2}\hat{n}^i+\frac{G\alpha m}{r^2}(A_{PN}\hat{n}^i+B_{PN}\dot{r}v^i)+\frac{8}{5}\eta\frac{(G\alpha m)^2}{r^3}(A_{1.5PN}\dot{r}\hat{n}^i-B_{1.5PN}v^i)+\frac{G\alpha m}{r^2}(A_{2PN}\hat{n}^i+B_{2PN}\dot{r}v^i) \, ,
\end{equation}
where $A_{PN}$, $B_{PN}$, $A_{1.5PN}$, $B_{1.5PN}$, $A_{2PN}$, and $B_{2PN}$ are given in MW (1.5).  Of these, $A_{PN}$, $A_{2PN}$, and $B_{2PN}$ depend on time, while the others are constant.  It turns out that only $A_{PN}$ is differentiated in the process of calculating the GWs.  From $x^i = r\hat{n}^i$, we can also find

\begin{equation}
\ddot{r}=\frac{v^2}{r}-\frac{\dot{r}^2}{r}-\frac{G\alpha m}{r^2}+\frac{G\alpha m}{r^2}(A_{PN}+\dot{r}^2B_{PN})+\frac{8}{5}\eta\frac{(G\alpha m)^2}{r^3}\dot{r}(A_{1.5PN}-B_{1.5PN})+\frac{G\alpha m}{r^2}(A_{2PN}+\dot{r}^2B_{2PN}) \, .
\end{equation}
While the Newtonian, 1PN, and 2PN terms are modified from the GR equations of motion, the 1.5PN terms, caused by dipole radiation reaction, are entirely new.  They will cause the introduction of more new terms in the waveform, arising from the two-index EW moment (at 1.5PN order) and the three-index EW moment (at 2PN order).  We hold off on presenting the final results until Sec.\ \ref{sec:results}.

\section{Radiation-zone integrals}
\label{sec:RZintegrals}

So far, our calculation of the gravitational waveform has only considered the contribution from the near zone.  We must also calculate the contribution from the radiation zone.  To do so, we will need to evaluate \eqref{eq:hRR} for $\tilde{h}^{ij}$, dropping all terms which fall off faster than $1/R$.

The first step is to derive an expression for the source $\tau^{ij}$ in the radiation zone.  Since there are no compact sources at $R > \Rs$, $\tau^{ij}$ is composed purely of field terms.  The fields $N = \tilde{h}^{00}$, $K^i = \tilde{h}^{0i}$, $B^{ij} = \tilde{h}^{ij}$, and $\Psi = \varphi-1$ themselves can be found by summing near-zone and radiation-zone contributions.  

\subsection{Radiation-zone fields: Near-zone contributions}
\label{sec:RNfields}
The near-zone contributions can be found using \eqref{eq:hRN} and \eqref{eq:phiRN}.  Recall that the Epstein-Wagoner construction \eqref{eq:EWwaveform} is a special case of \eqref{eq:hRN} for $\tilde{h}^{ij}$ in the far-away zone.  Now we want expressions for all of the fields at arbitrary $R$.

We need to evaluate the various moments defined in \eqref{eq:hM} and \eqref{eq:phiM}.  We begin with the moments of $\tau^{00}$.  We use the expression in \eqref{eq:tau00}, but we only need it to $O(\rho\epsilon)$.  [Actually, all the sources need to be evaluated to $O(\rho \epsilon^{3/2})$, but only $\tau^{0i}$ has terms at that order.]  The integration over the field terms uses the same strategies we used in calculating the EW moments.  We find

\begin{equation}
\begin{split}
M^{00} &= G(1-\zeta)\sum_A m_A\left\{1+\frac{1}{2}v_A^2-\frac{1}{2}\sum_{B\neq A}\frac{G\alpha_{AB}m_B}{r_{AB}}\right\}\\
&\equiv G(1-\zeta)(m+E) \, ,
\end{split}
\end{equation}
where $\alpha_{AB} \equiv 1-\zeta +\zeta(1-2s_A)(1-2s_B)$.  In the second step, we rewrite the moment in terms of $m \equiv \sum_A m_A$, the total mass of the system, and $E$, the (lowest order) conserved energy.  [See MW (6.4).]  The rest of the moments $M^{00Q}$ can be renamed $\Is^Q$, following the definition in MW (3.7a).  We have

\begin{equation}
M^{00i} \equiv \Is^{i} = G(1-\zeta)\sum_A m_Ax_A^i\left\{1+\frac{1}{2}v_A^2-\frac{1}{2}\sum_{B\neq A}\frac{G\alpha_{AB}m_B}{r_{AB}}\right\} \, ,
\end{equation}
which, to the required order, is equal to zero in the CM frame.  (This is the argument used earlier to reduce the order of $\sum_A m_Ax_A^i$ in surface integrals.)  Finally, the next few integrals are only needed to lowest order,

\begin{subequations}
\begin{align}
M^{00ij} &\equiv \Is^{ij} = G(1-\zeta)\sum_A m_Ax_A^{ij} \, ,\\
M^{00ijk} &\equiv \Is^{ijk} = G(1-\zeta)\sum_A m_Ax_A^{ijk} \, .
\end{align}
\end{subequations}

The $\tau^{0i}$ moments require the source to $O(\rho \epsilon^{3/2})$, \eqref{eq:tau0i}.  We find

\begin{equation}
M^{0i} = G(1-\zeta)\sum_A m_A\left\{v_A^i\left[1+\frac{1}{2}v_A^2-\frac{1}{2}\sum_{B\neq A}\frac{G\alpha_{AB}m_B}{r_{AB}}\right]-\frac{1}{2}\sum_{B\neq A}\frac{G\alpha_{AB}m_B}{r_{AB}}\hat{n}_{AB}^i(\mathbf{v}_A\cdot\mathbf{\hat{n}}_{AB})\right\} \, .
\end{equation}
This is proportional to the total momentum [see MW (6.5)], and we can set it equal to zero.  The other moments are only needed to lowest order,

\begin{subequations}
\begin{align}
M^{0ij} &= G(1-\zeta)\sum_A m_Av_A^ix_A^j = \frac{1}{2}(\dot{\Is}^{ij}-\epsilon^{ija}\Js^a) \, , \\
M^{0ijk} &= G(1-\zeta)\sum_A m_Av_A^ix_A^{jk} = \frac{1}{3}(\dot{\Is}^{ijk}-2\epsilon^{ika}\Js^{aj}) \, ,
\end{align}
\end{subequations}
where $\epsilon^{ijk}$ is the totally antisymmetric Levi-Civita symbol ($\epsilon^{123} = +1$).  The current moments $\Js^{iQ}$ are defined in MW (3.7b) as

\begin{equation}
\Js^{iQ} \equiv \epsilon^{iab}\int_\Ms \tau^{0b}x^{aQ}d^3x \, . 
\end{equation}
Note that, unlike the exact equivalence of $M^{00Q}$ and $\Is^Q$, the equality between $M^{0Q}$, $\Is^Q$, and $\Js^{Q-1}$ is valid only to lowest order in the post-Newtonian expansion.  

The $\tau^{ij}$ moments require the source to $O(\rho \epsilon)$, \eqref{eq:tauijlowest}.  We need them only to leading order,

\begin{subequations}
\begin{align}
M^{ij} &= G(1-\zeta)\sum_A m_A\left\{v_A^{ij}-\frac{1}{2}\sum_{B\neq A}\frac{G\alpha_{AB}m_B}{r_{AB}}\hat{n}_{AB}^{ij}\right\} = \frac{1}{2}\ddot{\Is}^{ij} \, ,\\
M^{ijk} &= G(1-\zeta)\sum_A m_A\left\{v_A^{ij}x_A^k-\frac{1}{2}\sum_{B\neq A}\frac{G\alpha_{AB}m_B}{r_{AB}}\hat{n}_{AB}^{ij}x_A^k\right\} = \frac{1}{6}\ddot{\Is}^{ijk}-\frac{2}{3}\epsilon^{(i|ka}\dot{\Js}^{a|j)} \, .
\end{align}
\end{subequations}
Here again, equality between $M^Q$, $\Is^Q$, and (potentially) $\Js^{Q-1}$ only holds to the lowest post-Newtonian order.  

Finally, the scalar moments require $\tau_s$ to $O(\rho\epsilon)$.  Taken from MW (4.9e) and converted to the $\rho^*$ density, it is given by

\begin{equation}
\begin{split}
\tau_s &= \rho^* G\zeta\left[(1-2s)-\frac{1}{2}(1-2s)v^2-G(1-\zeta)(1-2s)U+G\zeta(1-2s-4a_s)U_s-4G\lambda_1(1-2s)U_s\right]\\
&\quad+\frac{1}{2\pi}G^2\zeta(\lambda_1-\zeta)(\nabla U_s)^2 \, .
\end{split}
\end{equation}
The lowest order moment is

\begin{equation}
\begin{split}
M_s &= G\zeta\sum_A m_A\left\{\vphantom{-\sum_{B\neq A}\frac{Gm_B}{r_{AB}}[(1-\zeta)(1-2s_A)+\zeta(1-2s_A+4a_{sA})+2\lambda_1(1-2s_A)(1-2s_B)]}
(1-2s_A)-\frac{1}{2}(1-2s_A)v_A^2 \right. \\
&\left.\quad-\sum_{B\neq A}\frac{Gm_B}{r_{AB}}[(1-\zeta)(1-2s_A)+\zeta(1-2s_A+4a_{sA})(1-2s_B)+2\lambda_1(1-2s_A)(1-2s_B)]\right\}\\
&\equiv G\zeta(m_s+m_{s1}) \, ,
\end{split}
\end{equation}
where we have defined a ``scalar mass''

\begin{equation}
m_s = \sum_A m_A(1-2s_A)
\end{equation}
weighted by sensitivity factors.  We also let $m_{s1}$ be the rest of $M_s$ (when calculated to 1PN order).  The rest of the $M_s^Q$ can be renamed $\Is_s^Q$, following MW (3.7d).  To the orders we need, they are

\begin{subequations}
\begin{align}
\begin{split}
M_s^i &\equiv \Is_s^i = G\zeta\sum_A m_Ax_A^i\left\{\vphantom{-\sum_{B\neq A}\frac{Gm_B}{r_{AB}}(1-\zeta)(1-2s_A)+\zeta(1-2s_A+4a_{sA})(1-2s_B)+2\lambda_1(1-2s_A)(1-2s_B)}
(1-2s_A)-\frac{1}{2}(1-2s_A)v_A^2\right.\\
&\left.\qquad\quad-\sum_{B\neq A}\frac{Gm_B}{r_{AB}}[(1-\zeta)(1-2s_A)+\zeta(1-2s_A+4a_{sA})(1-2s_B)+2\lambda_1(1-2s_A)(1-2s_B)]\right\} \, ,
\end{split}\\
M_s^{ij} &\equiv \Is_s^{ij} = G\zeta \sum_A m_A(1-2s_A)x_A^{ij} \, ,\\
M_s^{ijk} &\equiv \Is_s^{ijk} = G\zeta \sum_A m_A(1-2s_A)x_A^{ijk} \, .
\end{align}
\end{subequations}
All of these moments can be plugged back into \eqref{eq:hRN} and \eqref{eq:phiRN} to produce the radiation-zone fields, as calculated from near-zone integrals,

\begin{subequations}
\begin{align}
N_\Ns  &= \tilde{h}_\Ns ^{00} = 4G(1-\zeta)\frac{m+E}{R}+2\left(\frac{\Is^{ij}}{R}\right)_{,ij}-\frac{2}{3}\left(\frac{\Is^{ijk}}{R}\right)_{,ijk}+\cdots \label{eq:NRN} \, ,\\
K^i_\Ns  &= \tilde{h}_\Ns ^{0i} = -2\left(\frac{\dot{\Is}^{ij}-\epsilon^{ija}\Js^a}{R}\right)_{,j}+\frac{2}{3}\left(\frac{\dot{\Is}^{ijk}-2\epsilon^{ika}\Js^{aj}}{R}\right)_{,jk} + \cdots \label{eq:KRN}\, ,\\
B^{ij}_\Ns  &= \tilde{h}_\Ns ^{ij} = 2\frac{\ddot{\Is}^{ij}}{R}-\frac{2}{3}\left(\frac{\ddot{\Is}^{ijk}-4\epsilon^{(i|ka}\dot{\Js}^{a|j)}}{R}\right)_{,k}\label{eq:BRN} + \cdots\, ,\\
\Psi_\Ns  &= 2G\zeta\frac{m_s+m_{s1}}{R}-2\left(\frac{\Is_s^i}{R}\right)_{,i}+\left(\frac{\Is_s^{ij}}{R}\right)_{,ij}-\frac{1}{3}\left(\frac{\Is_s^{ijk}}{R}\right)_{,ijk}+\cdots \, .
\label{eq:PsiRN}
\end{align}
\end{subequations}
It is worth pointing out the relative post-Newtonian orders of the terms in these expressions.  The lowest order terms, which will serve as the reference, are those involving $m$ and $m_s$.  Relative to those, the other terms in $N_\Ns$ enter at 1PN (those containing $E$ and $\Is^{ij}$) and 1.5PN ($\Is^{ijk}$) order.  The first term in $K^i_\Ns$ and $B^{ij}_\Ns$ is 1PN, while the second in each is 1.5PN beyond the leading order.  For $\Psi_\Ns$, the $m_{s1}$ piece is 1PN.  The dipole term has contributions at both 0.5PN and 1.5PN order.  The quadrupole piece is relative 1PN order, and the octupole is 1.5PN.

\subsection{Radiation-zone fields: Radiation-zone contributions}
\label{sec:RRfields}
We must also calculate the radiation-zone contributions to the radiation-zone fields using \eqref{eq:hRR} and \eqref{eq:phiRR}.  In the radiation zone, there are no compact sources, so $\tau^{\mu \nu}$ and $\tau_s$ are made up purely of field terms.  For $\tau^{\mu \nu}$, we can use MW (3.4) and (3.5), which write $\Lambda^{\mu \nu}$ and $\Lambda_s^{\mu \nu}$ generically in terms of $N$, $K^i$, $B^{ij}$, $B$, and $\Psi$.  For these, we substitute \eqref{eq:NRN}--\eqref{eq:PsiRN}.

We start with $\tilde{h}^{00}$.  To lowest order, we find

\begin{equation}
\tau^{00} = -\frac{7}{8\pi}G^2(1-\zeta)^2\frac{m^2}{R^4}+\frac{1}{8\pi}G^2\zeta(1-\zeta)\frac{m_s^2}{R^4} \, .
\end{equation}
This source has the generic form

\begin{equation}
\tau^{\mu \nu}(l, n) = \frac{1}{4\pi}\frac{f^{\mu\nu}(\tau)}{R^n}\hat{N}^{\langle L\rangle} \, ,
\label{eq:stdsource}
\end{equation}
with $l = 0$ and $n = 4$.  If we restrict ourselves to sources of this form, \eqref{eq:hRR} can be rewritten as

\begin{equation}
\tilde{h}^{\mu \nu}_{\Cs-\Ns } = \frac{4}{R}\hat{N}^{\langle L\rangle}\left[\int_0^\Rs f^{\mu \nu}(\tau-2s)A(s,R)\ ds + \int_\Rs^\infty f^{\mu \nu}(\tau-2s)B(s,R)\ ds\right] \, ,
\label{eq:ABintegrals}
\end{equation}
where

\begin{subequations}
\begin{align}
A(s,R) &\equiv \int_\Rs^{R+s} \frac{P_l(\xi)}{p^{n-1}}dp \, , \label{eq:A} \\
B(s,R) &\equiv \int_s^{R+s} \frac{P_l(\xi)}{p^{n-1}}dp \, ,
\label{eq:B}
\end{align}
\end{subequations}
and

\begin{equation}
\xi \equiv \frac{R+2s}{R}-\frac{2s(R+s)}{Rp} \, .
\end{equation}
In the $l = 0$ case, this is easy, and we find

\begin{equation}
N_{\Cs-\Ns} = \tilde{h}_{\Cs-\Ns}^{00} = 7G^2(1-\zeta)^2\frac{m^2}{R^2}-G^2\zeta(1-\zeta)\frac{m_s^2}{R^2} \, .
\label{eq:NRR}
\end{equation}
These terms are 1.5PN beyond the leading order term in \eqref{eq:NRN}.  Since we kept no higher order terms in that expression, we can also stop here.  The total field $N = \tilde{h}^{00}$ in the radiation zone is just the sum of \eqref{eq:NRN} and \eqref{eq:NRR}.  The second term in \eqref{eq:NRR} is obviously new to scalar-tensor theory.  As we shall see, it turns out not to have an impact on the final gravitational waves.  Otherwise, $N$ in the radiation zone has exactly the same form as in GR, modified only by the addition of the factors $G(1-\zeta)$, both explicitly and in the definitions of $\Is^Q$.

It turns out that there are no relevant radiation-zone contributions to $K^i$; to the order we need, \eqref{eq:KRN} is the complete expression in the radiation zone.  To calculate $B^{ij}_{\Cs-\Ns}$, we use

\begin{equation}
\tau^{ij} = \frac{1}{4\pi}\left[G^2(1-\zeta)^2\frac{m^2}{R^4}+G^2\zeta(1-\zeta)\frac{m_s^2}{R^4}\right]\left(\hat{N}^{\langle ij\rangle}-\frac{1}{6}\delta^{ij}\right) \, .
\end{equation}
We use the same method as for $N_{\Cs-\Ns}$, except that here we have two different $(l,n)$ combinations, (2, 4) and (0, 4), and thus two integrals \eqref{eq:ABintegrals} to evaluate.  Doing so gives

\begin{equation}
B^{ij}_{\Cs-\Ns} = \tilde{h}^{ij}_{\Cs-\Ns} = G^2(1-\zeta)^2\frac{m^2}{R^2}\hat{N}^{ij}+G^2\zeta(1-\zeta)\frac{m_s^2}{R^2}\hat{N}^{ij} \, .
\end{equation}
These terms are added to \eqref{eq:BRN} to produce the final expression for $B^{ij}$.  They enter at 1.5PN beyond the leading order.  Finally, for the scalar, we need $\tau_s$ to lowest order.  It is given by

\begin{equation}
\tau_s = \frac{1}{16\pi}\left(\frac{2\omega_0'}{3+2\omega_0}-2\right)(\nabla \Psi)^2 = \frac{1}{2\pi}G^2\zeta(\lambda_1-\zeta)\frac{m_s^2}{R^4} \, .
\end{equation}
For a source of the form \eqref{eq:stdsource}, \eqref{eq:phiRR} reduces to 

\begin{equation}
\Psi_{\Cs-\Ns} = \frac{2}{R}\hat{N}^{\langle L\rangle}\left(\int_0^\Rs f(\tau-2s)A(s,R)\ ds + \int_\Rs^\infty f(\tau-2s)B(s,R)\ ds\right) \, ,
\end{equation}
with $A(s,R)$ and $B(s,R)$ defined just as in \eqref{eq:A} and \eqref{eq:B}.  The source has $(l,m) = (4,0)$, and the result is

\begin{equation}
\Psi_{\Cs-\Ns} = -2G^2\zeta(\lambda_1-\zeta)\frac{m_s^2}{R^2} \, ,
\end{equation}
which adds to $\Psi_\Ns$, \eqref{eq:PsiRN}, at 1.5PN order.

\subsection{Radiation-zone contributions to the GWs}
\label{sec:RZGWs}
Now that we have the fields in the radiation zone, we need to evaluate the gravitational waveform.  Fundamentally, this just means continuing the procedure for calculating $\tilde{h}^{ij}$ in Sec.\ \ref{sec:RRfields} to higher order.  While doing so, we drop non-TT terms and terms that drop off faster than $1/R$.

For simplicity, we will treat contributions from $\Lambda^{ij}$ and $\Lambda_s^{ij}$ separately.  We will also do the calculation order by order.  We must be careful doing so: The expressions for $\Lambda^{ij}$ and $\Lambda_s^{ij}$ in MW are sorted by post-Newtonian order in the near zone.  In the radiation zone, the ordering can be slightly different.  For instance, while $K^i$ is at 0.5PN order relative to $N$ in the near zone, it is at 1PN relative order in the radiation zone.  Furthermore, time derivatives in the radiation zone do not increase the post-Newtonian order relative to spatial derivatives.

The lowest order term in $\Lambda^{ij}$ is proportional to $N^{,i}N^{,j}$.  (Now that we are dealing with GWs, we can ignore the non-TT $\delta^{ij}$ term.)  We first plug in the 0PN monopole pieces of $N$.  As seen in Sec.\ \ref{sec:RRfields}, the resulting $\tilde{h}^{ij}$ scales like $1/R^2$ and, as such, dies off in the far-away zone.  At the next order, $N^{,i}N^{,j}$ generates cross terms between the 0PN piece of $N$ and the 1PN pieces (including both the energy $E$ and the mass quadrupole $\Is^{ij}$).  Other terms which contribute at the same order are $N^{,(i}\dot{K}^{j)}$, $N^{,(i}B^{,j)}$, and $N\ddot{B}^{ij}$, all featuring the 0PN piece of $N$ and the 1PN piece of the other field.  The last term is found in (4.4c) of \cite{pw00}; its two time derivatives place it at higher order in the near-zone counting scheme.  

Our expressions for the fields have several unevaluated spatial derivatives; the expression for $\Lambda^{ij}$ adds more.  When we evaluate them, we must remember that the moments are functions of retarded time, $\tau = t-R$, so that, for instance, $\partial_c\Is^{ab} = -\dot{\Is}^{ab}\hat{N}^c$.  Completing all the derivatives and converting products of $\hat{N}^i$ to STF products, we find

\begin{equation}
\begin{split}
\tau^{ij} &= \cdots + \frac{1}{2\pi}G^2(1-\zeta)^2\frac{mE}{R^4}\hat{N}^{\langle ij\rangle}+\frac{1}{4\pi}G(1-\zeta)\frac{m}{R^2}\left[\left(15\frac{\Is^{ab}}{R^4}+15\frac{\dot{\Is}^{ab}}{R^3}+6\frac{\ddot{\Is}^{ab}}{R^2}+\frac{\dddot{\Is}^{ab}}{R}\right)\hat{N}^{\langle abij\rangle}\right.\\
&\quad+\left(-\frac{6}{7}\frac{\Is^{aa}}{R^4}-\frac{6}{7}\frac{\dot{\Is}^{aa}}{R^3}+\frac{6}{7}\frac{\ddot{\Is}^{aa}}{R^2}+\frac{8}{7}\frac{\dddot{\Is}^{aa}}{R}\right)\hat{N}^{\langle ij\rangle}+\left(\frac{18}{7}\frac{\Is^{a(i}}{R^4}+\frac{18}{7}\frac{\dot{\Is}^{a(i}}{R^3}-\frac{18}{7}\frac{\ddot{\Is}^{a(i}}{R^2}-\frac{24}{7}\frac{\dddot{\Is}^{a(i}}{R}\right)\hat{N}^{\langle j)a\rangle}\\
&\left.\quad-\frac{6}{5}\frac{\ddot{\Is}^{ij}}{R^2}-\frac{6}{5}\frac{\dddot{\Is}^{ij}}{R}-2\overset{(4)}{\Is}\vphantom{\Is}^{ij}\right]+\cdots \, .
\end{split}
\end{equation}
Here we have ignored $\delta^{ij}$ terms, which are non-TT.  Careful examination shows that we also should ignore terms proportional to $\hat{N}^{\langle ij\rangle}$, which will only produce non-TT terms in $\tilde{h}^{ij}$.  This eliminates the term with monopole-monopole coupling (i.e., the one proportional to $mE$) and leaves only those with monopole-mass quadrupole couplings.  In the end, we have to evaluate \eqref{eq:ABintegrals} for $l = 4, n =$ 3--6; $l = 2, n =$ 3--6; and $l = 0, n =$ 2--4.  Adding everything up, changing from $\hat{N}^{\langle L\rangle}$ back to $\hat{N}^L$, and discarding more non-TT terms which arise along the way, we find

\begin{equation}
\tilde{h}^{ij} = \frac{4G(1-\zeta)m}{R}\left[\frac{11}{12}\dddot{\Is}^{ij}+\int_0^\infty ds\; \overset{(4)}{\Is}\vphantom{\Is}^{ij}(\tau-s)\ln \frac{s}{2R+s}\right] \, .
\label{eq:tail1}
\end{equation}
We have made a slight change of variable from \eqref{eq:ABintegrals}: $s \rightarrow s/2$.  This is done to match the first term in WW (5.8); we see that the two expressions are equal except for a factor $G^2(1-\zeta)^2$ in the scalar-tensor case.  [Recall that the moment $\Is^{ij}$ contains the factor $G(1-\zeta)$.]  Later, we will also bring the first term under the integral, as in WW, but for now we leave it separate.  This is to emphasize the difference in the two pieces: The first relies on the instantaneous (but retarded) value of $\dddot{\Is}^{ij}$, while the second requires a weighted integral of $\overset{(4)}{\Is}\vphantom{\Is}^{ij}$ over the entire past history of the source.  It is the lowest order ``tail'' term.

The expression \eqref{eq:tail1} turns out to be 1.5PN order beyond the Newtonian quadrupole.  Therefore, we only need to go one-half order higher to find the 2PN waveform.  In this case, the source will include contributions from the same four terms in $\Lambda^{ij}$ as before, but they will now be cross terms between the 0PN piece of $N$ and the 1.5PN pieces of $N$, $K^i$, $B$, and $B^{ij}$.  Note that all of the 1.5PN pieces coming from radiation-zone integrals (those proportional to $m^2$ or $m_s^2$) will generate only non-TT terms in $\tilde{h}^{ij}$.   Similarly, the term $NN^{,i}N^{,j}$ in $\Lambda^{ij}$ should contribute at this order, but any piece of $\tilde{h}^{ij}$ it generates will also be non-TT.  In the end, the relevant piece of the source will contain only monopole-mass octupole and monopole-current quadrupole couplings.  After a great deal of algebra, we find it to be

\begin{align}
\tau^{ij} &= \cdots + \frac{1}{4\pi}G(1-\zeta)\frac{m}{R^2}\left[\left(35\frac{\Is^{abc}}{R^5}+35\frac{\dot{\Is}^{abc}}{R^4}+15\frac{\ddot{\Is}^{abc}}{R^3}+\frac{10}{3}\frac{\dddot{\Is}^{abc}}{R^2}+\frac{1}{3}\frac{\overset{(4)}{\Is}\vphantom{\Is}^{abc}}{R}\right)\hat{N}^{\langle abcij\rangle}\right. \nonumber \\
&\quad + \left(\frac{25}{3}\frac{\Is^{ab(i}}{R^5}+\frac{25}{3}\frac{\dot{\Is}^{ab(i}}{R^4}-\frac{25}{9}\frac{\dddot{\Is}^{ab(i}}{R^2}-\frac{10}{9}\frac{\overset{(4)}{\Is}\vphantom{\Is}^{ab(i}}{R}\right)\hat{N}^{\langle j)ab\rangle} \nonumber \\
&\quad + \left(-\frac{10}{7}\frac{\ddot{\Is}^{ija}}{R^3}-\frac{10}{7}\frac{\dddot{\Is}^{ija}}{R^2}-\frac{8}{7}\frac{\overset{(4)}{\Is}\vphantom{\Is}^{ija}}{R}-\frac{2}{3}\overset{(5)}{\Is}\vphantom{\Is}^{ija}\right)\hat{N}^a \nonumber \\
&\quad+\left(8\frac{\dot{\Js}^{ca}}{R^3}+8\frac{\ddot{\Js}^{ca}}{R^2}+\frac{8}{3}\frac{\dddot{\Js}^{ca}}{R}\right)\hat{N}^{\langle ab(i\rangle}\epsilon^{j)bc} \nonumber \\
&\left.\quad+ \left(\frac{8}{5}\frac{\dot{\Js}^{c(i}}{R^3}+\frac{8}{5}\frac{\ddot{\Js}^{c(i}}{R^2}+\frac{16}{5}\frac{\dddot{\Js}^{c(i}}{R}+\frac{8}{3}\overset{(5)}{\Js}\vphantom{\Js}^{c(i}\right)\epsilon^{j)ac}\hat{N}^a\right]+\cdots \, .
\end{align}
We must evaluate \eqref{eq:ABintegrals} for $l = 5, n =$ 3--7; $l = 3, n =$ 3--7; and $l = 1, n=$ 2--5.  The final result for the GWs is

\begin{equation}
\begin{split}
\tilde{h}^{ij} &= \frac{4G(1-\zeta)m}{R}\left[\frac{97}{180}\overset{(4)}{\Is}\vphantom{\Is}^{ija}\hat{N}^a+\frac{1}{3}\hat{N}^a\int_0^\infty ds\ \overset{(5)}{\Is}\vphantom{\Is}^{ija}(\tau-s)\ln \frac{s}{2R+s}-\frac{14}{9}\epsilon^{(i|ab}\dddot{\Js}^{b|j)}\hat{N}^a\right.\\
&\left.\quad-\frac{4}{3}\hat{N}^a\epsilon^{(i|ab}\int_0^\infty ds\ \overset{(4)}{\Js}\vphantom{\Js}^{b|j)}(\tau-s)\ln \frac{s}{2R+s}\right] \, .
\end{split}
\label{eq:tail2}
\end{equation}
These terms are again identical to the GR results [the remainder of WW (5.8)], except for a factor of $G^2(1-\zeta)^2$.

Now we examine the waveform produced by the $\Lambda_s^{ij}$ piece of $\tau^{ij}$.  We will mainly be concerned with the term in $\Lambda_s^{ij}$ which is proportional to $\Psi^{,i}\Psi^{,j}$.  The lowest order contribution from this term involves the 0PN monopole pieces of $\Psi$.  The resulting $\tilde{h}^{ij}$ was calculated in Sec. \ref{sec:RRfields} and shown to scale like $1/R^2$, making it irrelevant in the far-away zone.  Because $\Psi$ contains nonvanishing dipole terms, the next highest order source is only 0.5PN beyond the leading order.  It consists of monopole-dipole couplings,

\begin{equation}
\tau^{ij} = \cdots + \frac{1}{4\pi}G(1-\zeta)\frac{m_s}{R^2}\left[\left(6\frac{\Is_s^a}{R^3}+6\frac{\dot{\Is}_s^a}{R^2}+2\frac{\ddot{\Is}_s^a}{R}\right)\hat{N}^{\langle aij\rangle}+\left(\frac{2}{5}\frac{\Is_s^{(i}}{R^3}+\frac{2}{5}\frac{\dot{\Is}_s^{(i}}{R^2}+\frac{4}{5}\frac{\ddot{\Is}_s^{(i}}{R}\right)\hat{N}^{j)}\right]+\cdots \, .
\label{eq:monopoledipole}
\end{equation}
However, each piece of this source will generate a term in $\tilde{h}^{ij}$ proportional to $\hat{N}^i$, $\hat{N}^j$, or $\delta^{ij}$, all of which are non-TT.  Moving on to the next order, there are three types of terms in $\tau^{ij}$: a monopole-monopole coupling (the cross term of $m_s$ and $m_{s1}$), monopole-quadrupole couplings, and dipole-dipole couplings.  The first generates only non-TT terms in $\tilde{h}^{ij}$.  The second is analogous to the monopole-quadrupole terms in $\Lambda^{ij}$, while the third has no counterpart there.  The total source at this order (without non-TT terms) can be written as

\begin{equation}
\begin{split}
\tau^{ij} &= \cdots + \frac{1}{4\pi}G(1-\zeta)\frac{m_s}{R^2}\left[\left(15\frac{\Is_s^{ab}}{R^4}+15\frac{\dot{\Is}_s^{ab}}{R^3}+6\frac{\ddot{\Is}_s^{ab}}{R^2}+\frac{\dddot{\Is}_s^{ab}}{R}\right)\hat{N}^{\langle abij\rangle}\right.\\
&\left.\qquad+\left(\frac{18}{7}\frac{\Is_s^{a(i}}{R^4}+\frac{18}{7}\frac{\dot{\Is}_s^{a(i}}{R^3}+\frac{10}{7}\frac{\ddot{\Is}_s^{a(i}}{R^2}+\frac{4}{7}\frac{\dddot{\Is}_s^{a(i}}{R}\right)\hat{N}^{\langle j)a\rangle}+\frac{2}{15}\frac{\ddot{\Is}_s^{ij}}{R^2}+\frac{2}{15}\frac{\dddot{\Is}_s^{ij}}{R}\right]\\
&\quad+ \frac{1}{4\pi}\frac{1-\zeta}{\zeta}\left[\left(9\frac{\Is_s^a\Is_s^b}{R^6}+18\frac{\Is_s^a\dot{\Is}_s^b}{R^5}+6\frac{\Is_s^a\ddot{\Is}_s^b}{R^4}+9\frac{\dot{\Is}_s^a\dot{\Is}_s^b}{R^4}+6\frac{\dot{\Is}_s^a\ddot{\Is}_s^b}{R^3}+\frac{\ddot{\Is}_s^a\ddot{\Is}_s^b}{R^2}\right)\hat{N}^{\langle abij\rangle}\right.\\
&\qquad + \left(-\frac{6}{7}\frac{\Is_s^a\Is_s^{(i}}{R^6}-\frac{6}{7}\frac{\Is_s^a\dot{\Is}_s^{(i}}{R^5}-\frac{6}{7}\frac{\dot{\Is}_s^a\Is_s^{(i}}{R^5}+\frac{12}{7}\frac{\Is_s^a\ddot{\Is}_s^{(i}}{R^4}-\frac{2}{7}\frac{\ddot{\Is}_s^a\Is_s^{(i}}{R^4}-\frac{6}{7}\frac{\dot{\Is}_s^a\dot{\Is}_s^{(i}}{R^4}+\frac{12}{7}\frac{\dot{\Is}_s^a\ddot{\Is}_s^{(i}}{R^3}-\frac{2}{7}\frac{\ddot{\Is}_s^a\dot{\Is}_s^{(i}}{R^3}+\frac{4}{7}\frac{\ddot{\Is}_s^a\ddot{\Is}_s^{(i}}{R^2}\right)\hat{N}^{\langle j)a\rangle}\\
&\left.\qquad +\frac{1}{5}\frac{\Is_s^{(i}\Is_s^{j)}}{R^6}+\frac{2}{5}\frac{\Is_s^{(i}\dot{\Is}_s^{j)}}{R^5}+\frac{2}{15}\frac{\Is_s^{(i}\ddot{\Is}_s^{j)}}{R^4}+\frac{1}{5}\frac{\dot{\Is}_s^{(i}\dot{\Is}_s^{j)}}{R^4}+\frac{2}{15}\frac{\dot{\Is}_s^{(i}\ddot{\Is}_s^{j)}}{R^3}+\frac{2}{15}\frac{\ddot{\Is}_s^{(i}\ddot{\Is}_s^{j)}}{R^2}\right]+\cdots \, .
\end{split}\end{equation}
This requires integrals with $l = 4, n =$ 2--6; $l = 2, n =$ 2--6; and $l = 0, n =$ 2--6.  Many of these can be reused from earlier calculations; however, some are brand new.  The final expression for the GWs is

\begin{equation}
\tilde{h}^{ij} = \frac{4G(1-\zeta)m_s}{R}\left(-\frac{1}{12}\dddot{\Is}_s^{ij}\right)+ \frac{4}{R}\frac{1-\zeta}{\zeta}\left(\frac{1}{6}\int_{-\infty}^\tau \ddot{\Is}_s^i(\tau')\ddot{\Is}_s^j(\tau')d\tau'-\frac{1}{6}\dot{\Is}_s^{(i}\ddot{\Is}_s^{j)}-\frac{1}{18}\Is_s^{(i}\dddot{\Is}_s^{j)}\right) \, .
\label{eq:memory1}
\end{equation}
(Recall that the scalar dipole moment $\Is_s^i$ includes a factor of $\zeta$, which will cancel the $1/\zeta$ in front of the second term.)  This is an entirely new contribution to the GWs at 1.5PN order, featuring a new type of hereditary integral without a logarithmic factor.  We will discuss it further in Sec.\ \ref{sec:discussion}.  

At the next order, the $\Psi \Psi^{,i} \Psi^{,j}$ term in $\Lambda_s^{ij}$ begins to contribute to the source.  However, like the $NN^{,i}N^{,j}$ piece of $\Lambda^{ij}$, it produces only non-TT terms in $\tilde{h}^{ij}$ and can be ignored.  In the $\Psi^{,i}\Psi^{,j}$ term, we have monopole-dipole couplings (0PN monopole--1.5PN dipole and 1PN monopole--0.5PN dipole), a triple-monopole coupling (from $\Psi_{\Cs-\Ns}$), monopole-octupole couplings, and dipole-quadrupole couplings.  The first and second of these produce only non-TT terms in the final waveform.  The other two types of terms give

\begin{align}
\tau^{ij} &= \cdots + \frac{1}{4\pi}G(1-\zeta)\frac{m_s}{R^2}\left[\left(35\frac{\Is_s^{abc}}{R^5}+35\frac{\dot{\Is}_s^{abc}}{R^4}+15\frac{\ddot{\Is}_s^{abc}}{R^3}+\frac{10}{3}\frac{\dddot{\Is}_s^{abc}}{R^2}+\frac{1}{3}\frac{\overset{(4)}{\Is}\vphantom{\Is}_s^{abc}}{R}\right)\hat{N}^{\langle abcij\rangle}\right. \nonumber \\
&\qquad + \left(\frac{25}{3}\frac{\Is_s^{ab(i}}{R^5}+\frac{25}{3}\frac{\dot{\Is}_s^{ab(i}}{R^4}+4\frac{\ddot{\Is}_s^{ab(i}}{R^3}+\frac{11}{9}\frac{\dddot{\Is}_s^{ab(i}}{R^2}+\frac{2}{9}\frac{\overset{(4)}{\Is}\vphantom{\Is}_s^{ab(i}}{R}\right)\hat{N}^{\langle j)ab\rangle} \nonumber \\
&\left.\qquad + \left(\frac{6}{35}\frac{\ddot{\Is}_s^{ija}}{R^3}+\frac{6}{35}\frac{\dddot{\Is}_s^{ija}}{R^2}+\frac{2}{35}\frac{\overset{(4)}{\Is}\vphantom{\Is}_s^{ija}}{R}\right)\hat{N}^a\right] \nonumber \\
&\quad + \frac{1}{4\pi}\frac{1-\zeta}{\zeta}\left[\left(45\frac{\Is_s^a\Is_s^{bc}}{R^7}+45\frac{\Is_s^a\dot{\Is}_s^{bc}}{R^6}+45\frac{\dot{\Is}_s^a\Is_s^{bc}}{R^6}+18\frac{\Is_s^a\ddot{\Is}_s^{bc}}{R^5}+45\frac{\dot{\Is}_s^a\dot{\Is}_s^{bc}}{R^5}+15\frac{\ddot{\Is}_s^a\Is_s^{bc}}{R^5}+3\frac{\Is_s^a\dddot{\Is}_s^{bc}}{R^4}+18\frac{\dot{\Is}_s^a\ddot{\Is}_s^{bc}}{R^4}\right.\right. \nonumber \\
&\left.\qquad+15\frac{\ddot{\Is}_s^a\dot{\Is}_s^{bc}}{R^4}+3\frac{\dot{\Is}_s^a\dddot{\Is}_s^{bc}}{R^3}+6\frac{\ddot{\Is}_s^a\ddot{\Is}_s^{bc}}{R^3}+\frac{\ddot{\Is}_s^{a}\dddot{\Is}_s^{bc}}{R^2}\right)\hat{N}^{\langle abcij\rangle} \nonumber \\
&\qquad+\left(2\frac{\Is_s^a\Is_s^{b(i}}{R^7}-5\frac{\Is_s^{ab}\Is_s^{(i}}{R^7}+2\frac{\Is_s^a\dot{\Is}_s^{b(i}}{R^6}-5\frac{\dot{\Is}_s^{ab}\Is_s^{(i}}{R^6}+2\frac{\dot{\Is}_s^a\Is_s^{b(i}}{R^6}-5\frac{\Is_s^{ab}\dot{\Is}_s^{(i}}{R^6}+2\frac{\Is_s^a\ddot{\Is}_s^{b(i}}{R^5}-2\frac{\ddot{\Is}_s^{ab}\Is_s^{(i}}{R^5}+2\frac{\dot{\Is}_s^a\dot{\Is}_s^{b(i}}{R^5}-5\frac{\dot{\Is}_s^{ab}\dot{\Is}_s^{(i}}{R^5}\right. \nonumber \\
&\qquad+\frac{2}{3}\frac{\ddot{\Is}_s^a\Is_s^{b(i}}{R^5}+\frac{10}{3}\frac{\Is_s^{ab}\ddot{\Is}_s^{(i}}{R^5}+\frac{4}{3}\frac{\Is_s^a\dddot{\Is}_s^{b(i}}{R^4}-\frac{1}{3}\frac{\dddot{\Is}_s^{ab}\Is_s^{(i}}{R^4}+2\frac{\dot{\Is}_s^a\ddot{\Is}_s^{b(i}}{R^4}-2\frac{\ddot{\Is}_s^{ab}\dot{\Is}_s^{(i}}{R^4}+\frac{2}{3}\frac{\ddot{\Is}_s^a\dot{\Is}_s^{b(i}}{R^4}+\frac{10}{3}\frac{\dot{\Is}_s^{ab}\ddot{\Is}_s^{(i}}{R^4} \nonumber \\
&\left.\qquad+\frac{4}{3}\frac{\dot{\Is}_s^a\dddot{\Is}_s^{b(i}}{R^3}-\frac{1}{3}\frac{\dddot{\Is}_s^{ab}\dot{\Is}_s^{(i}}{R^3}+\frac{2}{3}\frac{\ddot{\Is}_s^a\ddot{\Is}_s^{b(i}}{R^3}+\frac{4}{3}\frac{\ddot{\Is}_s^{ab}\ddot{\Is}_s^{(i}}{R^3}+\frac{4}{9}\frac{\ddot{\Is}_s^a\dddot{\Is}_s^{b(i}}{R^2}+\frac{2}{9}\frac{\dddot{\Is}_s^{ab}\ddot{\Is}_s^{(i}}{R^2}\right)\hat{N}^{\langle j)ab\rangle} \nonumber \\
&\qquad + \left(-\frac{36}{35}\frac{\Is_s^a\Is_s^{ij}}{R^7}+\frac{54}{35}\frac{\Is_s^{(i}\Is_s^{j)a}}{R^7}-\frac{36}{35}\frac{\Is_s^a\dot{\Is}_s^{ij}}{R^6}+\frac{54}{35}\frac{\Is_s^{(i}\dot{\Is}_s^{j)a}}{R^6}-\frac{36}{35}\frac{\dot{\Is}_s^a\Is_s^{ij}}{R^6}+\frac{54}{35}\frac{\dot{\Is}_s^{(i}\Is_s^{j)a}}{R^6}-\frac{6}{35}\frac{\Is_s^a\ddot{\Is}_s^{ij}}{R^5}+\frac{16}{35}\frac{\Is_s^{(i}\ddot{\Is}_s^{j)a}}{R^5}\right. \nonumber \\
&\qquad -\frac{36}{35}\frac{\dot{\Is}_s^a\dot{\Is}_s^{ij}}{R^5}+\frac{54}{35}\frac{\dot{\Is}_s^{(i}\dot{\Is}_s^{j)a}}{R^5}-\frac{12}{35}\frac{\ddot{\Is}_s^a\Is_s^{ij}}{R^5}+\frac{18}{35}\frac{\ddot{\Is}_s^{(i}\Is_s^{j)a}}{R^5}+\frac{6}{35}\frac{\Is_s^a\dddot{\Is}_s^{ij}}{R^4}-\frac{2}{35}\frac{\Is_s^{(i}\dddot{\Is}_s^{j)a}}{R^4}-\frac{6}{35}\frac{\dot{\Is}_s^a\ddot{\Is}_s^{ij}}{R^4}+\frac{16}{35}\frac{\dot{\Is}_s^{(i}\ddot{\Is}_s^{j)a}}{R^4} \nonumber \\
&\left.\left.\qquad-\frac{12}{35}\frac{\ddot{\Is}_s^a\dot{\Is}_s^{ij}}{R^4}+\frac{18}{35}\frac{\ddot{\Is}_s^{(i}\dot{\Is}_s^{j)a}}{R^4}+\frac{6}{35}\frac{\dot{\Is}_s^a\dddot{\Is}_s^{ij}}{R^3}-\frac{2}{35}\frac{\dot{\Is}_s^{(i}\dddot{\Is}_s^{j)a}}{R^3}-\frac{2}{35}\frac{\ddot{\Is}_s^a\ddot{\Is}_s^{ij}}{R^3}+\frac{2}{7}\frac{\ddot{\Is}_s^{(i}\ddot{\Is}_s^{j)a}}{R^3}+\frac{2}{35}\frac{\ddot{\Is}_s^a\dddot{\Is}_s^{ij}}{R^2}+\frac{4}{35}\frac{\ddot{\Is}_s^{(i}\dddot{\Is}_s^{j)a}}{R^2}\right)\hat{N}^a\right] \nonumber \\
&\quad+\cdots \, .
\end{align}
This requires integrals with $l = 5, n =$ 2--7; $l = 3, n =$ 2--7; and $l = 1, n =$ 2--7.  The final answer is

\begin{align}
\tilde{h}^{ij} &= \frac{4G(1-\zeta)m_s}{R}\left(-\frac{1}{60}\overset{(4)}{\Is}\vphantom{\Is}_s^{ija}\hat{N}^a\right)+\frac{4}{R}\frac{1-\zeta}{\zeta}\left[\frac{1}{10}\hat{N}^a\int_{-\infty}^\tau \ddot{\Is}_s^{(a}(\tau')\dddot{\Is}_s^{ij)}(\tau') d\tau'+\left(-\frac{1}{15}\Is_s^a\overset{(4)}{\Is}\vphantom{\Is}_s^{ij}-\frac{1}{60}\dot{\Is}_s^a\dddot{\Is}_s^{ij}\right.\right. \nonumber \\
&\left.\left.\quad-\frac{1}{60}\ddot{\Is}_s^a\ddot{\Is}_s^{ij}+\frac{1}{60}\dddot{\Is}_s^a\dot{\Is}_s^{ij}-\frac{1}{60}\overset{(4)}{\Is}\vphantom{\Is}_s^a\Is_s^{ij}+\frac{1}{30}\Is_s^{(i}\overset{(4)}{\Is}\vphantom{\Is}_s^{j)a}-\frac{1}{30}\dot{\Is}_s^{(i}\dddot{\Is}_s^{j)a}-\frac{1}{30}\ddot{\Is}_s^{(i}\ddot{\Is}_s^{j)a}-\frac{1}{10}\dddot{\Is}_s^{(i}\dot{\Is}_s^{j)a}\right)\hat{N}^a\right] \, .
\label{eq:memory2}
\end{align}
This is a new piece of the waveform at 2PN order.

\section{Results}
\label{sec:results}

\subsection{Final tensor waveform}
To find the final tensor waveform, we add the contributions from the near and radiation zones.  The near-zone contribution is found by inserting the differentiated two-body Epstein-Wagoner moments into \eqref{eq:EWwaveform}.  For the radiation-zone pieces \eqref{eq:tail1}, \eqref{eq:tail2}, \eqref{eq:memory1}, and \eqref{eq:memory2}, we bring the instantaneous terms inside the integrals.  Then we write the moments $\Is^Q$, $\Js^Q$, $\Is_s^Q$, and their derivatives explicitly in terms of relative two-body variables.  Finally, we sum the four pieces to find the complete radiation-zone contribution. 

The final tensor waveform can be written as a post-Newtonian expansion,

\begin{equation}
\tilde{h}^{ij} = \frac{2G(1-\zeta)\mu}{R}[\tilde{Q}^{ij}+P^{1/2}Q^{ij}+PQ^{ij}+P^{3/2}Q_\Ns^{ij}+P^{3/2}Q_{\Cs-\Ns}^{ij}+P^2Q_\Ns^{ij}+P^2Q_{\Cs-\Ns}^{ij}+O(\epsilon^{5/2})]_\text{TT} \, ,
\end{equation}
where the superscripts on $P$ denote the PN order of each term.  For clarity, we have separated out the 1.5PN and 2PN near-zone terms from the radiation-zone terms at the same order.  We find

\begin{subequations}
\begin{equation}
\tilde{Q}^{ij} = 2\left(v^{ij}-\frac{G\alpha m}{r}\hat{n}^{ij}\right) \, ,
\label{eq:P0}
\end{equation}

\begin{equation}
P^{1/2}Q^{ij} = \frac{\delta m}{m}\left\{3(\mathbf{\hat{N}}\cdot \mathbf{\hat{n}})\frac{G\alpha m}{r}[2\hat{n}^{(i}v^{j)}-\dot{r}\hat{n}^{ij}]+(\mathbf{\hat{N}}\cdot \mathbf{v})\left[\frac{G\alpha m}{r}\hat{n}^{ij}-2v^{ij}\right]\right\} \, ,
\label{eq:P05}
\end{equation}

\begin{equation}
\begin{split}
PQ^{ij} &= \frac{1}{3}(1-3\eta)\left\{(\mathbf{\hat{N}}\cdot \mathbf{\hat{n}})^2\frac{G\alpha m}{r}\left[\left(3v^2-15\dot{r}^2+7\frac{G\alpha m}{r}\right)\hat{n}^{ij}+30\dot{r}\hat{n}^{(i}v^{j)}-14v^{ij}\right]\right.\\
&\left.\qquad+(\mathbf{\hat{N}}\cdot \mathbf{\hat{n}})(\mathbf{\hat{N}}\cdot \mathbf{v})\frac{G\alpha m}{r}[12\dot{r}\hat{n}^{ij}-32\hat{n}^{(i}v^{j)}]+(\mathbf{\hat{N}}\cdot \mathbf{v})^2\left[6v^{ij}-2\frac{G\alpha m}{r}\hat{n}^{ij}\right]\right\}\\
&\quad+\frac{1}{3}\left\{\left[3(1-3\eta)v^2-2(2-3\eta)\frac{G\alpha m}{r}\right]v^{ij}+4\frac{G\alpha m}{r}\dot{r}(5+3\eta+3\bar{\gamma})\hat{n}^{(i}v^{j)}\right.\\
&\left.\qquad+\frac{G\alpha m}{r}\left[3(1-3\eta)\dot{r}^2-(10+3\eta+6\bar{\gamma})v^2+\left(29+12\bar{\gamma}+12\bar{\beta}_+-12\frac{\delta m}{m}\bar{\beta}_-\right)\frac{G\alpha m}{r}\right]\hat{n}^{ij}\right\} \, ,
\end{split}
\label{eq:P1}
\end{equation}

\begin{align}
P^{3/2}Q_\Ns^{ij} &= \frac{\delta m}{m}(1-2\eta)\left\{(\mathbf{\hat{N}}\cdot \mathbf{\hat{n}})^3\frac{G\alpha m}{r}\left[\frac{5}{4}\left(3v^2-7\dot{r}^2+6\frac{G\alpha m}{r}\right)\dot{r}\hat{n}^{ij}-\frac{17}{2}\dot{r}v^{ij}\right.\right. \nonumber \\
&\left.\qquad-\frac{1}{6}\left(21v^2-105\dot{r}^2+44\frac{G\alpha m}{r}\right)\hat{n}^{(i}v^{j)}\right]+\frac{1}{4}(\mathbf{\hat{N}}\cdot \mathbf{\hat{n}})^2(\mathbf{\hat{N}}\cdot \mathbf{v})\frac{G\alpha m}{r}\left[\vphantom{\frac{G\alpha m}{r}}58v^{ij}\right. \nonumber \\
&\left.\qquad+\left(45\dot{r}^2-9v^2-28\frac{G\alpha m}{r}\right)\hat{n}^{ij}-108\dot{r}\hat{n}^{(i}v^{j)}\right]+\frac{3}{2}(\mathbf{\hat{N}}\cdot \mathbf{\hat{n}})(\mathbf{\hat{N}}\cdot \mathbf{v})^2\frac{G\alpha m}{r}[10\hat{n}^{(i}v^{j)}-3\dot{r}\hat{n}^{ij}] \nonumber \\
&\left.\qquad+\frac{1}{2}(\mathbf{\hat{N}}\cdot \mathbf{v})^3\left[\frac{G\alpha m}{r}\hat{n}^{ij}-4v^{ij}\right]\right\}  \nonumber \\
&\quad + \frac{1}{12}\frac{\delta m}{m}(\mathbf{\hat{N}}\cdot \mathbf{\hat{n}})\frac{G\alpha m}{r}\left\{2\hat{n}^{(i}v^{j)}\left[\dot{r}^2(63+54\eta+36\bar{\gamma})-\frac{G\alpha m}{r}\left(\vphantom{\frac{\delta m}{m}}128-36\eta+48\bar{\gamma}+72\bar{\beta}_+\right.\right.\right. \nonumber \\
&\left.\left.\qquad-72\frac{\delta m}{m}\bar{\beta}_-\right)+v^2(33-18\eta+24\bar{\gamma})\right]+\hat{n}^{ij}\dot{r}\left[\vphantom{\frac{G\alpha m}{r}}\dot{r}^2(15-90\eta)-v^2(63-54\eta+36\bar{\gamma})\right. \nonumber \\
&\left.\left.\qquad+\frac{G\alpha m}{r}\left(242-24\eta+96\bar{\gamma}+96\bar{\beta}_+-96\frac{\delta m}{m}\bar{\beta}_-\right)\right]-\dot{r}v^{ij}(186+24\eta+96\bar{\gamma})\vphantom{\frac{G\alpha m}{r}}\right\} \nonumber \\
&\quad+\frac{\delta m}{m}(\mathbf{\hat{N}}\cdot \mathbf{v})\left\{\frac{1}{2}v^{ij}\left[\frac{G\alpha m}{r}(3-8\eta)-2v^2(1-5\eta)\right]-\hat{n}^{(i}v^{j)}\frac{G\alpha m}{r}\dot{r}(7+4\eta+4\bar{\gamma}) \right. \nonumber \\
&\left.\qquad-\hat{n}^{ij}\frac{G\alpha m}{r}\left[\frac{3}{4}(1-2\eta)\dot{r}^2+\frac{1}{3}\left(26-3\eta+12\bar{\gamma}+6\bar{\beta}_+-6\frac{\delta m}{m}\bar{\beta}_-\right)\frac{G\alpha m}{r}-\frac{1}{4}(7-2\eta+4\bar{\gamma})v^2\right]\right\} \nonumber \\
&\quad+\frac{16}{3}\eta\left(\frac{G\alpha m}{r}\right)^2\zeta \Ss_-^2\left(\dot{r}\hat{n}^{ij}-\frac{1}{3}\hat{n}^{(i}v^{j)}\right) \, ,
\label{eq:P15N}
\end{align}

\begin{equation}
\begin{split}
P^{3/2}Q_{\Cs-\Ns}^{ij} &= 4m\int_0^\infty\left\{\frac{G\alpha m}{r^3}\left[\left(3v^2+\frac{G\alpha m}{r}-15\dot{r}^2\right)\hat{n}^{ij}+18\dot{r}\hat{n}^{(i}v^{j)}-4v^{ij}\right]\right\}_{\tau-s}\\
&\qquad \times \left\{G(1-\zeta)\left[\ln\left(\frac{s}{2R+s}\right)+\frac{11}{12}\right]-\frac{1}{12}G\alpha\zeta\left(\Ss_++\frac{\delta m}{m}\Ss_-\right)\left(\Ss_+-\frac{\delta m}{m}\Ss_-\right)\right\}ds \\
&\quad+8G\alpha\mu \zeta\Ss_-^2\int_0^\infty\left\{\frac{G\alpha m}{r^3}\left[\left(-\frac{1}{6}v^2+\frac{1}{9}\frac{G\alpha m}{r}+\frac{5}{6}\dot{r}^2\right)\hat{n}^{ij}-\dot{r}\hat{n}^{(i}v^{j)}+\frac{2}{9}v^{ij}\right]\right\}_{\tau-s}ds \, ,
\end{split}
\label{eq:P15R}
\end{equation}

\begin{align}
P^2Q_\Ns^{ij} &= \frac{1}{60}(1-5\eta+5\eta^2)\left\{\vphantom{\left(\frac{G\alpha m}{r}\right)^2}24(\mathbf{\hat{N}}\cdot\mathbf{v})^4\left[5v^{ij}-\frac{G\alpha m}{r}\hat{n}^{ij}\right]\right. \nonumber \\
&\qquad+\frac{G\alpha m}{r}(\mathbf{\hat{N}}\cdot\mathbf{\hat{n}})^4\left[\vphantom{\left(\frac{G\alpha m}{r}\right)^2}2\left(175\frac{G\alpha m}{r}-465\dot{r}^2+93v^2\right)v^{ij}+30\dot{r}\left(63\dot{r}^2-50\frac{G\alpha m}{r}-27v^2\right)\hat{n}^{(i}v^{j)}\right. \nonumber \\
&\left.\qquad+\left(1155\frac{G\alpha m}{r}\dot{r}^2-172\left(\frac{G\alpha m}{r}\right)^2-945\dot{r}^4-159\frac{G\alpha m}{r}v^2+630\dot{r}^2v^2-45v^4\right)\hat{n}^{ij}\right] \nonumber \\
&\qquad + 24\frac{G\alpha m}{r}(\mathbf{\hat{N}}\cdot\mathbf{\hat{n}})^3(\mathbf{\hat{N}}\cdot\mathbf{v})\left[87\dot{r}v^{ij}+5\dot{r}\left(14\dot{r}^2-15\frac{G\alpha m}{r}-6v^2\right)\hat{n}^{ij}\right. \nonumber \\
&\left.\qquad+16\left(5\frac{G\alpha m}{r}-10\dot{r}^2+2v^2\right)\hat{n}^{(i}v^{j)}\right]+288\frac{G\alpha m}{r}(\mathbf{\hat{N}}\cdot\mathbf{\hat{n}})(\mathbf{\hat{N}}\cdot\mathbf{v})^3[\dot{r}\hat{n}^{ij}-4\hat{n}^{(i}v^{j)}] \nonumber \\
&\left.\qquad+24\frac{G\alpha m}{r}(\mathbf{\hat{N}}\cdot\mathbf{\hat{n}})^2(\mathbf{\hat{N}}\cdot\mathbf{v})^2\left[\left(35\frac{G\alpha m}{r}-45\dot{r}^2+9v^2\right)\hat{n}^{ij}-76v^{ij}+126\dot{r}\hat{n}^{(i}v^{j)}\right]\vphantom{\left(\frac{G\alpha m}{r}\right)^2}\right\} \nonumber \\
&\quad +\frac{1}{15}(\mathbf{\hat{N}}\cdot\mathbf{v})^2\left\{\left[5\left(25-78\eta+12\eta^2+4(1-3\eta)\left(3\bar{\gamma}+\bar{\beta}_+-\frac{\delta m}{m}\bar{\beta}_-\right)\right)\frac{G\alpha m}{r}\right.\right. \nonumber \\
&\left.\qquad -(18-65\eta+45\eta^2+10(1-3\eta)\bar{\gamma})v^2+9(1-5\eta+5\eta^2)\dot{r}^2 \vphantom{\frac{G\alpha m}{r}}\right]\frac{G\alpha m}{r}\hat{n}^{ij} \nonumber \\
&\qquad+3\left[5(1-9\eta+21\eta^2)v^2-2(4-25\eta+45\eta^2)\frac{G\alpha m}{r}\right]v^{ij} \nonumber \\
&\left.\qquad+18\left[6-15\eta-10\eta^2+\frac{10}{3}(1-3\eta)\bar{\gamma}\right]\frac{G\alpha m}{r}\dot{r}\hat{n}^{(i}v^{j)}\right\} \nonumber \\
&\quad+\frac{1}{15}(\mathbf{\hat{N}}\cdot\mathbf{\hat{n}})(\mathbf{\hat{N}}\cdot\mathbf{v})\frac{G\alpha m}{r}\left\{\left[\vphantom{\frac{G\alpha m}{r}}3(36-145\eta+150\eta^2+20(1-3\eta)\bar{\gamma})v^2\right.\right. \nonumber \\
&\qquad -5\left(127-392\eta+36\eta^2+56(1-3\eta)\bar{\gamma}+32(1-3\eta)\left(\bar{\beta}_+-\frac{\delta m}{m}\bar{\beta}_-\right)\right)\frac{G\alpha m}{r} \nonumber \\
&\left.\qquad-15(2-15\eta+30\eta^2)\dot{r}^2\vphantom{\frac{G\alpha m}{r}}\right]\dot{r}\hat{n}^{ij}+6[98-295\eta-30\eta^2+50(1-3\eta)\bar{\gamma}]\dot{r}v^{ij} \nonumber \\
&\qquad+2\left[5\left(66-221\eta+96\eta^2+26(1-3\eta)\bar{\gamma}+32(1-3\eta)\left(\bar{\beta}_+-\frac{\delta m}{m}\bar{\beta}_-\right)\right)\frac{G\alpha m}{r}\right. \nonumber \\
&\left.\left.\qquad-9(18-45\eta-40\eta^2+10(1-3\eta)\bar{\gamma})\dot{r}^2-(66-265\eta+360\eta^2+50(1-3\eta)\bar{\gamma})v^2\vphantom{\frac{G\alpha m}{r}}\right]\hat{n}^{(i}v^{j)}\right\} \nonumber \\
&\quad+\frac{1}{60}(\mathbf{\hat{N}}\cdot\mathbf{\hat{n}})^2\frac{G\alpha m}{r}\left\{\left[\vphantom{\left(\frac{G\alpha m}{r}\right)^2}3(33-130\eta+150\eta^2+20(1-3\eta)\bar{\gamma})v^4+105(1-10\eta+30\eta^2)\dot{r}^4\right.\right. \nonumber \\
&\qquad+15\left(181-572\eta+84\eta^2+72(1-3\eta)\bar{\gamma}+64(1-3\eta)\left(\bar{\beta}_+-\frac{\delta m}{m}\bar{\beta}_-\right)\right)\frac{G\alpha m}{r}\dot{r}^2 \nonumber \\
&\qquad-\left(131-770\eta+930\eta^2-80(1-3\eta)\bar{\gamma}+160(1-3\eta)\left(\bar{\beta}_+-\frac{\delta m}{m}\bar{\beta}_-\right)\right)\frac{G\alpha m}{r}v^2 \nonumber \\
&\qquad-60(9-40\eta+60\eta^2+5(1-3\eta)\bar{\gamma})v^2\dot{r}^2 \nonumber \\
&\left.\qquad-8\left(131-390\eta+30\eta^2+60(1-3\eta)\bar{\gamma}+65(1-3\eta)\left(\bar{\beta}_+-\frac{\delta m}{m}\bar{\beta}_-\right)\right)\left(\frac{G\alpha m}{r}\right)^2\right]\hat{n}^{ij} \nonumber \\
&\qquad+4\left[\vphantom{\frac{G\alpha m}{r}}(12+5\eta-315\eta^2-10(1-3\eta)\bar{\gamma})v^2-9(39-115\eta-35\eta^2+20(1-3\eta)\bar{\gamma})\dot{r}^2 \right.\nonumber \\
&\left.\qquad+5\left(29-104\eta+84\eta^2+8(1-3\eta)\bar{\gamma}+28(1-3\eta)\left(\bar{\beta}_+-\frac{\delta m}{m}\bar{\beta}_-\right)\right)\frac{G\alpha m}{r}\right]v^{ij} \nonumber \\
&\qquad+4\left[\vphantom{\frac{G\alpha m}{r}}15(18-40\eta-75\eta^2+10(1-3\eta)\bar{\gamma})\dot{r}^2\right. \nonumber \\
&\qquad-5\left(197-640\eta+180\eta^2+76(1-3\eta)\bar{\gamma}+80(1-3\eta)\left(\bar{\beta}_+-\frac{\delta m}{m}\bar{\beta}_-\right)\right)\frac{G\alpha m}{r} \nonumber \\
&\left.\left.\qquad+3(21-130\eta+375\eta^2+20(1-3\eta)\bar{\gamma})v^2\vphantom{\frac{G\alpha m}{r}}\right]\dot{r}\hat{n}^{(i}v^{j)}\vphantom{\left(\frac{G\alpha m}{r}\right)^2}\right\} \nonumber \\
&\quad+\frac{1}{15}\eta\left(\frac{G\alpha m}{r}\right)^2\zeta\Ss_-\left\{\vphantom{\frac{G\alpha m}{r}}(\mathbf{\hat{N}}\cdot\mathbf{\hat{n}})\left[\vphantom{\frac{G\alpha m}{r}}192\left(\Ss_+-\frac{\delta m}{m}\Ss_-\right)\dot{r}\hat{n}^{(i}v^{j)}\right.\right. \nonumber \\
&\qquad+\left(-120\left(\Ss_+-\frac{\delta m}{m}\Ss_-\right)\dot{r}^2+24\left(\Ss_+-\frac{\delta m}{m}\Ss_-\right)v^2+4\left(11\Ss_++19\frac{\delta m}{m}\Ss_-\right)\frac{G\alpha m}{r}\right)\hat{n}^{ij} \nonumber \\
&\left.\left.\qquad-4\left(17\Ss_+-7\frac{\delta m}{m}\Ss_-\right)v^{ij}\right]+(\mathbf{\hat{N}}\cdot\mathbf{v})\left[-16\left(8\Ss_++7\frac{\delta m}{m}\Ss_-\right)\hat{n}^{(i}v^{j)}+12\left(7\Ss_++3\frac{\delta m}{m}\Ss_-\right)\dot{r}\hat{n}^{ij}\right]\right\} \nonumber \\
&\quad+\frac{1}{60}\left\{\left[\vphantom{\left(\frac{G\alpha m}{r}\right)^2}\left(\vphantom{\frac{\delta m}{m}}467+780\eta-120\eta^2+120(2+3\eta)\bar{\gamma}+10\bar{\gamma}^2+40\left(\bar{\delta}_++\frac{\delta m}{m}\bar{\delta}_-\right)\right.\right.\right. \nonumber \\
&\left.\quad \qquad-40(1-3\eta)\left(\bar{\beta}_+-\frac{\delta m}{m}\bar{\beta}_-\right)\right)\frac{G\alpha m}{r}v^2-15\left(61-96\eta+48\eta^2+\frac{8}{3}(7-12\eta)\bar{\gamma}-\frac{4}{3}\bar{\gamma}^2\right. \nonumber \\
&\left.\quad \qquad-\frac{16}{3}\left(\bar{\delta}_++\frac{\delta m}{m}\bar{\delta}_-\right)+\frac{32}{3}(2-3\eta)\left(\bar{\beta}_+ -\frac{\delta m}{m}\bar{\beta}_-\right)\right)\frac{G\alpha m}{r}\dot{r}^2 -(144-265\eta-135\eta^2+20(4-9\eta)\bar{\gamma})v^4 \nonumber \\
&\quad \qquad+6(24-95\eta+75\eta^2+10(1-3\eta)\bar{\gamma})v^2\dot{r}^2 -2\left(\vphantom{\frac{G\alpha m}{r}}642+545\eta+20(29+6\eta)\bar{\gamma}+15(9-2\eta)\bar{\gamma}^2 \right. \nonumber \\
&\left.\quad \qquad+80(8+6\eta+3\bar{\gamma})\bar{\beta}_+-80(8+3\bar{\gamma})\frac{\delta m}{m}\bar{\beta}_-+60(1-2\eta)(\bar{\delta}_+-2\bar{\chi}_+)+60\frac{\delta m}{m}(\bar{\delta}_-+2\bar{\chi}_-)\right. \nonumber \\
&\left.\left.\quad \qquad-1440\eta\frac{\bar{\beta}_1\bar{\beta}_2}{\bar{\gamma}}\right)\left(\frac{G\alpha m}{r}\right)^2 -45(1-5\eta+5\eta^2)\dot{r}^4\right]\frac{G\alpha m}{r}\hat{n}^{ij} \nonumber \\
&\qquad+\left[\vphantom{\left(\frac{G\alpha m}{r}\right)^2}4(69+10\eta-135\eta^2+10(4-3\eta)\bar{\gamma})\frac{G\alpha m}{r}v^2-12(3+60\eta+25\eta^2+40\eta\bar{\gamma})\frac{G\alpha m}{r}\dot{r}^2\right. \nonumber \\
&\quad \qquad+45(1-7\eta+13\eta^2)v^4-10\left(\vphantom{\frac{\delta m}{m}}56+165\eta-12\eta^2+4(7+24\eta)\bar{\gamma}+\bar{\gamma}^2\right. \nonumber \\
&\left.\left.\quad \qquad+4\left(\bar{\delta}_++\frac{\delta m}{m}\bar{\delta}_-\right)-4(1+3\eta)\bar{\beta}_++4(1-3\eta)\frac{\delta m}{m}\bar{\beta}_-\right)\left(\frac{G\alpha m}{r}\right)^2\right]v^{ij} \nonumber \\
&\qquad+4\left[\vphantom{\frac{G\alpha m}{r}}2(36+5\eta-75\eta^2+5(4-3\eta)\bar{\gamma})v^2-6(7-15\eta-15\eta^2+5(1-3\eta)\bar{\gamma})\dot{r}^2\right. \nonumber \\
&\quad \qquad+5\left(35+45\eta+36\eta^2+8(1+6\eta)\bar{\gamma}-\bar{\gamma}^2+16(1-3\eta)\bar{\beta}_+-8(2-3\eta)\frac{\delta m}{m}\bar{\beta}_-\right. \nonumber \\
&\left.\left.\left.\quad \qquad-4\left(\bar{\delta}_++\frac{\delta m}{m}\bar{\delta}_-\right)\right)\frac{G\alpha m}{r}\right]\frac{G\alpha m}{r}\dot{r}\hat{n}^{(i}v^{j)}\vphantom{\left(\frac{G\alpha m}{r}\right)^2}\right\} \, ,
\label{eq:P2N}
\end{align}

\begin{equation}
\begin{split}
P^2Q_{\Cs-\Ns}^{ij} &= 2m\int_0^\infty\left\{\frac{G\alpha m}{r^3}\left[15\left(3v^2+2\frac{G\alpha m}{r}-7\dot{r}^2\right)\dot{r}\hat{n}^{ij}(\mathbf{\hat{N}}\cdot\mathbf{\hat{n}})\right.\right.\\
&\qquad-\left(13v^2+\frac{22}{3}\frac{G\alpha m}{r}-65\dot{r}^2\right)(\hat{n}^{ij}(\mathbf{\hat{N}}\cdot\mathbf{v})+2\hat{n}^{(i}v^{j)}(\mathbf{\hat{N}}\cdot\mathbf{\hat{n}}))-40\dot{r}(v^{ij}(\mathbf{\hat{N}}\cdot\mathbf{\hat{n}})+2\hat{n}^{(i}v^{j)}(\mathbf{\hat{N}}\cdot\mathbf{v}))\\
&\left.\left.\qquad+20v^{ij}(\mathbf{\hat{N}}\cdot\mathbf{v})\vphantom{\frac{G\alpha m}{r}}\right]\right\}_{\tau-s}\\
&\qquad \times \left\{G(1-\zeta)\frac{\delta m}{m}\ln\left[\left(\frac{s}{2R+s}\right)+\frac{97}{60}\right]-\frac{1}{20}G\alpha\zeta\left(\Ss_++\frac{\delta m}{m}\Ss_-\right)\left(\frac{\delta m}{m}\Ss_+-(1-2\eta)\Ss_-\right)\right\}ds\\
&\quad+8G(1-\zeta)\delta m \int_0^\infty\left\{\frac{G\alpha m}{r^3}\left[\left(v^2-\frac{2}{3}\frac{G\alpha m}{r}-5\dot{r}^2\right)(\hat{n}^{ij}(\mathbf{\hat{N}}\cdot\mathbf{v})-\hat{n}^{(i}v^{j)}(\mathbf{\hat{N}}\cdot\mathbf{\hat{n}}))\right.\right.\\
&\left.\left.\qquad-2\dot{r}(v^{ij}(\mathbf{\hat{N}}\cdot\mathbf{\hat{n}})-\hat{n}^{(i}v^{j)}(\mathbf{\hat{N}}\cdot\mathbf{v}))\vphantom{\frac{G\alpha m}{r}}\right]\right\}_{\tau-s}\left[\ln\left(\frac{s}{2R+s}\right)+\frac{7}{6}\right]ds\\
&\quad + \frac{1}{15}G\alpha \mu \zeta\Ss_-\left(\Ss_+-\frac{\delta m}{m}\Ss_-\right)\int_0^\infty\left\{\frac{G\alpha m}{r^3}\left[\left(225v^2+18\frac{G\alpha m}{r}-525\dot{r}^2\right)\dot{r}\hat{n}^{ij}(\mathbf{\hat{N}}\cdot\mathbf{\hat{n}})\right.\right.\\
&\qquad+\left(-9v^2-6\frac{G\alpha m}{r}+45\dot{r}^2\right)\hat{n}^{ij}(\mathbf{\hat{N}}\cdot\mathbf{v})+\left(-162v^2+44\frac{G\alpha m}{r}+810\dot{r}^2\right)\hat{n}^{(i}v^{j)}(\mathbf{\hat{N}}\cdot\mathbf{\hat{n}})\\
&\left.\left.\qquad-144\dot{r}\hat{n}^{(i}v^{j)}(\mathbf{\hat{N}}\cdot\mathbf{v})-276\dot{r}v^{ij}(\mathbf{\hat{N}}\cdot\mathbf{\hat{n}})+56v^{ij}(\mathbf{\hat{N}}\cdot\mathbf{v})\vphantom{\frac{G\alpha m}{r}}\right]\right\}_{\tau-s}ds \, .
\end{split}
\label{eq:P2R}
\end{equation}
\end{subequations}

\subsection{Discussion}
\label{sec:discussion}
We can check that the expressions above reduce to the correct GR result, WW (6.11), by taking $\alpha = 1$ and $\zeta = \Ss_- = \bar{\gamma} = \bar{\beta}_i = \bar{\delta}_i = \bar{\chi}_i = 0$.  The expressions have been arranged to facilitate this comparison.  (For a complete comparison, we also set $G = 1$.)

At Newtonian order, the only difference from GR is the presence of $\alpha$ multiplying the total mass $m$ and the total factor of $1-\zeta$ out front.  The same is true at 0.5PN order.  More substantial differences begin at 1PN order, with the appearance of the parameters $\bar{\gamma}$ and $\bar{\beta}_i$.  The quantities $\Ss_+$ and $\Ss_-$ show up at 1.5PN order, with $\bar{\delta}_i$ and $\bar{\chi}_i$ appearing at 2PN order.  Although the expressions are very complicated, especially at 1.5PN order and above, they still depend only on this relatively small set of parameters.  The parameter set is also identical to that needed to describe the equations of motion; there are no additional dependences on the coupling $\omega(\phi)$ or the sensitivities of the bodies.

Most of the terms have the same form as in general relativity, albeit with highly modified coefficients.  The first exception occurs at 1.5PN order.  For $I_{\text{EW}}^{ij}$ to contribute to the GWs at this order, one of four pairings must exist: a 1.5PN term in the two-body EW moment with the 0PN equations of motion, a 1PN term with 0.5PN equations of motion, a 0.5PN term with 1PN equations of motion, or a 0PN term with 1.5PN equations of motion.  In general relativity, $I_{\text{EW}}^{ij}$ contains no terms at 0.5PN and 1.5PN order.  There are also no 0.5PN or 1.5 PN terms in the equations of motion, so $I_{\text{EW}}^{ij}$ does not contribute to the waves at 1.5PN order.  (Only the three- and five-index moments contribute.)  However, in scalar-tensor theory, $I_{\text{EW}}^{ij}$ contains a 1.5PN contribution arising from the surface moment.  This is a consequence of the $O(\rho \epsilon^{5/2})$ term in $\tau^{ij}$, which scales like three time derivatives of the scalar dipole moment $\Is_s^i$.  In addition, dipole radiation reaction introduces 1.5PN terms in the equations of motion.  Together, both of these effects produce a new 1.5PN term in the final gravitational waveform.  This is the last term in \eqref{eq:P15N}; note how it does not depend on the direction to the source $\mathbf{\hat{N}}$.  [If one constructs the waveform using \eqref{eq:tensorEW} instead of \eqref{eq:EWwaveform}, the same term comes solely from the $O(\rho \epsilon^{5/2})$ piece of $\tau^{ij}$.]

The scalar dipole moment also affects the contribution from $I_{\text{EW}}^{ijk}$, producing terms in the final waveform at 2PN order which were not present in GR.  They can be seen in \eqref{eq:P2N} as those which depend only on one power of $\mathbf{\hat{N}}$.  Here, the scalar dipole enters in three ways: First, the surface moment produces a 2PN term, again a consequence of $\dddot{\Is}_s^i$ in $\tau^{ij}$.  Second, as discussed in Sec.\ \ref{sec:twobody}, the radiation of linear momentum at 1.5PN order affects the conversion to relative coordinates, generating another 2PN piece in the two-body moment.  Finally, the 1.5PN piece of the equations of motion enters time derivatives of the lowest order (0.5PN) piece of $I_{\text EW}^{ijk}$.

Other interesting deviations from general relativity occur in \eqref{eq:P15R} and \eqref{eq:P2R}.  The radiation-zone integrals produce two types of terms, those which depend on the instantaneous (but retarded) values of the source moments and those which depend on the integrated history of the source up until the waves are emitted.  The latter terms are known as hereditary terms.  (The distinction is best seen in Sec. \ref{sec:RZGWs}, since the final results have the instantaneous terms brought inside the integrals.)  In GR, all hereditary terms up to 2PN order are so-called ``tail integrals,'' with logarithmic factors in their integrands.  Tails result from the scattering of the waves off the background curvature.  They arise from the final term in \eqref{eq:Lambdamunu}.  If moved to the left-hand side of the reduced wave equation, it represents a modification of the flat-spacetime wave operator.  The tails thus ensure that the waves propagate outward on the true null cones of the background spacetime, rather than the null cones of the fictitious flat spacetime used to formulate the relaxed field equations.  At 1.5PN order, the tail term arises from a coupling of the monopole and mass quadrupole.  The instantaneous terms at that order have the same coupling.  At 2PN order, both tail and instantaneous terms feature monopole-mass octupole and monopole-current quadrupole couplings.  

Scalar-tensor theory adds no new tail integrals to the ones already present in general relativity.  While we do find a new monopole-mass quadrupole coupling at 1.5PN order, as well as a new monopole-mass octupole coupling at 2PN order (this time involving the quantities $m_s$, $\Is_s^{ij}$, and $\Is_s^{ijk}$), these terms are all instantaneous.  Instead, we find an entirely different type of hereditary term at 1.5PN and 2PN orders, one which does not have a logarithmic factor in the integrand.  Terms like these are sourced by the energy of the gravitational waves themselves.  In general relativity, the first one appears at 2.5PN order, with a mass quadrupole-mass quadrupole coupling.  It contains the lowest order piece of the nonlinear gravitational-wave memory, or Christodoulou memory \cite{p83,c91,bd92,f09}.  Specifically, the multiplication of the two quadrupole moments leads to a contribution at zero frequency (a ``DC'' term), in addition to the usual oscillatory terms.  This DC term grows secularly throughout the inspiral of the system and causes a permanent change in a detector, a ``memory'' of the passing GW signal.  While the memory term formally enters the waveform at 2.5PN order, its effective post-Newtonian order is reduced by the integration over the entire history of the system.  In fact, because the memory integrand is approximately multiplied by the (2.5PN) radiation-reaction time scale, the lowest order memory term effectively enters the GR waveform at 0PN order.  With such a strong signal, it may be possible to detect the memory effect with gravitational-wave detectors \cite{f09,f09b}.  

In scalar-tensor theory, the 1.5PN nonlogarithmic integral contains a new, lower order memory effect with a mass dipole-mass dipole coupling.  Because the lowest order radiation reaction is now also at 1.5PN order, this term should effectively enter the waveform at 0PN order, with the quadrupole-quadrupole memory at higher order.  (The exact PN ordering of the memory terms will depend on the specifics of the scalar-tensor theory and the compact object sensitivities.)  Scalar-tensor theory also produces instantaneous dipole-dipole terms at 1.5PN order.  Like the hereditary term, they contain DC components; however, since they are not integrated over the binary's history, their effect remains at 1.5PN order.  They are equivalent to the nonhereditary, zero-frequency terms Arun {\it et al.} discovered at 2.5PN order in general relativity \cite{abiq04}.  

By contrast, the new 2PN hereditary term does not contain a memory effect.  While it has the same basic form as the 1.5PN term (i.e., no logarithm in the integrand), the beating between the mass dipole and mass quadrupole produces no DC component.  The same is true for the new instantaneous terms with this coupling.  This result is equivalent to the lack of a 0.5PN memory effect (appearing formally at 3PN order) in general relativity: In that case, the beating is between the mass quadrupole and mass octupole, resulting in no zero-frequency terms \cite{bfis08}.

It is instructive to examine the waveform in a few special cases.  For binary black holes, $s_1 = s_2 = 1/2$, and all sensitivity derivatives vanish.  This means that $\alpha = 1-\zeta$, and all of the rest of our scalar-tensor parameters ($\bar{\gamma}$, $\bar{\beta}_i$, $\bar{\delta}_i$, $\bar{\chi}_i$, $\Ss_+$, and $\Ss_-$) vanish.  The 2.5PN equations of motion derived in MW then have the exact same form as in general relativity, except for a factor $1-\zeta$ multiplying each instance of the total mass $m$.  That is, the equations of motion for a binary with masses $(m_1,m_2)$ in general relativity are identical to those for a binary with rescaled masses $(m_1/(1-\zeta),m_2/(1-\zeta))$ in scalar-tensor theory.  Since the masses of the bodies are defined by their Keplerian motion, this rescaling is unmeasurable.  Therefore, to 2.5PN order, the motion of two black holes in scalar-tensor theory is indistinguishable from the motion in GR.  MW predicted that the gravitational waves produced by binary black holes would be similarly indistinguishable from those produced in GR.  We see here that the conjecture is correct, at least to 2PN order in the tensor gravitational waves.  

As discussed in MW, this result is not surprising.  Hawking originally showed that stationary, asymptotically flat black holes in vacuum are identical in both theories \cite{h72}, leading to conjectures that the same might be true for black hole binaries.  Still, this work shows only that the theories are indistinguishable to 2.5PN order in the dynamics and 2PN order in the radiation.  It remains for future work to investigate whether indistinguishability holds to all post-Newtonian orders.  (See MW for precise details of a conjecture on this point.)  There is good evidence that it does: Yunes {\it et al.} \cite{ypc12} proved it, but only to lowest order in the mass ratio.  Healy {\it et al.} \cite{h11} used numerical relativity simulations to show that any initial scalar field in the system is quickly radiated away, after which the holes behave identically to the general relativity case.  Possible caveats which may break indistinguishability include the introduction of a potential for the scalar field or a time-varying scalar field at infinity \cite{j99,hb12}.  It would be interesting to investigate the dynamics and radiation in such scenarios.

For a system containing one neutron star (say, body 1) and one black hole (body 2), $s_2 = 1/2, s_2' = 0$, and $s_2'' = 0$.  Then $\alpha = 1-\zeta$, and all other parameters vanish except

\begin{equation}
\bar{\delta}_1 = \frac{\zeta}{1-\zeta}(1-2s_1)^2 \equiv Q 
\end{equation}
and

\begin{equation}
\Ss_+ = \Ss_- = \frac{1}{2}\alpha^{-1/2}(1-2s_1) \, .
\end{equation}
For the opposite choice of bodies, $\Ss_-$ has the opposite sign.  Notably, $\Ss_+^2 = \Ss_-^2 = \Ss_+\Ss_- = Q/(4\zeta)$.  Through 1PN order, only $\alpha$, $\bar{\gamma} = 0$, and $\bar{\beta_i} = 0$ appear in the expressions for the waves, and so the waves are identical to those in GR (after mass rescaling).  At 1.5PN order, deviations start to occur.  However, the deviations are always parametrized by $\bar{\delta}_1$, $\zeta \Ss_+^2$, $\zeta \Ss_-^2$, or $\zeta \Ss_+\Ss_-$.  Therefore, through 2PN order, the tensor waveform for a mixed black hole-neutron star system differs from the general relativity waveform only by the single parameter $Q$.  This is again equivalent to a result found by MW for the equations of motion.  Because $Q$ contains no information on the derivatives of the coupling function $\omega(\phi)$ (i.e., the parameters $\lambda_1$ and $\lambda_2$), we cannot, at 2PN order, formally distinguish the waveform produced in the Brans-Dicke theory [$\omega(\phi) = \omega_0$] from that produced in a general scalar-tensor theory of the type we consider.  The only difference will be that for a given neutron star of a certain central density and total number of baryons, different scalar-tensor theories will produce different results for the neutron star mass $m_1$ and sensitivity $s_1$.  One can imagine using gravitational waves to measure masses and sensitivities for a wide variety of sources and then producing a mass-sensitivity relation, much like the neutron-star mass-radius relations used to study the nuclear equation of state.  This relation could then be used to rule out various models of the coupling function $\omega(\phi)$.

Before we can completely understand what the measurement of gravitational waves from a compact binary will tell us about scalar-tensor theories of gravity, we must first derive the gravitational-wave phasing.  To do so, we will need the rate at which the binary loses energy to gravitational waves, both tensor and scalar.  The next paper in our series will derive the scalar waves.  The process is identical to that presented in this paper, with two complications.  First, we cannot eliminate non-TT terms in the scalar case; indeed, for a scalar, TT is not defined.  This will lead to more surviving terms and a need for care when reusing parts of this analysis.  Second, and more daunting, the scalar ``EW moments'' defined by \eqref{eq:scalarEW} begin with a monopole moment, which has relative order -1PN compared to the tensor two-index moment.  The next piece, a dipole moment, has order -0.5PN.  Therefore, to obtain the 2PN scalar waveform, we will need to compute the source $\tau_s(\tau,\mathbf{x}')$ to $O(\rho\epsilon^3)$, or 3PN order.

\acknowledgments

We are extremely grateful to Clifford Will for his many useful insights into the calculation and careful reading of the manuscript.  We also thank Saeed Mirshekari for useful discussions.  This work was supported in part by the National Science Foundation, Grants No.\ PHY-0965133, No.\ PHY-1260995, and No.\ PHY-1306069.  We acknowledge the hospitality of the Institut d'Astrophysique de Paris, where a portion of this work was completed.  The software \textsc{mathematica} was used to check or perform many of the calculations.

\appendix

\section{Potentials}
\label{app:potentials}

Here we present a list of potentials which appear in MW and this paper.  All potentials are defined in terms of the $\rho^*$ density.  The fundamental integrals are defined in \eqref{eq:Ppot} and \eqref{eq:sigmapot}--\eqref{eq:Ypot}.  Some of the potentials involve generalizations of these fundamental integrals, like $X_s$, or $X^{ij}$, which are defined in the obvious ways.

\begin{align}
U &\equiv \Sigma(1) \, , & U_s &\equiv \Sigma_s(1) \, ,  \nonumber \\
V^i &\equiv \Sigma^i(1) \, , & \Phi_1^{ij} &\equiv \Sigma^{ij}(1) \, ,  \nonumber \\
\Phi_1 &\equiv \Sigma^{ii}(1) \, , & \Phi^s_1 &\equiv \Sigma_s(v^2) \, ,  \nonumber \\
\Phi_2 &\equiv \Sigma(U) \, , & \Phi^s_2 &\equiv \Sigma_s(U) \, ,  \nonumber \\
\Phi_{2s} &\equiv \Sigma(U_s) \, , & \Phi^s_{2s} &\equiv \Sigma_s(U_s) \, ,  \nonumber \\
X &\equiv X(1) \, , & X_s &\equiv X_s(1) \, ,  \nonumber \\
V_2^i &\equiv \Sigma^i(U) \, ,  & V_{2s}^i &\equiv \Sigma^i(U_s) \, ,  \nonumber \\
\Phi_2^i &\equiv \Sigma(V^i) \, ,  & Y &\equiv Y(1) \, ,  \nonumber \\
X^i &\equiv X^i(1) \, ,  & X_1 &\equiv X^{ii}(1) \, ,  \nonumber \\
X_2 &\equiv X(U) \, , & X_{2s} &\equiv X(U_s) \, , \nonumber \\ 
X^s_2 &\equiv X_s(U) \, , & X^s_{2s} &\equiv X_s(U_s) \, ,  \nonumber \\
P_2^{ij} &\equiv P(U^{,i}U^{,j}) \, , & P_2 &\equiv P_2^{ii} = \Phi_2-\frac{1}{2}U^2 \, ,  \nonumber \\
P_{2s}^{ij} &\equiv P(U_s^{,i}U_s^{,j}) \, , & P_{2s} &\equiv P_{2s}^{ii} = \Phi^s_{2s}-\frac{1}{2}U_s^2 \, ,  \nonumber \\
G_1 &\equiv P(\dot{U}^2) \, , & G_{1s} &\equiv P(\dot{U}_s^2) \, ,  \nonumber \\
G_2 &\equiv P(U\ddot{U}) \, , & G_{2s} &\equiv P(U\ddot{U}_s) \, ,  \nonumber \\
G_3 &\equiv -P(\dot{U}^{,k}V^k) \, , & G_{3s} &\equiv -P(\dot{U}_s^{,k}V^k) \, ,  \nonumber \\
G_4 &\equiv P(V^{i,j}V^{j,i}) \, , & G_5 &\equiv -P(\dot{V}^kU^{,k}) \, ,  \nonumber \\
G_6 &\equiv P(U^{,ij}\Phi_1^{ij}) \, , & G_{6s} &\equiv P(U_s^{,ij}\Phi_1^{ij}) \, ,  \nonumber \\
G_7^i &\equiv P(U^{,k}V^{k,i})+\frac{3}{4}P(U^{,i}\dot{U}) \, ,  \nonumber \\
H &\equiv P(U^{,ij}P_2^{ij}) \, ,  & H_s &\equiv P(U^{,ij}P_{2s}^{ij}) \, ,  \nonumber \\
H^s &\equiv P(U_s^{,ij}P_2^{ij}) \, , & H_s^s &\equiv P(U_s^{,ij}P_{2s}^{ij}) \, .
\end{align}

\section{Products of unit vectors}
\label{app:unitvectors}

Both \eqref{eq:angularintegral} and \eqref{eq:ABintegrals} rely on the use of symmetric, trace-free (STF) products of unit vectors.  They can be found using the formula

\begin{equation}
\hat{n}^{\langle L\rangle} \equiv \sum_{p=0}^{[l/2]}(-1)^p\frac{(2l-2p-1)!!}{(2l-1)!!}[\hat{n}^{L-2P}\delta^P+\text{sym}(q)] \, .
\end{equation}
Angle braces on indices define a tensor as being STF.  Here we use the convention that capital letters denote the dimensionality of products: There are $l$ indices on the STF tensor, $p$ Kronecker deltas (with $2p$ total indices among them), and $l-2p$ unit vectors.  We use $[l/2]$ to denote the largest integer less than or equal to $l/2$.  The expression $\text{sym}(q)$ stands for all the other distinct terms which result from permuting the indices on $\hat{n}^{L-2P}\delta^P$.  There are a total of $q = l!/[(l-2p)!(2p)!!]$ terms, including the one shown.  The STF tensors we need are

\begin{subequations}
\begin{align}
\hat{n}^{\langle ij\rangle} &= \hat{n}^{ij}-\frac{1}{3}\delta^{ij} \, , \\
\hat{n}^{\langle ijk\rangle} &= \hat{n}^{ijk} - \frac{1}{5}(\hat{n}^i\delta^{jk}+\hat{n}^j\delta^{ik}+\hat{n}^k\delta^{ij}) \, , \\
\hat{n}^{\langle ijkl\rangle} &= \hat{n}^{ijkl}-\frac{1}{7}[\hat{n}^{ij}\delta^{kl}+\text{sym}(6)]+\frac{1}{35}(\delta^{ij}\delta^{kl}+\delta^{ik}\delta^{jl}+\delta^{il}\delta^{jk}) \, , \\
\hat{n}^{\langle ijkl m\rangle} &= \hat{n}^{ijkl m}-\frac{1}{9}[\hat{n}^{ijk}\delta^{l m}+\text{sym}(10)]+\frac{1}{63}[\hat{n}^i\delta^{jk}\delta^{l m}+\text{sym}(15)] \, , \\
\hat{n}^{\langle ijkl mn\rangle} &= \hat{n}^{ijkl mn}-\frac{1}{11}[\hat{n}^{ijkl}\delta^{mn}+\text{sym}(15)]+\frac{1}{99}[\hat{n}^{ij}\delta^{kl}\delta^{mn}+\text{sym}(45)]-\frac{1}{693}[\delta^{ij}\delta^{kl}\delta^{mn}+\text{sym}(15)] \, .
\end{align}
\end{subequations}
These expressions can be used to convert back and forth between STF and non-STF products as needed.

Many times in this work we need to evaluate averages of unit tensors over a spherical surface.  Defining

\begin{equation}
\langle\langle\Psi\rangle\rangle \equiv \frac{1}{4\pi}\int \Psi(\theta,\phi)\ d^2\Omega \, ,
\end{equation}
it can be shown that

\begin{equation}
\langle\langle\hat{n}^{\langle L\rangle}\rangle\rangle = 0 \, .
\end{equation}
Converting to non-STF tensors, we find

\begin{equation}
\langle\langle\hat{n}^L\rangle\rangle = \frac{1}{(l+1)!!}[\delta^{L/2}+\text{sym}(q)] \, ,
\end{equation}
where $q = (l-1)!!$, for $l$ even.  Specifically, we need the following: 

\begin{subequations}
\begin{align}
\langle\langle\hat{n}^{ij}\rangle\rangle &= \frac{1}{3}\delta^{ij} \, ,  \\
\langle\langle\hat{n}^{ijkl}\rangle\rangle &= \frac{1}{15}(\delta^{ij}\delta^{kl}+\delta^{ik}\delta^{jl}+\delta^{il}\delta^{jk}) \, .
\end{align}
\end{subequations}
For $l$ odd, $\langle\langle\hat{n}^L\rangle\rangle = 0$.  

\bibliographystyle{apsrev}
\bibliography{tensorwaves}

\begin{thebibliography}{65}
\expandafter\ifx\csname natexlab\endcsname\relax\def\natexlab#1{#1}\fi
\expandafter\ifx\csname bibnamefont\endcsname\relax
  \def\bibnamefont#1{#1}\fi
\expandafter\ifx\csname bibfnamefont\endcsname\relax
  \def\bibfnamefont#1{#1}\fi
\expandafter\ifx\csname citenamefont\endcsname\relax
  \def\citenamefont#1{#1}\fi
\expandafter\ifx\csname url\endcsname\relax
  \def\url#1{\texttt{#1}}\fi
\expandafter\ifx\csname urlprefix\endcsname\relax\def\urlprefix{URL }\fi
\providecommand{\bibinfo}[2]{#2}
\providecommand{\eprint}[2][]{\url{#2}}

\bibitem[{\citenamefont{{Harry} and {LIGO Scientific
  Collaboration}}(2010)}]{h10a}
\bibinfo{author}{\bibfnamefont{G.~M.} \bibnamefont{{Harry}}} \bibnamefont{and}
  \bibinfo{author}{\bibnamefont{{LIGO Scientific Collaboration}}},
  \bibinfo{journal}{Classical Quantum Gravity} \textbf{\bibinfo{volume}{27}},
  \bibinfo{eid}{084006} (\bibinfo{year}{2010}).

\bibitem[{\citenamefont{{Accadia {\it et al.}}}()}]{virgo}
\bibinfo{author}{\bibfnamefont{T.}~\bibnamefont{{Accadia {\it et al.}}}},
  \bibinfo{note}{{Virgo Document VIR-0128A-12 (2012),
  https://tds.ego-gw.it/ql/?c=8940}}.

\bibitem[{\citenamefont{{Abadie {\it et al.}}}(2010)}]{a10}
\bibinfo{author}{\bibfnamefont{J.}~\bibnamefont{{Abadie {\it et al.}}}},
  \bibinfo{journal}{Classical Quantum Gravity} \textbf{\bibinfo{volume}{27}},
  \bibinfo{eid}{173001} (\bibinfo{year}{2010}).

\bibitem[{\citenamefont{{Amaro-Seoane {\it et al.}}}(2013)}]{a13}
\bibinfo{author}{\bibfnamefont{P.}~\bibnamefont{{Amaro-Seoane {\it et al.}}}},
  \bibinfo{journal}{GW Notes} \textbf{\bibinfo{volume}{6}}, \bibinfo{pages}{4}
  (\bibinfo{year}{2013}).

\bibitem[{\citenamefont{{Hobbs {\it et al.}}}(2010)}]{h10b}
\bibinfo{author}{\bibfnamefont{G.}~\bibnamefont{{Hobbs {\it et al.}}}},
  \bibinfo{journal}{Classical Quantum Gravity} \textbf{\bibinfo{volume}{27}},
  \bibinfo{eid}{084013} (\bibinfo{year}{2010}).

\bibitem[{\citenamefont{{Blanchet} et~al.}(2008)\citenamefont{{Blanchet},
  {Faye}, {Iyer}, and {Sinha}}}]{bfis08}
\bibinfo{author}{\bibfnamefont{L.}~\bibnamefont{{Blanchet}}},
  \bibinfo{author}{\bibfnamefont{G.}~\bibnamefont{{Faye}}},
  \bibinfo{author}{\bibfnamefont{B.~R.} \bibnamefont{{Iyer}}},
  \bibnamefont{and} \bibinfo{author}{\bibfnamefont{S.}~\bibnamefont{{Sinha}}},
  \bibinfo{journal}{Classical Quantum Gravity} \textbf{\bibinfo{volume}{25}},
  \bibinfo{eid}{165003} (\bibinfo{year}{2008}).

\bibitem[{\citenamefont{{Faye} et~al.}(2012)\citenamefont{{Faye}, {Marsat},
  {Blanchet}, and {Iyer}}}]{fmbi12}
\bibinfo{author}{\bibfnamefont{G.}~\bibnamefont{{Faye}}},
  \bibinfo{author}{\bibfnamefont{S.}~\bibnamefont{{Marsat}}},
  \bibinfo{author}{\bibfnamefont{L.}~\bibnamefont{{Blanchet}}},
  \bibnamefont{and} \bibinfo{author}{\bibfnamefont{B.~R.}
  \bibnamefont{{Iyer}}}, \bibinfo{journal}{Classical Quantum Gravity}
  \textbf{\bibinfo{volume}{29}}, \bibinfo{eid}{175004} (\bibinfo{year}{2012}).

\bibitem[{\citenamefont{{Blanchet}}(2014)}]{b13}
\bibinfo{author}{\bibfnamefont{L.}~\bibnamefont{{Blanchet}}},
  \bibinfo{journal}{Living Rev. Relativity} \textbf{\bibinfo{volume}{17}},
  \bibinfo{pages}{2} (\bibinfo{year}{2014}).

\bibitem[{\citenamefont{{Pretorius}}(2005)}]{p05}
\bibinfo{author}{\bibfnamefont{F.}~\bibnamefont{{Pretorius}}},
  \bibinfo{journal}{\prl} \textbf{\bibinfo{volume}{95}},
  \bibinfo{pages}{121101} (\bibinfo{year}{2005}).

\bibitem[{\citenamefont{{Campanelli} et~al.}(2006)\citenamefont{{Campanelli},
  {Lousto}, {Marronetti}, and {Zlochower}}}]{clmz06}
\bibinfo{author}{\bibfnamefont{M.}~\bibnamefont{{Campanelli}}},
  \bibinfo{author}{\bibfnamefont{C.~O.} \bibnamefont{{Lousto}}},
  \bibinfo{author}{\bibfnamefont{P.}~\bibnamefont{{Marronetti}}},
  \bibnamefont{and}
  \bibinfo{author}{\bibfnamefont{Y.}~\bibnamefont{{Zlochower}}},
  \bibinfo{journal}{\prl} \textbf{\bibinfo{volume}{96}},
  \bibinfo{pages}{111101} (\bibinfo{year}{2006}).

\bibitem[{\citenamefont{{Baker} et~al.}(2006)\citenamefont{{Baker},
  {Centrella}, {Choi}, {Koppitz}, and {van Meter}}}]{bcckv06}
\bibinfo{author}{\bibfnamefont{J.~G.} \bibnamefont{{Baker}}},
  \bibinfo{author}{\bibfnamefont{J.}~\bibnamefont{{Centrella}}},
  \bibinfo{author}{\bibfnamefont{D.-I.} \bibnamefont{{Choi}}},
  \bibinfo{author}{\bibfnamefont{M.}~\bibnamefont{{Koppitz}}},
  \bibnamefont{and} \bibinfo{author}{\bibfnamefont{J.}~\bibnamefont{{van
  Meter}}}, \bibinfo{journal}{\prl} \textbf{\bibinfo{volume}{96}},
  \bibinfo{pages}{111102} (\bibinfo{year}{2006}).

\bibitem[{\citenamefont{{Lang} and {Hughes}}(2006)}]{lh06}
\bibinfo{author}{\bibfnamefont{R.~N.} \bibnamefont{{Lang}}} \bibnamefont{and}
  \bibinfo{author}{\bibfnamefont{S.~A.} \bibnamefont{{Hughes}}},
  \bibinfo{journal}{\prd} \textbf{\bibinfo{volume}{74}},
  \bibinfo{pages}{122001} (\bibinfo{year}{2006}).

\bibitem[{\citenamefont{{Lang} and {Hughes}}(2008)}]{lh08}
\bibinfo{author}{\bibfnamefont{R.~N.} \bibnamefont{{Lang}}} \bibnamefont{and}
  \bibinfo{author}{\bibfnamefont{S.~A.} \bibnamefont{{Hughes}}},
  \bibinfo{journal}{\apj} \textbf{\bibinfo{volume}{677}}, \bibinfo{pages}{1184}
  (\bibinfo{year}{2008}).

\bibitem[{\citenamefont{{Lang} et~al.}(2011)\citenamefont{{Lang}, {Hughes}, and
  {Cornish}}}]{lhc11}
\bibinfo{author}{\bibfnamefont{R.~N.} \bibnamefont{{Lang}}},
  \bibinfo{author}{\bibfnamefont{S.~A.} \bibnamefont{{Hughes}}},
  \bibnamefont{and} \bibinfo{author}{\bibfnamefont{N.~J.}
  \bibnamefont{{Cornish}}}, \bibinfo{journal}{\prd}
  \textbf{\bibinfo{volume}{84}}, \bibinfo{eid}{022002} (\bibinfo{year}{2011}).

\bibitem[{\citenamefont{{Holz} and {Hughes}}(2005)}]{hh05}
\bibinfo{author}{\bibfnamefont{D.~E.} \bibnamefont{{Holz}}} \bibnamefont{and}
  \bibinfo{author}{\bibfnamefont{S.~A.} \bibnamefont{{Hughes}}},
  \bibinfo{journal}{\apj} \textbf{\bibinfo{volume}{629}}, \bibinfo{pages}{15}
  (\bibinfo{year}{2005}).

\bibitem[{\citenamefont{{Will}}(2006)}]{w06}
\bibinfo{author}{\bibfnamefont{C.~M.} \bibnamefont{{Will}}},
  \bibinfo{journal}{Living Rev. Relativity} \textbf{\bibinfo{volume}{9}},
  \bibinfo{pages}{3} (\bibinfo{year}{2006}).

\bibitem[{\citenamefont{{Arun} et~al.}(2006)\citenamefont{{Arun}, {Iyer},
  {Qusailah}, and {Sathyaprakash}}}]{aiqs06}
\bibinfo{author}{\bibfnamefont{K.~G.} \bibnamefont{{Arun}}},
  \bibinfo{author}{\bibfnamefont{B.~R.} \bibnamefont{{Iyer}}},
  \bibinfo{author}{\bibfnamefont{M.~S.~S.} \bibnamefont{{Qusailah}}},
  \bibnamefont{and} \bibinfo{author}{\bibfnamefont{B.~S.}
  \bibnamefont{{Sathyaprakash}}}, \bibinfo{journal}{\prd}
  \textbf{\bibinfo{volume}{74}}, \bibinfo{eid}{024006} (\bibinfo{year}{2006}).

\bibitem[{\citenamefont{{Will}}(1994)}]{w94}
\bibinfo{author}{\bibfnamefont{C.~M.} \bibnamefont{{Will}}},
  \bibinfo{journal}{\prd} \textbf{\bibinfo{volume}{50}}, \bibinfo{pages}{6058}
  (\bibinfo{year}{1994}).

\bibitem[{\citenamefont{{Will}}(1998)}]{w98}
\bibinfo{author}{\bibfnamefont{C.~M.} \bibnamefont{{Will}}},
  \bibinfo{journal}{\prd} \textbf{\bibinfo{volume}{57}}, \bibinfo{pages}{2061}
  (\bibinfo{year}{1998}).

\bibitem[{\citenamefont{{Scharre} and {Will}}(2002)}]{sw02}
\bibinfo{author}{\bibfnamefont{P.~D.} \bibnamefont{{Scharre}}}
  \bibnamefont{and} \bibinfo{author}{\bibfnamefont{C.~M.}
  \bibnamefont{{Will}}}, \bibinfo{journal}{\prd} \textbf{\bibinfo{volume}{65}},
  \bibinfo{eid}{042002} (\bibinfo{year}{2002}).

\bibitem[{\citenamefont{{Will} and {Yunes}}(2004)}]{wy04}
\bibinfo{author}{\bibfnamefont{C.~M.} \bibnamefont{{Will}}} \bibnamefont{and}
  \bibinfo{author}{\bibfnamefont{N.}~\bibnamefont{{Yunes}}},
  \bibinfo{journal}{Classical Quantum Gravity} \textbf{\bibinfo{volume}{21}},
  \bibinfo{pages}{4367} (\bibinfo{year}{2004}).

\bibitem[{\citenamefont{{Berti}
  et~al.}(2005{\natexlab{a}})\citenamefont{{Berti}, {Buonanno}, and
  {Will}}}]{bbw05a}
\bibinfo{author}{\bibfnamefont{E.}~\bibnamefont{{Berti}}},
  \bibinfo{author}{\bibfnamefont{A.}~\bibnamefont{{Buonanno}}},
  \bibnamefont{and} \bibinfo{author}{\bibfnamefont{C.~M.}
  \bibnamefont{{Will}}}, \bibinfo{journal}{\prd} \textbf{\bibinfo{volume}{71}},
  \bibinfo{eid}{084025} (\bibinfo{year}{2005}{\natexlab{a}}).

\bibitem[{\citenamefont{{Berti}
  et~al.}(2005{\natexlab{b}})\citenamefont{{Berti}, {Buonanno}, and
  {Will}}}]{bbw05b}
\bibinfo{author}{\bibfnamefont{E.}~\bibnamefont{{Berti}}},
  \bibinfo{author}{\bibfnamefont{A.}~\bibnamefont{{Buonanno}}},
  \bibnamefont{and} \bibinfo{author}{\bibfnamefont{C.~M.}
  \bibnamefont{{Will}}}, \bibinfo{journal}{Classical Quantum Gravity}
  \textbf{\bibinfo{volume}{22}}, \bibinfo{pages}{S943}
  (\bibinfo{year}{2005}{\natexlab{b}}).

\bibitem[{\citenamefont{{Stavridis} and {Will}}(2009)}]{sw09}
\bibinfo{author}{\bibfnamefont{A.}~\bibnamefont{{Stavridis}}} \bibnamefont{and}
  \bibinfo{author}{\bibfnamefont{C.~M.} \bibnamefont{{Will}}},
  \bibinfo{journal}{\prd} \textbf{\bibinfo{volume}{80}}, \bibinfo{eid}{044002}
  (\bibinfo{year}{2009}).

\bibitem[{\citenamefont{{Arun} and {Will}}(2009)}]{aw09}
\bibinfo{author}{\bibfnamefont{K.~G.} \bibnamefont{{Arun}}} \bibnamefont{and}
  \bibinfo{author}{\bibfnamefont{C.~M.} \bibnamefont{{Will}}},
  \bibinfo{journal}{Classical Quantum Gravity} \textbf{\bibinfo{volume}{26}},
  \bibinfo{eid}{155002} (\bibinfo{year}{2009}).

\bibitem[{\citenamefont{{Sopuerta} and {Yunes}}(2009)}]{sy09}
\bibinfo{author}{\bibfnamefont{C.~F.} \bibnamefont{{Sopuerta}}}
  \bibnamefont{and} \bibinfo{author}{\bibfnamefont{N.}~\bibnamefont{{Yunes}}},
  \bibinfo{journal}{\prd} \textbf{\bibinfo{volume}{80}}, \bibinfo{eid}{064006}
  (\bibinfo{year}{2009}).

\bibitem[{\citenamefont{{Yunes} et~al.}(2012)\citenamefont{{Yunes}, {Pani}, and
  {Cardoso}}}]{ypc12}
\bibinfo{author}{\bibfnamefont{N.}~\bibnamefont{{Yunes}}},
  \bibinfo{author}{\bibfnamefont{P.}~\bibnamefont{{Pani}}}, \bibnamefont{and}
  \bibinfo{author}{\bibfnamefont{V.}~\bibnamefont{{Cardoso}}},
  \bibinfo{journal}{\prd} \textbf{\bibinfo{volume}{85}}, \bibinfo{eid}{102003}
  (\bibinfo{year}{2012}).

\bibitem[{\citenamefont{{Berti} et~al.}(2012)\citenamefont{{Berti},
  {Gualtieri}, {Horbatsch}, and {Alsing}}}]{bgha12}
\bibinfo{author}{\bibfnamefont{E.}~\bibnamefont{{Berti}}},
  \bibinfo{author}{\bibfnamefont{L.}~\bibnamefont{{Gualtieri}}},
  \bibinfo{author}{\bibfnamefont{M.}~\bibnamefont{{Horbatsch}}},
  \bibnamefont{and} \bibinfo{author}{\bibfnamefont{J.}~\bibnamefont{{Alsing}}},
  \bibinfo{journal}{\prd} \textbf{\bibinfo{volume}{85}}, \bibinfo{eid}{122005}
  (\bibinfo{year}{2012}).

\bibitem[{\citenamefont{{Yunes} and {Pretorius}}(2009)}]{yp09}
\bibinfo{author}{\bibfnamefont{N.}~\bibnamefont{{Yunes}}} \bibnamefont{and}
  \bibinfo{author}{\bibfnamefont{F.}~\bibnamefont{{Pretorius}}},
  \bibinfo{journal}{\prd} \textbf{\bibinfo{volume}{80}}, \bibinfo{eid}{122003}
  (\bibinfo{year}{2009}).

\bibitem[{\citenamefont{{Mishra} et~al.}(2010)\citenamefont{{Mishra}, {Arun},
  {Iyer}, and {Sathyaprakash}}}]{mais10}
\bibinfo{author}{\bibfnamefont{C.~K.} \bibnamefont{{Mishra}}},
  \bibinfo{author}{\bibfnamefont{K.~G.} \bibnamefont{{Arun}}},
  \bibinfo{author}{\bibfnamefont{B.~R.} \bibnamefont{{Iyer}}},
  \bibnamefont{and} \bibinfo{author}{\bibfnamefont{B.~S.}
  \bibnamefont{{Sathyaprakash}}}, \bibinfo{journal}{\prd}
  \textbf{\bibinfo{volume}{82}}, \bibinfo{eid}{064010} (\bibinfo{year}{2010}).

\bibitem[{\citenamefont{{Mirshekari} et~al.}(2012)\citenamefont{{Mirshekari},
  {Yunes}, and {Will}}}]{myw12}
\bibinfo{author}{\bibfnamefont{S.}~\bibnamefont{{Mirshekari}}},
  \bibinfo{author}{\bibfnamefont{N.}~\bibnamefont{{Yunes}}}, \bibnamefont{and}
  \bibinfo{author}{\bibfnamefont{C.~M.} \bibnamefont{{Will}}},
  \bibinfo{journal}{\prd} \textbf{\bibinfo{volume}{85}}, \bibinfo{eid}{024041}
  (\bibinfo{year}{2012}).

\bibitem[{\citenamefont{{Arun}}(2012)}]{a12}
\bibinfo{author}{\bibfnamefont{K.~G.} \bibnamefont{{Arun}}},
  \bibinfo{journal}{Classical Quantum Gravity} \textbf{\bibinfo{volume}{29}},
  \bibinfo{eid}{075011} (\bibinfo{year}{2012}).

\bibitem[{\citenamefont{{Will}}(1993)}]{w93}
\bibinfo{author}{\bibfnamefont{C.~M.} \bibnamefont{{Will}}},
  \emph{\bibinfo{title}{{Theory and Experiment in Gravitational Physics}}}
  (\bibinfo{publisher}{Cambridge University Press},
  \bibinfo{address}{Cambridge, UK}, \bibinfo{year}{1993}).

\bibitem[{\citenamefont{{Fujii} and {Maeda}}(2003)}]{fm03}
\bibinfo{author}{\bibfnamefont{Y.}~\bibnamefont{{Fujii}}} \bibnamefont{and}
  \bibinfo{author}{\bibfnamefont{K.-I.} \bibnamefont{{Maeda}}},
  \emph{\bibinfo{title}{{The Scalar-Tensor Theory of Gravitation}}}
  (\bibinfo{publisher}{Cambridge University Press},
  \bibinfo{address}{Cambridge, UK}, \bibinfo{year}{2003}).

\bibitem[{\citenamefont{{de Felice} and {Tsujikawa}}(2010)}]{dt10}
\bibinfo{author}{\bibfnamefont{A.}~\bibnamefont{{de Felice}}} \bibnamefont{and}
  \bibinfo{author}{\bibfnamefont{S.}~\bibnamefont{{Tsujikawa}}},
  \bibinfo{journal}{Living Rev. Relativity} \textbf{\bibinfo{volume}{13}},
  \bibinfo{pages}{3} (\bibinfo{year}{2010}).

\bibitem[{\citenamefont{{Damour} and {Esposito-Far{\`e}se}}(1992)}]{de92}
\bibinfo{author}{\bibfnamefont{T.}~\bibnamefont{{Damour}}} \bibnamefont{and}
  \bibinfo{author}{\bibfnamefont{G.}~\bibnamefont{{Esposito-Far{\`e}se}}},
  \bibinfo{journal}{Classical Quantum Gravity} \textbf{\bibinfo{volume}{9}},
  \bibinfo{pages}{2093} (\bibinfo{year}{1992}).

\bibitem[{\citenamefont{{Mirshekari} and {Will}}(2013)}]{mw13}
\bibinfo{author}{\bibfnamefont{S.}~\bibnamefont{{Mirshekari}}}
  \bibnamefont{and} \bibinfo{author}{\bibfnamefont{C.~M.}
  \bibnamefont{{Will}}}, \bibinfo{journal}{\prd} \textbf{\bibinfo{volume}{87}},
  \bibinfo{eid}{084070} (\bibinfo{year}{2013}).

\bibitem[{\citenamefont{{Epstein} and {Wagoner}}(1975)}]{ew75}
\bibinfo{author}{\bibfnamefont{R.}~\bibnamefont{{Epstein}}} \bibnamefont{and}
  \bibinfo{author}{\bibfnamefont{R.~V.} \bibnamefont{{Wagoner}}},
  \bibinfo{journal}{\apj} \textbf{\bibinfo{volume}{197}}, \bibinfo{pages}{717}
  (\bibinfo{year}{1975}).

\bibitem[{\citenamefont{{Wiseman}}(1992)}]{w92}
\bibinfo{author}{\bibfnamefont{A.~G.} \bibnamefont{{Wiseman}}},
  \bibinfo{journal}{\prd} \textbf{\bibinfo{volume}{46}}, \bibinfo{pages}{1517}
  (\bibinfo{year}{1992}).

\bibitem[{\citenamefont{{Will} and {Wiseman}}(1996)}]{ww96}
\bibinfo{author}{\bibfnamefont{C.~M.} \bibnamefont{{Will}}} \bibnamefont{and}
  \bibinfo{author}{\bibfnamefont{A.~G.} \bibnamefont{{Wiseman}}},
  \bibinfo{journal}{\prd} \textbf{\bibinfo{volume}{54}}, \bibinfo{pages}{4813}
  (\bibinfo{year}{1996}).

\bibitem[{\citenamefont{{Pati} and {Will}}(2000)}]{pw00}
\bibinfo{author}{\bibfnamefont{M.~E.} \bibnamefont{{Pati}}} \bibnamefont{and}
  \bibinfo{author}{\bibfnamefont{C.~M.} \bibnamefont{{Will}}},
  \bibinfo{journal}{\prd} \textbf{\bibinfo{volume}{62}}, \bibinfo{eid}{124015}
  (\bibinfo{year}{2000}).

\bibitem[{\citenamefont{{Pati} and {Will}}(2002)}]{pw02}
\bibinfo{author}{\bibfnamefont{M.~E.} \bibnamefont{{Pati}}} \bibnamefont{and}
  \bibinfo{author}{\bibfnamefont{C.~M.} \bibnamefont{{Will}}},
  \bibinfo{journal}{\prd} \textbf{\bibinfo{volume}{65}}, \bibinfo{eid}{104008}
  (\bibinfo{year}{2002}).

\bibitem[{\citenamefont{{Jaranowski} and {Sch{\"a}fer}}(1998)}]{js98}
\bibinfo{author}{\bibfnamefont{P.}~\bibnamefont{{Jaranowski}}}
  \bibnamefont{and}
  \bibinfo{author}{\bibfnamefont{G.}~\bibnamefont{{Sch{\"a}fer}}},
  \bibinfo{journal}{\prd} \textbf{\bibinfo{volume}{57}}, \bibinfo{pages}{7274}
  (\bibinfo{year}{1998}).

\bibitem[{\citenamefont{{Futamase} and {Itoh}}(2007)}]{fi07}
\bibinfo{author}{\bibfnamefont{T.}~\bibnamefont{{Futamase}}} \bibnamefont{and}
  \bibinfo{author}{\bibfnamefont{Y.}~\bibnamefont{{Itoh}}},
  \bibinfo{journal}{Living Rev. Relativity} \textbf{\bibinfo{volume}{10}},
  \bibinfo{pages}{2} (\bibinfo{year}{2007}).

\bibitem[{\citenamefont{{Goldberger} and {Rothstein}}(2006)}]{gr06}
\bibinfo{author}{\bibfnamefont{W.~D.} \bibnamefont{{Goldberger}}}
  \bibnamefont{and} \bibinfo{author}{\bibfnamefont{I.~Z.}
  \bibnamefont{{Rothstein}}}, \bibinfo{journal}{\prd}
  \textbf{\bibinfo{volume}{73}}, \bibinfo{eid}{104029} (\bibinfo{year}{2006}).

\bibitem[{\citenamefont{{Eardley}}(1975)}]{e75}
\bibinfo{author}{\bibfnamefont{D.~M.} \bibnamefont{{Eardley}}},
  \bibinfo{journal}{\apj} \textbf{\bibinfo{volume}{196}}, \bibinfo{pages}{L59}
  (\bibinfo{year}{1975}).

\bibitem[{\citenamefont{{Will} and {Zaglauer}}(1989)}]{wz89}
\bibinfo{author}{\bibfnamefont{C.~M.} \bibnamefont{{Will}}} \bibnamefont{and}
  \bibinfo{author}{\bibfnamefont{H.~W.} \bibnamefont{{Zaglauer}}},
  \bibinfo{journal}{\apj} \textbf{\bibinfo{volume}{346}}, \bibinfo{pages}{366}
  (\bibinfo{year}{1989}).

\bibitem[{\citenamefont{{Zaglauer}}(1992)}]{z92}
\bibinfo{author}{\bibfnamefont{H.~W.} \bibnamefont{{Zaglauer}}},
  \bibinfo{journal}{\apj} \textbf{\bibinfo{volume}{393}}, \bibinfo{pages}{685}
  (\bibinfo{year}{1992}).

\bibitem[{\citenamefont{{Barausse} et~al.}(2013)\citenamefont{{Barausse},
  {Palenzuela}, {Ponce}, and {Lehner}}}]{bppl13}
\bibinfo{author}{\bibfnamefont{E.}~\bibnamefont{{Barausse}}},
  \bibinfo{author}{\bibfnamefont{C.}~\bibnamefont{{Palenzuela}}},
  \bibinfo{author}{\bibfnamefont{M.}~\bibnamefont{{Ponce}}}, \bibnamefont{and}
  \bibinfo{author}{\bibfnamefont{L.}~\bibnamefont{{Lehner}}},
  \bibinfo{journal}{\prd} \textbf{\bibinfo{volume}{87}}, \bibinfo{eid}{081506}
  (\bibinfo{year}{2013}).

\bibitem[{\citenamefont{{Palenzuela} et~al.}(2014)\citenamefont{{Palenzuela},
  {Barausse}, {Ponce}, and {Lehner}}}]{pbpl13}
\bibinfo{author}{\bibfnamefont{C.}~\bibnamefont{{Palenzuela}}},
  \bibinfo{author}{\bibfnamefont{E.}~\bibnamefont{{Barausse}}},
  \bibinfo{author}{\bibfnamefont{M.}~\bibnamefont{{Ponce}}}, \bibnamefont{and}
  \bibinfo{author}{\bibfnamefont{L.}~\bibnamefont{{Lehner}}},
  \bibinfo{journal}{\prd} \textbf{\bibinfo{volume}{89}}, \bibinfo{eid}{044024}
  (\bibinfo{year}{2014}).

\bibitem[{\citenamefont{{Shibata} et~al.}()\citenamefont{{Shibata},
  {Taniguchi}, {Okawa}, and {Buonanno}}}]{stob13}
\bibinfo{author}{\bibfnamefont{M.}~\bibnamefont{{Shibata}}},
  \bibinfo{author}{\bibfnamefont{K.}~\bibnamefont{{Taniguchi}}},
  \bibinfo{author}{\bibfnamefont{H.}~\bibnamefont{{Okawa}}}, \bibnamefont{and}
  \bibinfo{author}{\bibfnamefont{A.}~\bibnamefont{{Buonanno}}},
  \bibinfo{note}{arXiv:1310.0627}.

\bibitem[{\citenamefont{{Damour} and {Esposito-Far{\`e}se}}(1993)}]{de93}
\bibinfo{author}{\bibfnamefont{T.}~\bibnamefont{{Damour}}} \bibnamefont{and}
  \bibinfo{author}{\bibfnamefont{G.}~\bibnamefont{{Esposito-Far{\`e}se}}},
  \bibinfo{journal}{\prl} \textbf{\bibinfo{volume}{70}}, \bibinfo{pages}{2220}
  (\bibinfo{year}{1993}).

\bibitem[{\citenamefont{{Damour} and {Esposito-Far{\`e}se}}(1996)}]{de96}
\bibinfo{author}{\bibfnamefont{T.}~\bibnamefont{{Damour}}} \bibnamefont{and}
  \bibinfo{author}{\bibfnamefont{G.}~\bibnamefont{{Esposito-Far{\`e}se}}},
  \bibinfo{journal}{\prd} \textbf{\bibinfo{volume}{54}}, \bibinfo{pages}{1474}
  (\bibinfo{year}{1996}).

\bibitem[{\citenamefont{{Payne}}(1983)}]{p83}
\bibinfo{author}{\bibfnamefont{P.~N.} \bibnamefont{{Payne}}},
  \bibinfo{journal}{\prd} \textbf{\bibinfo{volume}{28}}, \bibinfo{pages}{1894}
  (\bibinfo{year}{1983}).

\bibitem[{\citenamefont{{Christodoulou}}(1991)}]{c91}
\bibinfo{author}{\bibfnamefont{D.}~\bibnamefont{{Christodoulou}}},
  \bibinfo{journal}{\prl} \textbf{\bibinfo{volume}{67}}, \bibinfo{pages}{1486}
  (\bibinfo{year}{1991}).

\bibitem[{\citenamefont{{Blanchet} and {Damour}}(1992)}]{bd92}
\bibinfo{author}{\bibfnamefont{L.}~\bibnamefont{{Blanchet}}} \bibnamefont{and}
  \bibinfo{author}{\bibfnamefont{T.}~\bibnamefont{{Damour}}},
  \bibinfo{journal}{\prd} \textbf{\bibinfo{volume}{46}}, \bibinfo{pages}{4304}
  (\bibinfo{year}{1992}).

\bibitem[{\citenamefont{{Favata}}(2009{\natexlab{a}})}]{f09}
\bibinfo{author}{\bibfnamefont{M.}~\bibnamefont{{Favata}}},
  \bibinfo{journal}{\prd} \textbf{\bibinfo{volume}{80}}, \bibinfo{eid}{024002}
  (\bibinfo{year}{2009}{\natexlab{a}}).

\bibitem[{\citenamefont{{Favata}}(2009{\natexlab{b}})}]{f09b}
\bibinfo{author}{\bibfnamefont{M.}~\bibnamefont{{Favata}}},
  \bibinfo{journal}{\apj} \textbf{\bibinfo{volume}{696}}, \bibinfo{pages}{L159}
  (\bibinfo{year}{2009}{\natexlab{b}}).

\bibitem[{\citenamefont{{Blanchet} and {Damour}}(1989)}]{bd89}
\bibinfo{author}{\bibfnamefont{L.}~\bibnamefont{{Blanchet}}} \bibnamefont{and}
  \bibinfo{author}{\bibfnamefont{T.}~\bibnamefont{{Damour}}},
  \bibinfo{journal}{Ann. Inst. H. Poincar\'{e}, Phys. Theor.}
  \textbf{\bibinfo{volume}{50}}, \bibinfo{pages}{377} (\bibinfo{year}{1989}).

\bibitem[{\citenamefont{{Alsing} et~al.}(2012)\citenamefont{{Alsing}, {Berti},
  {Will}, and {Zaglauer}}}]{abwz12}
\bibinfo{author}{\bibfnamefont{J.}~\bibnamefont{{Alsing}}},
  \bibinfo{author}{\bibfnamefont{E.}~\bibnamefont{{Berti}}},
  \bibinfo{author}{\bibfnamefont{C.~M.} \bibnamefont{{Will}}},
  \bibnamefont{and}
  \bibinfo{author}{\bibfnamefont{H.}~\bibnamefont{{Zaglauer}}},
  \bibinfo{journal}{\prd} \textbf{\bibinfo{volume}{85}}, \bibinfo{eid}{064041}
  (\bibinfo{year}{2012}).

\bibitem[{\citenamefont{{Arun} et~al.}(2004)\citenamefont{{Arun}, {Blanchet},
  {Iyer}, and {Qusailah}}}]{abiq04}
\bibinfo{author}{\bibfnamefont{K.~G.} \bibnamefont{{Arun}}},
  \bibinfo{author}{\bibfnamefont{L.}~\bibnamefont{{Blanchet}}},
  \bibinfo{author}{\bibfnamefont{B.~R.} \bibnamefont{{Iyer}}},
  \bibnamefont{and} \bibinfo{author}{\bibfnamefont{M.~S.~S.}
  \bibnamefont{{Qusailah}}}, \bibinfo{journal}{Classical Quantum Gravity}
  \textbf{\bibinfo{volume}{21}}, \bibinfo{pages}{3771} (\bibinfo{year}{2004}).

\bibitem[{\citenamefont{{Hawking}}(1972)}]{h72}
\bibinfo{author}{\bibfnamefont{S.~W.} \bibnamefont{{Hawking}}},
  \bibinfo{journal}{Commun. Math. Phys.} \textbf{\bibinfo{volume}{25}},
  \bibinfo{pages}{167} (\bibinfo{year}{1972}).

\bibitem[{\citenamefont{{Healy} et~al.}(2012)\citenamefont{{Healy}, {Bode},
  {Haas}, {Pazos}, {Laguna}, {Shoemaker}, and {Yunes}}}]{h11}
\bibinfo{author}{\bibfnamefont{J.}~\bibnamefont{{Healy}}},
  \bibinfo{author}{\bibfnamefont{T.}~\bibnamefont{{Bode}}},
  \bibinfo{author}{\bibfnamefont{R.}~\bibnamefont{{Haas}}},
  \bibinfo{author}{\bibfnamefont{E.}~\bibnamefont{{Pazos}}},
  \bibinfo{author}{\bibfnamefont{P.}~\bibnamefont{{Laguna}}},
  \bibinfo{author}{\bibfnamefont{D.~M.} \bibnamefont{{Shoemaker}}},
  \bibnamefont{and} \bibinfo{author}{\bibfnamefont{N.}~\bibnamefont{{Yunes}}},
  \bibinfo{journal}{Classical Quantum Gravity} \textbf{\bibinfo{volume}{29}},
  \bibinfo{pages}{232002} (\bibinfo{year}{2012}).

\bibitem[{\citenamefont{{Jacobson}}(1999)}]{j99}
\bibinfo{author}{\bibfnamefont{T.}~\bibnamefont{{Jacobson}}},
  \bibinfo{journal}{\prl} \textbf{\bibinfo{volume}{83}}, \bibinfo{pages}{2699}
  (\bibinfo{year}{1999}).

\bibitem[{\citenamefont{{Horbatsch} and {Burgess}}()}]{hb12}
\bibinfo{author}{\bibfnamefont{M.~W.} \bibnamefont{{Horbatsch}}}
  \bibnamefont{and} \bibinfo{author}{\bibfnamefont{C.~P.}
  \bibnamefont{{Burgess}}}, \bibinfo{note}{{J. Cosmol. Astropart. Phys. 05
  (2012) 010}}.

\end{thebibliography}

\end{document}